\documentclass[11pt, fleqn]{paper}

\usepackage[utf8]{inputenc}
\usepackage[T1]{fontenc}

\usepackage[a4paper,margin=2.5cm]{geometry}

\usepackage{amsmath, amsfonts}
\usepackage{tensor, mathrsfs, upgreek, stmaryrd}
\SetSymbolFont{stmry}{bold}{U}{stmry}{m}{n}

\usepackage[hidelinks]{hyperref}
\usepackage{cleveref}

\usepackage[backref=true,sorting=none,url=false,backend=biber,sortcites=true,style=numeric-comp]{biblatex}

\bibliography{bibliography}

\numberwithin{equation}{section}
\usepackage{mathtools}
	\mathtoolsset{centercolon}
\usepackage[onehalfspacing]{setspace}
\usepackage{booktabs}

\usepackage{lmodern}

\newcommand{\checkedtogether}[1]{}

\def\GW{\textsc{gw}}
\def\GWs{\textsc{gw}'s}
\def\TT{\textsc{tt}}

\def\half{\tfrac{1}{2}}
\def\ihalf{\tfrac{i}{2}}
\def\quarter{\tfrac{1}{4}}

\let\t\tensor
\let\p\partial
\let\wop\square

\def\gs{\mathrm g}	
\def\dd{\mathrm{d}}	

\def\pl{\parallel}
\def\xp{x_\pl}
\def\Zz{{\mathcal Z}}
\def\Rho{{\mathrm P}}

\newcommand*{\bessel}[2][]{\mathrm{B}^{#1}_{#2}}

\def\o#1{{(#1)}}

\def\omegaG{\omega_g}

\def\Fb{{\bar{F}}}	
\def\permeability{\upmu}
\def\permittivity{\upepsilon}

\def\k{{\hat\kappa}}

\def\f{\mathfrak e}

\newcommand*{\jump}[1]{\left\llbracket{#1}\right\rrbracket}

\DeclareMathOperator{\trace}{tr}
\DeclareMathOperator{\adjugate}{adj}
\DeclareMathOperator{\diag}{diag}

\let\bs\boldsymbol

\newcommand{\HE}[2]{\text{HE}_{#1,#2}}
\newcommand{\LP}[3]{\text{LP}_{#1,#2,#3}}
\newcommand{\LPp}[3]{\widetilde{\text{LP}}_{#1,#2,#3}}

\begin{document}
\pagestyle{plain}
\title{The Response of Optical Fibres to Gravitational Waves}
\author{\upshape Thomas B. Mieling\thanks{
	\textsc{orcid:} \href{https://orcid.org/0000-0002-6905-0183}{0000-0002-6905-0183},
	{email:} \href{mailto:thomas.mieling@univie.ac.at}{thomas.mieling@univie.ac.at}
	}
}
\date{\today}
\institution{University of Vienna, Faculty of Physics, Vienna, Austria,\\
TURIS Research Platform, University of Vienna, Austria}
\maketitle\vspace{-1.2em} 
\noindent\today

\begin{abstract}
	The response of optical fibre modes to plane gravitational waves of low frequency is computed.
	By solving perturbatively the Maxwell equations for step-index optical fibres in a gravitational wave background and implementing appropriate boundary conditions to describe single-mode fibres, explicit formulae for the perturbations of the phase and the polarisation of the fibre modes are obtained.
\end{abstract}

\smalltableofcontents

\section{Introduction and Summary}

While free space laser interferometric detectors of gravitational waves (\GWs) such as \textsc{ligo} and \textsc{virgo} are mainly analysed in the low-frequency regime $\omegaG \ll \omega$ where the \GW\ frequency is much smaller than the laser frequency (using either the geodesic deviation equation \cite{Forward1978, Schutz1987, Faraoni2007}, geometrical optics \cite{Sazhin1997, Rudenko1980, Finn2009, Rakhmanov2009, Cornish2009, Koop2014} or Maxwell’s equations directly \cite{Cooperstock1968, Lobo1992, Cooperstock1993}), \GW\ detectors employing electromagnetic waveguides are typically discussed in the high-frequency regime \cite{Braginsky1972, Lobo1992, Cruise2000, Tang2000, Mensky2009}.
In this work, we bring together aspects of both of these kinds of analyses by studying an electromagnetic waveguide in the low-frequency regime.

One of our main motivations for such calculations is to study the effect of \GWs\ on light in regimes which are beyond a description in terms of geometrical optics. While such descriptions suffice to describe light propagation in vacuum or in dielectrics of large extensions (generally, whenever the wavelength is small compared to all other relevant length scales), light propagating in waveguides is only consistently described by wave optics, since the diameter of optical fibres is comparable to the optical wavelength. As a consequence, polarisation-dependent effects are not necessarily as strongly suppressed as in geometrical optics, where the transport equation for the polarisation arises only at second order in the $1/\lambda$-expansion, see e.g.\ the general discussion in Ref.~\cite[Chapter III]{Born2019}.

Further motivation for such investigations is provided by recent analyses of the perturbation of fibre-modes by Earth’s gravitational field \cite{Beig_2018} and by Earth’s rotation \cite{Mieling2020}. These calculations assume time-independent metrics, and the study of the influence of \GWs\ illustrates how optical fibres react to time-dependent perturbations of the space-time metric.

Of course, there are still many technical challenges still to be overcome before fibre-optic interferometers could be used in experimental gravitational wave detection, see e.g.\ the discussion in Ref.~\cite[Appendix~B]{Linsay1983}.
Apart from the limited power handling capacity of optical fibres due to intensity damage and the emergence of non-linear effects at high power levels, the main obstructions seem to be shot noise (which is much more pronounced than in free-space interferometers due to fibre attenuation, cf.\ Refs.~\cites[Sect.~5.1]{Saulson1994}[Sect.~2.2 and 4.2]{Adhikari2004}), as well as thermal and mechanical noises inside the fibre (both of which are absent in free-space interferometers).
As these noise sources are more pronounced for long interferometer arms, we expect the detection of \GWs\ in fibre-optic interferometers to be viable only for signals of higher frequencies than those measured in \textsc{ligo} and Virgo (which are in the frequency range of roughly $30\,\text{to}\,250\,\mathrm{Hz}$, see e.g.\ Refs.~\cite{Abbott2016, Abbott2017, Abbott2017a, Abbott2020}).
Consequently, it is to be expected that fibre-optic interferometers would be most suitable for searching for continuous signals (recent searches for continuous \GWs\ in \textsc{ligo} data can be found e.g.\ in Refs.~\cite{Abbott2019, Dergachev2020a, Dergachev2020b}).
Moreover, since such interferometers can be made much smaller than the hitherto constructed free-space interferometers, it is conceivable to construct fibre-optic interferometers of multiple arms, providing more detailed position information on the \GW\ source than the standard two-arm Michelson interferometers.
We plan to assess the experimental feasibility of such fibre-optic interferometers in the near future.

To summarise our findings: we consider a gravitational wave in transverse-traceless gauge of the form
\begin{equation}
	\t g{_\mu_\nu}
		= \t \eta{_\mu_\nu} + \varepsilon \t A{_\mu_\nu} \cos(\omegaG(t-\k_i x^i))\,,
\end{equation}
propagating in an arbitrary direction $\k$. (Here, we restrict the discussion to monochromatic plane waves. For arbitrary waveforms, one may use the Fourier transform and apply the analysis provided here to every Fourier component separately.)
Within this \GW\ background, we solve perturbatively the Maxwell equations for (single-mode) step-index optical fibres and obtain the following result.
Choosing the $z$-axis to be the symmetry axis of the waveguide, we find the  perturbed optical phase for fibres much shorter than the \GW\ wavelength ($\omegaG z \ll 1$) to be
\begin{equation}
	\label{eq:summary phase}
	\psi
		= \omega t - \beta z + m \theta
		- \half \varepsilon c_1 \t A{_z_z}\, \omega z \cos(\omegaG t) \,,
\end{equation}
where $\omega$ is the angular frequency of the mode, $\beta$ the propagation constant and $m$ the azimuthal mode index.
The Jones matrix, describing the perturbation of the polarisation (see e.g.\ Refs.~\cites[Chap.~3]{Hodgson2005}[Sect.~1.4]{Damask2004} for an introduction to the Jones formalism), is found to be
\begin{equation}
	\label{eq:summary polarisation}
	M
		= \mathbf 1 - i \varepsilon c_2 \varDelta^2\, \omega z \cos(\omegaG t) \begin{pmatrix}
			+ \half (\t A{_x_x} - \t A{_y_y})	&
			\t A{_x_y}						\\
			\t A{_x_y}						&
			- \half (\t A{_x_x} - \t A{_y_y})
		\end{pmatrix}\,,
\end{equation}
where $\varDelta = (n_1 - n_2)/n_1$ is the relative difference of the refractive indices in the core ($n_1$) and the cladding ($n_2$).
The coefficients $c_1$ and $c_2$ arising here are computed numerically for typical single-mode fibres.
We find that $c_1 \approx \bar n$, where $\bar n$ is the (frequency dependent) effective refractive index of the fibre, and $c_2 \approx 0.3$.
The diagonal terms (of opposite sign) describe birefringence, while the off-diagonal terms (of equal sign) describe a deformation of linear polarisation to elliptic polarisation.

\begin{figure}[ht!]
	\centering
	\includegraphics[width=0.5\columnwidth]{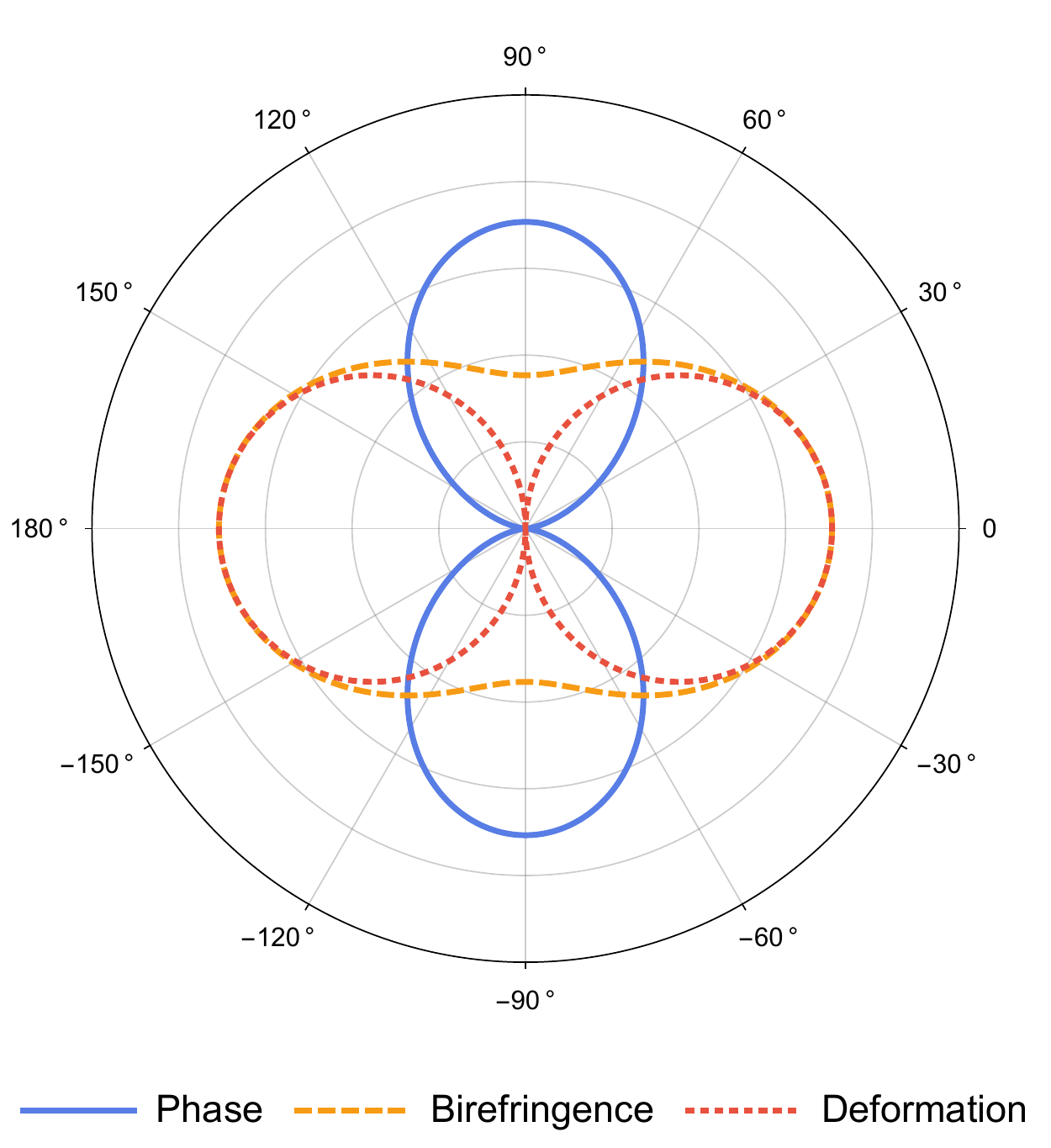}
	\caption{Dependence of three terms $\t A{_z_z}$, $\half(\t A{_x_x}-\t A{_y_y})$ and $\t A{_x_y}$ (in absolute value) on the angle $\vartheta$ formed by the optical and the gravitational wave-vectors.
	The blue curve (“phase”) describes the angular dependence of the overall phase shift, the dashed orange line (“birefringence”) describes the difference in phase shift between the two polarisation states of light, and the dotted red curve (“deformation”) quantifies the mixing of the polarisation states.}
	\label{fig:angular dependence 0}
\end{figure}

\Cref{fig:angular dependence 0} shows how these three effects (phase shift, birefringence and polarisation deformation) depend on the angle $\vartheta$ between the direction of light propagation and the direction of the gravitational wave. The precise dependence on this angle and on the polarisation of the gravitational wave is discussed in detail in \Cref{s:numerical examples}.

It is seen that the phase shift is maximal when the gravitational wave propagates orthogonally to the fibre axis, and vanishes when it propagates collinearly with the light ray (i.e.\ parallel or antiparallel). For collinear propagation, the polarisation-dependent effects are maximal: the birefringence effect is then sensitive to the $+$ polarisation of the gravitational wave, while the deformation effect is sensitive to the $\times$ polarisation.

\section{Statement of the Problem}

Consider the gravitational wave (\GW) metric
\begin{equation}
	\label{eq:metric g}
	\t g{_\mu_\nu}
		= \t \eta{_\mu_\nu}
		+ \varepsilon \t A{_\mu_\nu} \cos(\kappa.x)\,,
\end{equation}
where $\kappa.x = \t \kappa{_\mu} \t x{^\mu}$, $\kappa = \omegaG(\dd t - \k_i \dd x^i)$ with $\k_i = \k^i$ being a “spatial” unit vector (normalised with respect to the unperturbed metric) determining the direction of wave propagation, $\varepsilon$ is the amplitude of the gravitational wave (we assume $\varepsilon \ll 1$), and $A$ is a symmetric matrix of constant entries which is transverse ($\t A{_\mu_0} = \t A{_\mu_i} \t\k{^i} = 0$) and traceless (${\t \eta{^\mu^\nu} \t A{_\mu_\nu} = 0}$), so we are using \TT\ coordinates.

We wish to describe the propagation of light in a cylindrical waveguide at rest in this coordinate system.
More precisely, we consider a cylindrical step-index fibre consisting of a linear dielectric with constant refractive index $n_1$ in the core and $n_2$ in the cladding (with $n_1 > n_2$). Such waveguides are typically non-magnetic (i.e.\ of permeability $\permeability = 1$) and hence the permittivity is $\permittivity = \sqrt{n}$.

We will neglect boundary effects from the ends of the waveguide and the outer boundary of the cladding (the waveguide is thus modelled as infinitely long and the cladding as infinitely thick), and we shall neglect deformations of the fibre, so that in cylindrical coordinates $r, \theta, z$ (related to the above \TT\ coordinates as in flat space) the core-cladding interface is located at $r = \rho$.
A more accurate model would have to take into account the perturbation of the spatial metric when formulating the boundary conditions, but at such a level of accuracy we estimate that one would also have to consider elastic deformations of the waveguide due to the gravitational wave, which is beyond the scope of this article.

Being concerned with \emph{weak} gravitational waves of \emph{low} frequency, we work in a perturbative setting with two expansion parameters
\begin{align}
	\varepsilon &\ll 1\,,
	&
	\Omega &:= \omegaG/\omega \ll 1\,,
\end{align}
where $\varepsilon$ is the \GW\ amplitude, $\omegaG$ its frequency, and $\omega$ is the laser frequency.

\section{Decomposition of the  Maxwell Equations}

In this section, we write the Maxwell equations in the \GW\ background \eqref{eq:metric g} in a 3+1 form, following Ref.~\cite{Beig_2018} with $c = 1$.
To this end, set
\begin{equation}
	\t \gs{_i_j}
		= \t g{_i_j}
		= \t \delta{_i_j} + \varepsilon \t A{_i_j} \cos(\kappa.x)
		\,,
\end{equation}
which is the time-dependent “spatial” metric induced on the hypersurfaces of constant time $t$, and denote by $\nabla$ the associated (spatial) Levi-Civita derivative.

Since $\t g{_0_0} = - 1 + O(\varepsilon^2)$ and $\p_0 (\det \gs) = O(\varepsilon^2)$, the space-time split of the Faraday two-form $F$ and the displacement bivector $\Fb$ given in Ref.~\cite{Beig_2018} yields
\begin{align}
	\t D{^i}
		&= \t \Fb{^0^i}\,,
	&
	\t H{_i}
		&= \half \t \epsilon{_i_j_k} \t \Fb{^j^k}\,,
	&
	\t E{_i}
		&= \t F{_i_0}\,,
	&
	\t B{^i}
		&= \half \t \epsilon{^i^j^k} \t F{_j_k}\,,
\end{align}
where $\epsilon$ is the spatial volume form (not to be confused with the gravitational wave amplitude $\varepsilon$).
Here and henceforward, we neglect terms of second and higher order in $\varepsilon$ and, for brevity, we will not write the error term $O(\varepsilon^2)$ explicitly.
The field equations then take the standard form
\begin{align}
	\label{eq:Maxwell standard form}
	\t {\dot B}{^i} + \t \epsilon{^i^j^k} \t \nabla{_j} \t E{_k} &= 0\,,
	&
	\t {\dot D}{^i} - \t \epsilon{^i^j^k} \t \nabla{_j} \t H{_k} &= 0\,,
	&
	\t \nabla{_i} \t B{^i} &= 0\,,
	&
	\t \nabla{_i} \t D{^i} &= 0\,,
\end{align}
where an overset dot indicates a time derivative.
Following Ref.~\cite{Gordon1923}, we define the optical metric
\begin{equation}
	\label{eq:optical metric def}
	\t \gamma{^\mu^\nu}
		= \t g{^\mu^\nu}
		+ (1 - n^2) \t u{^\mu} \t u{^\nu}\,,
\end{equation}
where $u$ is the four-velocity of the dielectric and $n$ its refractive index.
This allows to write the constitutive equation in the form
\begin{equation}
	\permeability \t \Fb{^\alpha^\beta}
		= \t \gamma{^\alpha^\rho} \t \gamma{^\beta^\sigma} \t F{_\rho_\sigma}\,.
\end{equation}
In the considered case, we have $\t u{^\mu} = \t* \delta{^\mu_0}$ and
\begin{equation}
	\t g{^\mu^\nu}
		= \t \eta{^\mu^\nu}
		- \varepsilon \t A{^\mu^\nu} \cos(\kappa.x)\,,
\end{equation}
where the indices of $A$ were raised with the unperturbed metric $\eta$.
The constitutive equations then reduce to
\begin{align}
	\t D{^i}
		&= \permittivity \t\gs{^i^j} \t E{_j}\,,
	&
	\t B{^i}
		&= \permeability \t\gs{^i^j} \t H{_j}\,.
\end{align}
As the spatial metric is perturbed by the gravitational wave, this is the only part where the field equations differ from those of flat space.

To further simplify the equations, we define the Riemann-Silberstein vector
\begin{equation}
	\label{eq:complex field def}
	\t Z{^i} := \permeability \t D{^i} + j n \t B{^i}\,,
\end{equation}
where $j$ is an imaginary unit ($j^2 = -1$), such that the Maxwell equations take the form
\begin{align}
	\label{eq:maxwell evolution covariant}
	n \t {\dot Z}{^i}
		+ j \t \epsilon{^i^j^k} \t \nabla{_j}( \t \gs{_k_l} \t Z{^l})
		= 0\,,
		\\
	\label{eq:maxwell constraint covariant}
	\t \nabla{_i} \t Z{^i} = 0\,,
\end{align}
wherever $n$ is locally constant.

In the following calculations, we will use \emph{two} independent (commuting) imaginary units $i$ and $j$, where $i$ is reserved for the usual complex description of waves (i.e.\ the physical fields are the $i$-real parts) and $j$ is used for this compact formulation of the field equations.

As a side remark: the algebraic structure here is that of the complex vector space $\mathbf{C}^2$ with $1$ identified with $(1,0)$ and $j$ identified with $(0,1)$, and the product $(a,b) \times (c,d) := (ac -bd, ad+bc)$, where $a,b,c,d$ are usual complex numbers. In particular, the product $ij = ji$ is represented by $(i,0)\times (0,1) = (0, i)$, which is not equal to $-1$, here identified with $(-1, 0)$.
\checkedtogether{14.8.}

Using only a single complex unit, it would not be possible to use this complex description of waves while simultaneously encoding the entire electromagnetic field in a single vector field. Indeed, if $D$ and $B$ are complex, then a field of the form $Z' = \permeability D + i n B$ would only provide incomplete information about $D$ and $B$, as it mixes the real part of $D$ with the imaginary part of $B$, and vice versa. To recover those two fields entirely, one would have to use a second complex field $Z'' = \permeability D - i n B$. In contrast, the field $Z$ defined in \eqref{eq:complex field def} allows obtaining the $i$-complex fields $D$ and $B$ simply by separating the $j$-real and $j$-imaginary parts.

\section{The Unperturbed Modes}

In this section, we briefly review the unperturbed solution to the problem in a notation which is convenient for the following calculations.

\subsection{Adapted Basis}
To describe the modes in a cylindrical waveguide, it is useful to use the complex frame
\begin{align}
	\f_\pl
		&\equiv \f_z\,,
	&
	\f_+
		&= \tfrac{1}{\sqrt{2}}(\f_r - i \f_\theta)\,,
	&
	\f_-
		&= \tfrac{1}{\sqrt{2}}(\f_r + i \f_\theta)\,,
\end{align}
where $(\f_r, \f_\theta, \f_z)$ is the standard orthonormal frame adapted to cylindrical coordinates $(r, \theta, z)$.
The associated coframe is given by
\begin{align}
	\f^\pl
		&= \dd z\,,
	&
	\f^+
		&= \tfrac{1}{\sqrt 2}(\dd r + i r \dd \theta )\,,
	&
	\f^-
		&= \tfrac{1}{\sqrt 2}(\dd r - i r \dd \theta )\,.
\end{align}
The metric of flat space can then be written as
\begin{equation}
	\delta
		= \f^+ \otimes \f^-
		+ \f^- \otimes \f^+
		+ \f^\pl \otimes \f^\pl
		\,,
\end{equation}
so raising and lowering of indices interchanges $+$ and $-$, e.g.\ for a vector $v$ one has ${\t v{^\pm} = \t v{_\mp}}$ but $\t v{^\pl} = \t v{_\pl}$.
The only non-vanishing components of the spatial Levi-Civita connection one-form are
\begin{align}
	\t \omega{^+_+}
		&= + i \dd \theta
		= \tfrac{1}{\sqrt{2} r}\left( \f^+ - \f^- \right)\,,
	&
	\t \omega{^-_-}
		&= - i \dd \theta
		= \tfrac{1}{\sqrt{2} r} \left( \f^- - \f^+ \right)\,.
\end{align}
We find the cross-products to be
\begin{equation}
	\f_\pl \times \f_\pm = \pm i \f_\pm\,,
	\qquad\text{and}\qquad
	\f_+ \times \f_- = i \f_\pl\,,
\end{equation}
and consequently
\begin{equation}
	\f_+ \cdot (\f_- \times \f_\pl) = +i\,,
\end{equation}
so that for the Levi-Civita tensor (the volume form) we obtain
\begin{equation}
	\label{eq:volume form}
	\t \epsilon{_+_-_\pl}
	:= 	\epsilon(\f_+, \f_-, \f_\pl)  = +i\,,
	\qquad\text{and}\qquad
	\t \epsilon{^+^-^\pl} = -i\,.
\end{equation}
\checkedtogether{14.8.}

\subsection{Unperturbed Bessel Modes}

The computation of the electromagnetic modes in a cylindrical step-index waveguide is a standard calculation which is discussed in many textbooks, e.g.\ Refs.~\cites[Chapter~8]{Jackson1998}[Section~3.1]{Liu2005}[Section~16.10]{Davis2009}.
Here, we merely state the result, expressed in terms of the $j$-complex field $Z$ and referred to the $i$-complex basis introduced above.

The unperturbed Bessel modes take the concise form
\begin{align}
	\label{eq:unperturbed parallel}
	\t Z{^{\o0}^\pl}
		&= f_0(\alpha; r) e^{i (\omega t - \beta z + m \theta)}\,,
		\\
	\label{eq:unperturbed pm}
	i \zeta \t Z{^{\o0}^\pm}
		&= \tfrac{1}{\sqrt{2}} \rho^2(\beta \pm i j n \omega) c_m^\pm \t Z{^{\o0}^\pl}\,,
		\\
\shortintertext{where}
	\label{eq:zeta}
	\zeta	
		&= \rho^2(n^2 \omega^2 - \beta^2)\,,
		\\
\shortintertext{and}
	\label{eq:cm operators}
	c_m^\pm
		&= \p_r \mp \tfrac{m}{r}\,.
\end{align}
Here, $\omega$ is the angular frequency of the wave, $\beta$ the propagation constant, $m$ is the azimuthal mode index, and $\rho$ is the radius of the fibre core. The radial function $f_0$ is given by
\begin{equation}
	\label{eq:unperturbed radial f0}
	f_0(\alpha_1, \alpha_2; r)
		= \begin{cases}
			\alpha_1 J_m(U r/\rho)\,,	&	r < \rho\,,\\
			\alpha_2 K_m(W r/\rho)\,,	&	r > \rho\,,
		\end{cases}
\end{equation}
where $\alpha_1, \alpha_2$ are $j$-complex constants (related by appropriate continuity conditions discussed below), and
\begin{equation}
	\label{eq:UW def}
	U = \sqrt{\rho^2(n_1^2 \omega^2 - \beta^2)}\,,
	\qquad\text{and}\qquad
	W = \sqrt{\rho^2(\beta^2 - n_2^2 \omega^2)}\,.
\end{equation}
The function $f_0$ satisfies the Bessel equation
\begin{equation}
	\label{eq:unperturbed radial ode}
	\bessel{0} f_0(\alpha; r)
	\equiv
	\left[
		\frac{\p^2}{\p r^2}
		+ \frac{1}{r} \frac{\p}{\p r}
		- \frac{m^2}{r^2}
		+ \frac{\zeta}{\rho^2}
	\right] f_0(\alpha; r)
	= 0\,,
\end{equation}
both in the core and the cladding of the fibre.

In passing, let us note that the $c_m^\pm$ act as “ladder operators”, i.e.\ for any real number $\lambda$ one has the identities
\begin{equation}
	\label{eq:Bessel recursion}
	c_m^\pm J_m(\lambda r) = \mp \lambda  J_{m \pm 1}(\lambda r)\,,
	\qquad\text{and}\qquad
	c_m^\pm K_m(\lambda r) = - \lambda  K_{m \pm 1}(\lambda r)\,,
\end{equation}
see e.g.\ Ref.~\cite[Eq.~9.1.27 on p.~361 and Eq.~9.6.26 on p.~376]{Olver1964}, so that we may write the $\pm$-components of the field as
\begin{equation}
	i \zeta \t Z{^{\o0}^\pm}
		= \tfrac{1}{\sqrt 2}\rho (\beta \pm ij n \omega) f_\pm(\alpha; r) e^{i (\omega t - \beta z + m \theta)}\,,
\end{equation}
where
\begin{equation}
	f_\pm(\alpha_1, \alpha_2; r)
		= \rho c_m^\pm f_0(\alpha_1, \alpha_2; r)
		= \begin{cases}
			\mp \alpha_1 U J_{m \pm 1}(U r/\rho)	\,, &	r < \rho\,,\\
			-   \alpha_2 W K_{m \pm 1}(W r/\rho)	\,, &	r > \rho\,.
		\end{cases}
\end{equation}

The dispersion relation is then obtained by imposing continuity conditions at the core-cladding interface.
As we will discuss these conditions extensively for the perturbed problem, we merely outline the basic reasoning here.
\begin{enumerate}
	\item Maxwell’s equations imply that the field components $\t D{^r}, \t E{_\theta}, \t E{_z}, \t B{^r}, \t H{_\theta}, \t H{_z}$ are continuous at the core-cladding interface (at $r = \rho$).
	\item Only four of these six continuity conditions are linearly independent.
	\item Since the fields have two $j$-complex, and thus four $j$-real parameters, there is a $j$-real equation of the form $\text{(jumps)} = \Pi_0 \text{(parameters)}$, where $\Pi_0$ is a $4 \times 4$ matrix.
	\item The continuity conditions are thus only solvable if $\det \Pi_0 = 0$, which constitutes the dispersion relation for $\beta$ and $\omega$. This equation is transcendental (as it contains various Bessel functions) and must thus be solved numerically.
\end{enumerate}
\checkedtogether{19.8.}

\section{The Perturbed Modes}

As we shall see in this section, the effect of weak \GWs\ on light in waveguides is twofold:
(i) the phase is modulated by the \GW, and
(ii) due to the tensorial nature of the \GW\ amplitude, optical sidebands with an angular dependence of the form $e^{i (m + k) \theta}$, with $k \neq 0$, arise.
We find the dominant effects to be (a) the correction of the phase ($k = 0$) and (b) the perturbation of the polarisation, which arises when $m + k = -m$.

\paragraph{Outline}

Let us first give an outline of the calculation.
We start by deriving and solving the first order correction to the wave equation for the longitudinal field component.
From this, we derive the transverse components of the electromagnetic field, finding structures similar to the ones of the unperturbed case.
Schematically, the correction to the central mode can be written as
\begin{equation}
	Z^\o1_0
		= [f^+(\alpha) + g^+(\sigma^+)] e^{i (\omega t - \beta z + m \theta + \kappa.x)}
		+ [f^-(\alpha) + g^-(\sigma^-)] e^{i (\omega t - \beta z + m \theta - \kappa.x)}
		\,,
\end{equation}
where the coefficients $\alpha$ describe the unperturbed field (they act as source terms in the wave equation for $Z^\o1$) and the coefficients $\sigma^\pm$ parametrise homogeneous solutions of the wave equation.
The continuity condition then takes the schematic form
\begin{equation}
	\Pi_0 \alpha
	+ \varepsilon [\Rho \bs \alpha + \Sigma^+ \bs \sigma^+] e^{+i \kappa.x}
	+ \varepsilon [\Rho \bs \alpha + \Sigma^- \bs \sigma^-] e^{-i \kappa.x}
	= 0\,,
\end{equation}
where $\bs \alpha, \bs \sigma^\pm$ are column vectors, each comprising four $j$-real numbers which parametrise $\alpha$ and $\sigma^\pm$, and $\Pi_0, \Rho, \Sigma^\pm$ are $j$-real $4 \times 4$-matrices.

The first term is the same as in the unperturbed problem, so the dispersion relation $\det \Pi_0 = 0$ and the coefficients $\bs \alpha$ remain unchanged.
The coefficients $\bs \sigma^\pm$ are then determined from
\begin{equation}
	\label{eq:outline continuity sigma}
	\Sigma^\pm \bs \sigma^\pm
	+ \Rho \bs \alpha = 0\,.
\end{equation}
For the central mode, we find that $\Sigma^\pm$ is close to the singular matrix $\Pi_0$ and its inverse has the form $(\Sigma^\pm)^{-1} = \pm \Omega^{-1} \Sigma' + O(\Omega^0)$, so that the parameters $\bs \sigma$ are given by
\begin{equation}
	\bs \sigma^\pm
		= \mp \Omega^{-1} \Sigma' \Rho \bs \alpha
		+ O(\Omega^0)\,.
\end{equation}
In \Cref{s:phase shift}, we show that this correction of order $1/\Omega$ amounts to a perturbation to the phase of the electromagnetic wave.

For the sidebands, we obtain similar equations. In almost all cases ($m + k \neq - m$), the coefficient matrices are \emph{not} close to being singular, and hence no terms of order $1/\Omega$ arise there.
However, in the exceptional case $m + k = - m$ we have $\det \Pi_k = \det \Pi_0 = 0$, so that we again obtain terms of order $1/\Omega$ there. We show in \Cref{s:polarisation} that these terms describe the perturbation of the polarisation of the electromagnetic wave.

\paragraph{The Polarisation of the Gravitational Wave}
We start by writing the polarisation tensor $A$ of the \GW, as defined in \eqref{eq:metric g}, in the complex frame introduced above.
Using
\begin{align}
	\label{eq:differential dx dy}
	\dd x
		&= \tfrac{1}{\sqrt{2}}[
			\f^+ e^{+i \theta} + \f^- e^{-i \theta}
		]\,,
	&
	\dd y
		&= \tfrac{-i}{\sqrt{2}}[
			\f^+ e^{+i \theta} - \f^- e^{-i \theta}
		]\,,
\end{align}
one finds the components of
\begin{equation}
	A
		= \t A{_\mu_\nu} \dd x^\mu \otimes \dd x^\nu
		= \t A{_i_j} \f^i \otimes \f^j
\end{equation}
in the basis $\f_+,\f_-,\f_\pl$ to be
\begin{align}
	A_{\parallel\parallel}
		&= A_{zz}\,,
		\\
	A_{+-}
		&= A_{-+}
		= \half(A_{xx} + A_{yy}) \,,
		\\
	A_{\pm\parallel}
		&= A_{\parallel\pm}
		= \tfrac{1}{\sqrt{2}}(A_{xz} \mp i A_{yz}) e^{\pm i \theta} \,,
		\\
	\label{eq:A++}
	A_{++}
		&= \half(A_{xx} - A_{yy} - 2 i A_{xy})e^{+ 2 i \theta} \,,
		\\
	\label{eq:A--}
	A_{--}
		&= \half(A_{xx} - A_{yy} + 2 i A_{xy})e^{- 2 i \theta} \,.
\end{align}
The condition that $A$ is traceless now translates to
\begin{equation}
	\label{eq:traceless complex frame}
	2 A_{+-} + A_{\parallel\parallel} = 0\,,
\end{equation}
while the transversality condition takes the form
\begin{equation}
	\t A{_i_+} \t\k{_-}
	+ \t A{_i_-} \t\k{_+}
	+ \t A{_i_\pl} \t\k{_\pl}
	= 0\,,
\end{equation}
where the frame components of $\k$ are
\begin{equation}
	\t\k{_\pm}
		= \tfrac{1}{\sqrt{2}}( \t\k{_x} \mp i \t\k{_y}) e^{\pm i \theta}\,,
	\qquad\text{and}\qquad
	\t\k{_\pl}
		= \t\k{_z}\,.
\end{equation}
To separate the amplitudes of $\t A{_i_j}$ from their $\theta$-dependence, we write
\begin{align}
	\t A{_\pl_\parallel}
		&= a_0\,,
	&
	\t A{_\pl_\pm}
		&= a_{\pm 1} e^{\pm i \theta}\,,
	&
	\t A{_\pm_\pm}
		&= a_{\pm 2} e^{\pm 2 i \theta}\,,
\end{align}
and due to the trace condition we have $\t A{_+_-} = \t A{_-_+} = - \half a_0$.
\checkedtogether{24.8.}

\paragraph{The Wave Equation}
As shown in \Cref{appendix:wave equation}, in regions where $n$ is constant, the field equations~\eqref{eq:maxwell evolution covariant} and \eqref{eq:maxwell constraint covariant} imply the wave equation
\checkedtogether{in appendix A}
\begin{equation}
	\label{eq:maxwell wave eq full}
	n^2 \t {\ddot Z}{^i} - \t {(\Delta Z)}{^i}
		+ \t R{^i_j} Z^j
		+ j n \t \epsilon{^i^j^k} \t \nabla{_j}(
			\t{\dot \gs}{_k_l} \t Z{^l}
		) = 0\,,
\end{equation}
where $\Delta$ is the \emph{perturbed} spatial vector Laplacian (i.e.\ associated to the perturbed spatial metric~$\gs$) and $\t R{^i_j}$ is the spatial Ricci tensor.
As the exterior derivative is independent of the connection (provided it is torsion-free), it suffices to use the unperturbed Levi-Civita connection here: $\nabla \equiv \nabla^{\o0}$.

Let us now expand the electromagnetic field $Z = Z^\o0 + \varepsilon Z^\o1$ as well as the Laplacian ${\Delta = \Delta^\o0 + \varepsilon \Delta^\o1}$.
Since $Z^\o0$ satisfies the unperturbed equation $n^2 \ddot Z^\o0 - \Delta^\o0 Z^\o0 = 0$, only terms of order $\varepsilon$ remain, so we obtain at first order
\begin{equation}
	\label{eq:maxwell wave eq full 2}
	\wop \t Z{^{\o1}^i}
	- (\Delta^\o1 Z^\o0)^i
		+ \t R{^{\o1}^i_j} \t Z{^{\o0}^j}
		+ j n \t \epsilon{^i^j^k} \t \nabla{_j}(
			\t*{\dot\gs}{^{\o1}_k_l} \t Z{^{\o0}^l}
		) = 0\,,
\end{equation}
where we have set
\begin{equation}
	\wop \t Z{^i}
	:= n^2 \t {\ddot Z}{^i} - (\Delta^\o0 Z)^i\,.
\end{equation}
The derivatives of the unperturbed field are of order $\omega$, which is much larger than the terms arising from derivatives of the metric or the spatial Ricci tensor, so we neglect the last two terms in \eqref{eq:maxwell wave eq full 2}, which are suppressed by $\Omega = \omegaG/\omega$ or more.
At this level of accuracy, we also have
\begin{equation}
	\label{eq:vector laplacian approx}
	\Delta^\o1 \t Z{^{\o0}^i}
	\approx \t \gs{^{\o1}^j^k} \nabla_j \nabla_k \t Z{^{\o0}^i}
	= - \cos(\kappa.x) \t A{^j^k} \nabla_j \nabla_k \t Z{^{\o0}^i}\,.
\end{equation}
With these approximations, we arrive at the wave equation
\begin{equation}
	\wop \t Z{^{\o1}^i}
		+ \cos(\kappa.x) \t A{^j^k} \nabla_j \nabla_k \t Z{^{\o0}^i}
	= 0\,.
\end{equation}
We shall use one further natural approximation:
\begin{equation}
	\cos(\kappa.x)
	= \cos(\omegaG(t - \t\k{_z} z - r (\t\k{_x} \cos\theta + \t\k{_y} \sin \theta)) )
	\approx
	\cos(\omegaG(t - \t\k{_\pl} z ))\,,
\end{equation}
which is admissible since the diameter of the waveguide is negligible compared to the \GW\ wavelength ($\rho \omegaG \ll 1$).
As an illustration: for typical optical fibres, $\rho$ measures a couple of micrometers and \GW\ wavelengths detected by \textsc{ligo} measure hundreds of kilometres: $\rho \omegaG \sim 10^{-10}$, so the error terms here will be much larger than those arising from  quadratic terms in the \GW\ amplitude $\varepsilon \sim 10^{-21}$.
At this level of accuracy, we may thus replace $\kappa.x$ by $\kappa.\xp := \omegaG(t - \t\k{_\pl} z )$ to obtain
\begin{equation}
	\label{eq:wave equation pl}
	\wop \t Z{^{\o1}^i}
		+ \cos(\kappa.\xp) \t A{^j^k} \nabla_j \nabla_k \t Z{^{\o0}^i}
	= 0\,.
\end{equation}
Using the fact that the connection one-form satisfies $\t\omega{^\pl_i_j} = 0$, the second covariant derivative in this equation evaluates to
\begin{equation}
	\t\nabla{_j} \t\nabla{_k} Z^\pl
		= \f_j(\f_k(Z^\pl))
		- \t\omega{^l_k_j} \f_l(Z^\pl)\,.
\end{equation}
Note that the unperturbed field $Z^\o0$ depends on the angle $\theta$ only as $e^{i m \theta}$.
Since $A$ has components which vary with $\theta$ as $e^{\pm i \theta}$ and $e^{\pm 2 i \theta}$, the first order correction will have sidebands with $m$ shifted by one and two in either direction.
Moreover, we find it convenient to separate terms which oscillate like the metric perturbation (i.e.\ trigonometric functions of $\kappa.\xp$) into terms with complex exponentials and thus decompose the perturbation of the electromagnetic field (and other quantities if needed) as
\begin{equation}
	\label{eq:field decomposition}
	\t Z{^{\o1}^a}
		= \sum_{k=-2}^2 \sum_{l = \pm 1} \t \Zz{^a_k_l}(r) e^{i(\omega t - \beta z + (m+k) \theta + l \, \kappa.\xp)}\,.
\end{equation}
Thus, in $\t \Zz{^a_k_l}$ the first index refers to the vector component (typically taken with respect to the complex frame $\f_\pl, \f_\pm$), the first lower index determines the angular mode number, and the second lower index determines the sign in the complex exponential $\exp(\pm i \kappa.\xp )$.

For the wave equation, we discuss the central mode ($k = 0$) and the sidebands ($k \neq 0$) separately.

\paragraph{Central Mode}

For the central mode ($k = 0$), the only contributions come from $\t A{^\pl^\pl}$ and $\t A{^+^-} = \t A{^-^+}$, so we have
\begin{equation}
	(\t A{^j^k} \nabla_j \nabla_k \t Z{^{\o0}^\pl})_{0}
		= \t A{^\pl^\pl} \p_z \p_z \t Z{^{\o0}^\pl}
		+ \t A{^+^-} [ \f_+ \f_- + \f_- \f_+ + \tfrac{1}{\sqrt 2 r}(\f_+ + \f_-) ] \t Z{^{\o0}^\pl}\,,
\end{equation}
where the subscript indicates $k = 0$.
Now, $\f_\pm$ acts on $\t Z{^{\o0}^\pl}$ as $\tfrac{1}{\sqrt 2} c_m^\pm$, as defined in \eqref{eq:cm operators}, so
\begin{equation}
	[ \f_+ \f_- + \f_- \f_+ + \tfrac{1}{\sqrt 2 r}(\f_+ + \f_-) ] \t Z{^{\o0}^\pl}
	= ( \p_r^2 + r^{-1} \p_r - m^2 r^{-2} ) \t Z{^{\o0}^\pl}\,.
\end{equation}
Using the unperturbed radial equation \eqref{eq:unperturbed radial ode} as well as the trace condition \eqref{eq:traceless complex frame}, we thus obtain
\begin{equation}
	(\t A{^j^k} \nabla_j \nabla_k \t Z{^{\o0}^\pl})_0
		= \half a_0 (\zeta/\rho^2 - 2\beta^2) \t Z{^{\o0}^\pl}\,.
\end{equation}

\paragraph{Sidebands}
For the sidebands with $m$ shifted by $\pm 1$, the only relevant terms are those involving $\t A{^\mp^\pl} = \t A{_\pm_\pl} = a_{\pm 1} e^{\pm i \theta}$, so we find
\begin{equation}
	(\t A{^j^k} \nabla_j \nabla_k \t Z{^{\o0}^\pl})_{\pm 1}
		= - \sqrt 2 i \beta a_{\pm 1} c_m^\pm \t Z{^{\o0}^\pl}\,,
\end{equation}
and for the sidebands with $m$ shifted by $\pm 2$, the relevant terms come from $\t A{^\mp^\mp} = \t A{_\pm_\pm} = a_{\pm 2} e^{\pm 2 i \theta}$, so
\begin{equation}
	(\t A{^j^k} \nabla_j \nabla_k \t Z{^{\o0}^\pl})_{\pm 2}
		= \half a_{\pm 2} ( c_m^\pm c_m^\pm - \tfrac{1}{r} c_m^\pm) \t Z{^{\o0}^\pl}
		= \half a_{\pm 2} c_{m \pm 1}^\pm c_m^\pm \t Z{^{\o0}^\pl}\,.
\end{equation}
\checkedtogether{16.10.}

\subsection{The Radial Equations}

Using separation of variables, the wave equations just derived are reduced to radial equations.
We give explicit solutions Bessel equations arising this way.

\paragraph{Central Mode}

Using the decomposition \eqref{eq:field decomposition}, the inhomogeneous wave equation \eqref{eq:wave equation pl} yields
\begin{align}
	\label{eq:radial ode k=0}
	\bessel[\pm]{0} \t \Zz{^\pl_0_\pm}
		&= \quarter a_0 (\zeta/\rho^2 - 2\beta^2) f_0(\alpha, r)
		 \,,
		 \\
\shortintertext{where}
	\bessel[\pm]{0}
		&= \frac{\p^2}{\p r^2}
			+ \frac{1}{r} \frac{\p}{\p r}
			+ \frac{\zeta_\pm}{\rho^2}
			- \frac{m^2}{r^2}\,,
		\\
	\label{eq:zeta ±}
	\zeta_\pm
		&= \rho^2 (n^2 \omega_\pm^2 - \beta_\pm^2)\,,
		\\
	\label{eq:omega ±}
	\omega_\pm
		&= \omega \pm \omegaG\,,
		\\
	\label{eq:beta ±}
	\beta_\pm
		&= \beta \pm \t\k{_\pl} \omegaG\,.
\end{align}
The homogeneous solutions to this equation which are regular at the origin and vanish at infinity are evidently
\begin{equation}
	\label{eq:f^± def}
	f_0^\pm(\sigma, r)
	= \begin{cases}
		\sigma_1 J_{m}(U_\pm r/\rho)\,,
			&	r < \rho\,,
			\\
		\sigma_2 K_{m}(W_\pm r/\rho)\,,
			&	r > \rho\,,
	\end{cases}
\end{equation}
where $\sigma_1, \sigma_2$ are constants (continuity conditions relating them are discussed in a later section), and where $U_\pm, W_\pm$ are defined in analogy to $U,W$ in \eqref{eq:UW def}:
\begin{equation}
	\label{eq:UW ±}
	U_\pm 
		= \sqrt{\rho^2(n_1^2 \omega_\pm^2 - \beta_\pm^2)}\,,
	\qquad\text{and}\qquad
	W_\pm
		= \sqrt{\rho^2(\beta_\pm^2 - n_2^2 \omega_\pm^2)}\,.
\end{equation}
\checkedtogether{23.10.}
To find particular solutions, we consider the more general problem
\begin{equation}
	\label{eq:Bessel inhomogeneous}
	\rho^2 \bessel[\pm]{\nu} p_\nu^\pm = \begin{cases}
		\alpha_1 J_{m+\nu}(Ur/\rho)\,,
			&	r < \rho\,,
		\\
		\alpha_2 K_{m+\nu}(Wr/\rho)\,,
			&	r > \rho\,,
	\end{cases}
\end{equation}
where $\alpha_1, \alpha_2$ are arbitrary constants and $\bessel[\pm]{\nu}$ is a Bessel operator of arbitrary order $\nu$:
\begin{equation}
	\label{eq:Bessel operator general}
	\bessel[\pm]{\nu}
		= \frac{\p^2}{\p r^2}
			+ \frac{1}{r} \frac{\p}{\p r}
			+ \frac{\zeta_\pm}{\rho^2}
			- \frac{(m+\nu)^2}{r^2}\,.
\end{equation}
Particular solutions to this equation are given by
\begin{align}
	\label{eq:radial particular p}
	p_\nu^\pm(\alpha, r)
	&= \begin{cases}
		\alpha_1 [
			Y_{m+\nu}(U_\pm r/\rho) \Gamma^\pm_{m+\nu}(r/\rho)
			+ J_{m+\nu}(U_\pm r/\rho) \tilde \Gamma^\pm_{m+\nu}(r/\rho)
		]\,,
		&	r < \rho\,,
		\\
		\alpha_2 [
			I_{m+\nu}(W_\pm r/\rho) \Delta^\pm_{m+\nu}(r/\rho)
			+ K_{m+\nu}(W_\pm r/\rho) \tilde \Delta^\pm_{m+\nu}(r/\rho)
		]\,,
		&	r > \rho\,,
	\end{cases}\\
\shortintertext{where}
	\label{eq:integral Gamma def}
	\Gamma_\nu^\pm(z)
		&= \frac{\pi}{2} \int_0^z J_\nu(U_\pm z') J_{\nu}(U z') z' \, \dd z'\,,\\
	\label{eq:integral Gamma tilde def}
	\tilde \Gamma_\nu^\pm(z)
		&= \frac{\pi}{2} \int_z^1 Y_{\nu}(U_\pm z') J_{\nu}(U z') z' \, \dd z'\,,\\
	\label{eq:integral Delta def}
	\Delta_\nu^\pm(z)
		&= - \int_z^\infty K_{\nu}(W_\pm z') K_{\nu}(W z') z' \, \dd z'\,,\\
	\label{eq:integral Delta tilde def}
	\tilde \Delta_\nu^\pm(z)
		&= -\int_1^z I_{\nu}(W_\pm z') K_{\nu}(W z') z' \, \dd z'\,.
\end{align}
Equation~\eqref{eq:Bessel inhomogeneous} is readily verified using the Wrońskians
\begin{equation}
	\begin{vmatrix}
		J_m(r)	&	J_m'(r)\\
		Y_m(r)	&	Y_m'(r)
	\end{vmatrix}
	= \frac{2}{\pi} \frac{1}{r}\,,
	\qquad\text{and}\qquad
	\begin{vmatrix}
		K_m(r)	&	K_m'(r)\\
		I_m(r)	&	I_m'(r)
	\end{vmatrix}
	= \frac{1}{r}\,,
\end{equation}
cf.\ e.g.\ Ref.~\cite[eqns.~9.1.15 on p.~360 and 9.6.15 on p.~375]{Olver1964}.
The integral ranges were chosen as follows.
The function $\Gamma_\nu^\pm$ multiplies $Y_\nu$ which diverges at the origin, so we chose the region of integration such that $\Gamma_\nu^\pm$ vanishes there.
The region of integration for $\Delta_\nu^\pm$ was chosen for a similar reason, as $I_\nu$ diverges at large distances.
Finally, the ranges for $\tilde \Gamma_\nu^\pm$ and $\tilde \Delta_\nu^\pm$ were chosen such that these functions vanish at the core-cladding interface $r = \rho$, where certain components of the electromagnetic field are required to be continuous.

Using the function $p_0$, one obtain the general solution to \eqref{eq:radial ode k=0} which is regular at the origin and vanishes at infinity:
\begin{equation}
	\label{eq:Z perturbed parallel m±0}
	\t \Zz{^\pl_0_\pm}
		= a_0[
			\quarter (\zeta - 2\rho^2 \beta) p_0^\pm(\alpha)
			+ f_0^\pm(\sigma_{0, \pm})
		]\,,
\end{equation}
where $a_0$ was factored out for convenience.

\paragraph{Sidebands}

Using the decomposition \eqref{eq:field decomposition} for the sidebands also, the inhomogeneous wave equation \eqref{eq:wave equation pl} reduces to
\begin{align}
	\bessel[+]{\pm 1} \t \Zz{^\pl_{\pm 1}_+}
		= \bessel[-]{\pm 1}  \t \Zz{^\pl_{\pm 1}_-}
		&= \tfrac{i}{\sqrt 2} a_{\pm 1} \beta \rho^{-1} f_{\pm1}(\alpha)
		\,,
		\\
	\bessel[+]{\pm 2} \t \Zz{^\pl_{\pm 2}_+}
		= \bessel[-]{\pm 2}  \t \Zz{^\pl_{\pm 2}_-}
		&= \quarter a_{\pm 2} \rho^{-2} f_{\pm 2}(\alpha)
		\,,
\end{align}
where $\bessel[\pm]{\nu}$ is as in \eqref{eq:Bessel operator general}.
Proceeding in the same way as for the central mode, one obtains
\begin{align}
	\label{eq:Z perturbed parallel m±1 +}
	\t \Zz{^\pl_{\pm 1}_+}
		&= a_{\pm 1} \left(
			\tfrac{i}{\sqrt 2} \rho \beta \, p^+_{\pm 1}(\mp U \alpha_1, - W \alpha_2, r)
			 + f^+_{\pm 1}(\sigma_{\pm 1,+}, r)
		\right)
		\,,
		\\
	\label{eq:Z perturbed parallel m±1 -}
	\t \Zz{^\pl_{\pm 1}_-}
		&= a_{\pm 1} \left(
			\tfrac{i}{\sqrt 2} \rho \beta \, p^-_{\pm 1}(\mp U \alpha_1, - W \alpha_2, r)
		 	+ f^-_{\pm 1}(\sigma_{\pm 1, -}, r)
		 \right)
		 \,,
		\\
	\label{eq:Z perturbed parallel m±2 +}
	\t \Zz{^\pl_{\pm 2}_+}
		&= a_{\pm 2} \left(
			\quarter p^+_{\pm 2}(U^2 \alpha_1, W^2 \alpha_2, r)
		 	+ f^+_{\pm 2}(\sigma_{\pm 2, +}, r)
		 \right)
		 \,,
		\\
	\label{eq:Z perturbed parallel m±2 -}
	\t \Zz{^\pl_{\pm 2}_-}
		&= a_{\pm 2} \left(
			\quarter p^-_{\pm 2}(U^2 \alpha_1, W^2 \alpha_2, r)
			 + f^-_{\pm 2}(\sigma_{\pm 2, -}, r)
		\right)
		\,,
\end{align}
where the various $\sigma$’s are parameters to be determined from continuity conditions, and the functions $f^\pm_k$ are derived from $f^\pm_0$, defined in \eqref{eq:f^± def}, by successive application of the ladder operators $c_m^\pm$ defined in \eqref{eq:cm operators}:
\begin{align}
	f^+_{\pm 1}
		&= \rho c_m^\pm f^+_0\,,
	&
	f^+_{\pm 2}
		&= \rho^2 c_{m \pm 1}^\pm c_m^\pm f^+_0\,,
	\\
	f^-_{\pm 1}
		&= \rho c_m^\pm f^-_0\,,
	&
	f^-_{\pm 2}
		&= \rho^2 c_{m \pm 1}^\pm c_m^\pm f^-_0\,.
\end{align}

Having found the longitudinal components of the electromagnetic field, one can now compute the remaining (transverse) components.

\subsection{The Transverse Components}

To obtain the transverse components of $Z$, consider the $\pm$ components of \eqref{eq:maxwell evolution covariant}.
Expanding $\gs = \gs^\o0 + \varepsilon \gs^\o1$, $Z = Z^\o0 + \varepsilon Z^\o1$, and using the unperturbed equation, one obtains at first order
\begin{equation}
	\label{eq:transverse components full}
	n \p_0 Z^{\o1 i}
	+ j \t \epsilon{^i^j^k} \nabla_j (\t* \gs{^{\o0}_k_l} \t Z{^{\o1}^l})
	+ j \t \epsilon{^i^j^k} \nabla_j (\t* \gs{^{\o1}_k_l} \t Z{^{\o0}^l})
	= 0\,.
\end{equation}
In the last term, we may neglect derivatives of the metric perturbation (of order $\omegaG$) compared to derivatives of the unperturbed electromagnetic field (of order $\omega$) to arrive at
\begin{equation}
	\label{eq:transverse eq}
	(n \p_0 \pm ij \p_z)\t* Z{^{\o1}^\pm}
	= \pm i j [
		\f_\mp(\t* Z{^{\o1}^\pl})
		+ \chi^\pm \cos(\kappa.\xp)
	]\,,
\end{equation}
where we have used \eqref{eq:volume form} and have set
\begin{equation}
	\chi^\pm
		= \t A{_\pl_l} \nabla_\mp \t Z{^{\o0}^l}
		+ i \beta \t A{_\mp_l} \t Z{^{\o0}^l} \,.
\end{equation}
Using the notation of \eqref{eq:field decomposition}, a direct calculation shows that
%
\begin{align}
	\chi^+_{-2}
		&= i \beta a_{-2} \t Z{^{\o0}^-}\,,
	&
	\chi^-_{+2}
		&= i \beta a_{+2} \t Z{^{\o0}^+}\,,
	\\
	\chi^+_{-1}
		&= \tfrac{i}{2} a_{-1} (3 \beta - i j n \omega) \t Z{^{\o0}^\pl}
		\,,
	&
	\chi^-_{+1}
		&= \tfrac{i}{2} a_{+1} (3 \beta + i j n \omega) \t Z{^{\o0}^\pl}\,,
		\\
	\chi^+_{\phantom{+}0}
		&= -i a_0 (\tfrac{3}{2} \beta - i j n \omega) \t Z{^{\o0}^+}\,,
	&
	\chi^-_{\phantom{+}0}
		&= -i a_{0} (\tfrac{3}{2} \beta + i j n \omega) Z^{\o0 -}\,,
		\\
	\chi^+_{+1}
		&= \tfrac{1}{\sqrt 2} a_{+1} c_{m+1}^+ \t Z{^{\o0}^+}\,,
	&
	\chi^-_{-1}
		&= \tfrac{1}{\sqrt 2} a_{-1} c_{m-1}^- \t Z{^{\o0}^-}\,,
		\\
	\chi^+_{+2}
		&= 0\,,
	&
	\chi^-_{-2}
		&= 0\,.
\end{align}
To arrive at the stated formula for $\chi^\pm_0$, we have used $\f_{\mp}(Z^{\o0 \parallel}) = -i (\beta \mp i j n \omega) Z^{\o0 \pm}$ which follows from \eqref{eq:unperturbed pm}, and the trace condition \eqref{eq:traceless complex frame}.
For $\chi^+_{-1}$ and $\chi^-_{+1}$ we have used
\begin{equation}
	\rho^2 c_{m+1}^- c_m^+ f_0 = \rho^2 c_{m-1}^+ c_m^- f_0 = - \zeta f_0\,,
\end{equation}
which can be shown either using the recursion relations \eqref{eq:Bessel recursion} or by noting that $c_{m+1}^- c_m^+ = c_{m-1}^+ c_m^- = \bessel{0} - \zeta/\rho^2 $ and using the Bessel equation $\bessel{0} f_0 = 0$.
These formulae have the general structure that the radial dependence of $\chi^a_b$ is proportional to $a_b f_{a+b}$.

Using the notation \eqref{eq:field decomposition}, as well as the abbreviations $\omega_\pm$ and $\beta_\pm$ as defined in \eqref{eq:omega ±} and \eqref{eq:beta ±}, equation \eqref{eq:transverse eq} reduces to
\begin{align}
	i ( n \omega_+ \mp ij \beta_+ ) \t \Zz{^\pm_k_+}
		&= \pm ij [ \tfrac{1}{\sqrt 2} c_{m+k}^\pm \t \Zz{^\pl_k_+} + \half \t* \chi{^\pm_k} ]\,,
		\\
	i ( n \omega_- \mp ij \beta_- ) \t \Zz{^\pm_k_-}
		&= \pm ij [ \tfrac{1}{\sqrt 2} c_{m+k}^\pm \t \Zz{^\pl_k_-} + \half \t* \chi{^\pm_k} ]\,.
\end{align}
%
Multiplying the first equation by $n \omega_+ \pm i j \beta_+$ and the second one by $n \omega_- \pm i j \beta_-$, and using the definition of $\zeta_\pm$ given in \eqref{eq:zeta ±}, one obtains
\begin{align}
	\label{eq:transverse components +}
	i \zeta_+ \t \Zz{^\pm_k_+}
		&= \tfrac{1}{\sqrt 2} \rho^2 [ \beta_+ \pm i j n \omega_+ ] ( c_{m+k}^\pm \t \Zz{^\pl_k_+} + \tfrac{1}{\sqrt 2}  \t* \chi{^\pm_k} )
		\,,\\
	\label{eq:transverse components -}
	i \zeta_- \t \Zz{^\pm_k_-}
		&= \tfrac{1}{\sqrt 2} \rho^2 [ \beta_- \pm i j n \omega_- ] ( c_{m+k}^\pm \t \Zz{^\pl_k_-} + \tfrac{1}{\sqrt 2}  \t* \chi{^\pm_k} )
		\,,
\end{align}
which is structurally similar to the unperturbed equation \eqref{eq:unperturbed pm}.

Now that the field equations are solved in the core and cladding separately, we consider how these solutions match at the core-cladding interface of the waveguide.

\subsection{Continuity Conditions}

From \eqref{eq:Maxwell standard form}, one finds that the field components $D^r, E_\theta, E_z, B^r, H_\theta, H_z$ must be continuous at the core-cladding interface $r = \rho$.
As is well-known, these six continuity conditions are linearly dependent, and it suffices to impose continuity of the four components
\begin{align}
	\label{eq:continuous fields}
	D^r
		&= \Re Z^r\,,
	&
	B^r
		&= n^{-1} \Im Z^r\,,
	&
	H_\pl
		&= n^{-1} \t \gs{_\pl_i} \Im Z^i\,,
	&
	E_\pl
		&= n^{-2} \t \gs{_\pl_i} \Re Z^i\,,
\end{align}
where $\Re$ and $\Im$ refer to the $j$-real and $j$-imaginary parts, respectively.
To assess the jumps of the fields, we introduce the notation
\begin{equation}
	\jump{f}
		:= \lim_{r \nearrow \rho} f(r)
		-  \lim_{r \searrow \rho} f(r)\,,
\end{equation}
and write symbolically
\begin{equation}
	\label{eq:jump vector}
	\jump{Z}
	= \begin{pmatrix}
		\jump{D^r}	&
		\jump{E_\pl}	&
		\jump{B^r}	&
		\jump{H_\pl}
	\end{pmatrix}^{\mathsf T}\,,
\end{equation}
which conflates the jumps of all relevant fields in one \emph{column} vector.

Since the jumps arising in \eqref{eq:continuous fields} are determined by the jumps of the two components
\begin{align}
	\label{eq:Z continuous r}
	\t Z{^r}
		&= \tfrac{1}{\sqrt 2}( \t Z{^+} + \t Z{^-} )\,,
		\\
	\label{eq:Z continuous z}
	\t \gs{_\pl_i} \t Z{^i}
		&= \t Z{^{\o0}^\pl}
		+ \varepsilon \t Z{^{\o1}^\pl}
		+ \varepsilon \t A{_\pl_i} \t Z{^{\o0}^i} \cos(\kappa.\xp)\,,
\end{align}
for the perturbed problem it suffices to consider the jumps of
\begin{align}
	\label{eq:Z r}
	\t \Zz{^r_k_l}
		&= \tfrac{1}{\sqrt{2}}( \t \Zz{^+_k_l} + \t \Zz{^-_k_l} )\,,
		\\
	\label{eq:Z pl}
	\t \Zz{_\pl_k_l}
		&= \t \Zz{^\pl_k_l}
			+ \half \t* \chi{^\pl_k}\,,
\end{align}
where the $\t \Zz{^a_k_l}$ are the components from the decomposition \eqref{eq:field decomposition}, and where
\begin{align}
	\t* \chi{^\pl_{\phantom\pm}_0}
		&= a_0 \t Z{^{\o0}^\pl}\,,
		&
	\t* \chi{^\pl_{\pm 1}}
		&= a_{\pm 1} \t Z{^{\o0}^\pm}\,,
		&
	\t* \chi{^\pl_{\pm 2}}
		&= 0\,.
\end{align}

Similar to the decomposition \eqref{eq:field decomposition}, we write
\begin{equation}
	\label{eq:jumps decomposition}
	\jump{Z}
	\equiv
	\left(
		\jump{Z}^\o0
		+ \varepsilon \sum_{k=-2}^{+2} \sum_{l = \pm }
		\jump{Z}^\o1_{kl} e^{i (k \theta + l \kappa.\xp)}
	\right) e^{i (\omega t - \beta z + m \theta)}
	= 0
		\,,
\end{equation}
where the first term does not contain any $\varepsilon$ corrections because all components of $Z^\o1$ oscillate with $\exp(\pm i \kappa.\xp)$.

The four $j$-real components of $\jump{Z}^\o1_{kl}$ are parametrised by $\alpha$ (coming from the particular solutions) and $\sigma_{kl}$ (homogeneous solutions), each having four $j$-real parameters. Hence, one can write
\begin{equation}
	\jump{Z}_{kl}^\o1
		= a_k (M_{kl}\, \bs \alpha
		+ \Sigma_{kl}\, \bs \sigma_{kl})
	\qquad\text{(no summation implied)}
	\,,
\end{equation}
where $\bs \alpha$ and $\bs \sigma_{kl}$ are $j$-real vectors:
\begin{align}
	\bs \alpha
		&= (\Re \alpha_1, \Re \alpha_2, \Im \alpha_1, \Im \alpha_2)^{\mathsf T}\,,
	&
	\bs \sigma_{k l}
		&\text{ similar},
\end{align}
and $M_{kl}$ and $\Sigma_{kl}$ are $j$-real $4 \times 4$ matrices. Note that the jumps are \emph{not} simply linear in the complex coefficients $\alpha, \sigma_{k l}$ since \eqref{eq:continuous fields} discriminates between the real and complex parts.

In \Cref{appendix:jmp symmetries}, it is shown that because the electric and magnetic fields are derived from a single complex field, these matrices are fully determined by their first two rows, which motivates the following abbreviating notation:
\begin{equation}
	\label{eq:jmp shorthand}
	\begin{vmatrix}
		a	&	b	&	c	&	d\\
		e	&	f	&	g	&	h
	\end{vmatrix}
	:=
	\begin{pmatrix}
		a		&	b		&	c		&	d		\\
		e		&	f		&	g		&	h		\\
		-c/n_1	&	-d/n_2	&	a/n_1	&	b/n_2	\\
		-g\;n_1	&	-h\;n_2	&	e\;n_1	&	f\;n_2
	\end{pmatrix}\,.
\end{equation}
As many matrices considered here have similar structures, we define the following abbreviation
\begin{multline}
	\label{eq:jmp matrix structure}
	\begin{bmatrix}
		a_1		&	a_2		&	a_3	&	a_4	\\
		\beta	&	\omega	&	U	&	W
	\end{bmatrix}_\nu
	:=\\
	\begin{vmatrix}
		-i a_1 \rho \beta J_{\nu}'(U)	&
		-i a_2 \rho \beta K_{\nu}'(W)	&
		a_3 n_1 \nu \rho \omega J_{\nu}(U)&
		a_4 n_2 \nu \rho \omega K_{\nu}(W)\\
		\frac{1}{n_1^2} J_{\nu}(U)	&
		-\frac{1}{n_2^2} K_{\nu}(W)	&
		0	&
		0
	\end{vmatrix}
	\,,
\end{multline}
where the parameters $a_1, \ldots, a_4$ and $\beta, \omega$ determine the numerical coefficients in the first row of the matrix, and the last two parameters $U,W$ determine the arguments of the Bessel functions, whose order is given by the integer $\nu$ which is given as a general subscript.

Within this class of matrices, we identify the following families, which are useful for the problem considered here:
\begin{align}
	\label{eq:jmp Pi}
	\Pi_k &= 
		\begin{bmatrix}
			1/U		&	1/W		&	1/U^2	&	1/W^2	\\
			\beta	&	\omega	&	U		&	W
		\end{bmatrix}_{m+k}\,,
		\\
	\label{eq:jmp Sigma}
	\Sigma^\pm_k &= 
		\begin{bmatrix}
			1/U_\pm		&	1/W_\pm		&	1/U_\pm^2	&	1/W_\pm^2	\\
			\beta_\pm	&	\omega_\pm	&	U_\pm		&	W_\pm
		\end{bmatrix}_{m+k}\,,
\end{align}
where $\beta_\pm,\omega_\pm$ are defined in \eqref{eq:omega ±} and \eqref{eq:beta ±}, and $U_\pm, W_\pm$ are defined in \eqref{eq:UW ±}.
The $\Pi$ family is used to describe the central mode and to formulate the dispersion relation, and the $\Sigma$ matrices arise when considering the jumps of the homogeneous solutions to the radial equations.

Similar to the family \eqref{eq:jmp matrix structure}, we define
\begin{multline}
	\label{eq:jmp matrix structure 2}
	\begin{Bmatrix}
		a_1		&	a_2		&	a_3	&	a_4	\\
		\beta	&	\omega	&	U	&	W
	\end{Bmatrix}_\nu
	:=\\
	\begin{vmatrix}
		-i a_1 \rho \beta Y_{\nu}'(U) &
		-i a_2 \rho \beta I_{\nu}'(W) &
		a_3 n_1 \nu \rho \omega Y_{\nu}(U) &
		a_4 n_2 \nu \rho \omega I_{\nu}(W) \\
		\frac{1}{n_1^2} Y_{\nu}(U)	&
		-\frac{1}{n_2^2} I_{\nu}(W) &
		0	&
		0
	\end{vmatrix}
	\,,
\end{multline}
where the Bessel functions $J_\nu$ and $K_\nu$ are replaced by $Y_\nu$ and $I_\nu$.
Using this notation, let
\begin{align}
	\label{eq:jmp Rho}
	\Rho^\pm_k
		&= 
		\begin{Bmatrix}
			U/U_\pm^2		&	W/W_\pm^2		&	1/U_\pm^2	&	1/W_\pm^2	\\
			\beta_\pm	&	\omega_\pm	&	U		&	W
		\end{Bmatrix}_{m+k} \Lambda_{m+k}^\pm\,,
	\\
\shortintertext{where}
	\Lambda_\nu^\pm
		&= \diag\left(
			\Gamma^\pm_\nu(1), \Delta^\pm_\nu(1), \Gamma^\pm_\nu(1), \Delta^\pm_\nu(1)
		\right)\,.
\end{align}
\checkedtogether{30.10 rough reading} %
This is used to describe the jumps of the particular solutions $p_\nu^\pm$ defined in \eqref{eq:radial particular p}.
The functions $\Gamma^\pm_\nu, \Delta^\pm_\nu$ appearing here are defined in \eqref{eq:integral Gamma def} and \eqref{eq:integral Delta def}, while the functions $\tilde \Gamma^\pm_\nu, \tilde \Delta^\pm_\nu$ defined in \eqref{eq:integral Gamma tilde def} and \eqref{eq:integral Delta tilde def} do not contribute, since they vanish at the core-cladding interface $r = \rho$.

Having prepared the notation, we now analyse \eqref{eq:jumps decomposition}. By linear independence of the exponential functions, all terms $\jump{Z}^\o0$ and $\jump{Z}^\o1_{kl}$ must vanish separately.

\paragraph{Dispersion Relation}

Because all corrections coming from the gravitational wave oscillate as $\exp(\pm i \kappa.\xp)$, the first term in \eqref{eq:jumps decomposition} is the same as for the unperturbed problem.
As shown explicitly in \Cref{appendix:jmp unperturbed}, the continuity condition for this part is given by
\begin{equation}
	\label{eq:continuity alpha}
	\Pi_0 \bs \alpha = 0\,,
\end{equation}
where $\Pi_0$ is given by \eqref{eq:jmp Pi}.
To obtain a non-trivial solution, the coefficient matrix must be singular:
\begin{equation}
	\label{eq:dispersion det abstract}
	\det \Pi_0 = 0\,,
\end{equation}
which is equivalent to
\begin{equation}
	\label{eq:dispersion det explicit}
	\left(\mathscr J_m + \mathscr K_m \right)
		\left( n_1^2 \mathscr J_m + n_2^2 \mathscr K_m \right)
	=
	\left( m \beta/\omega \right)^2 \left( U^{-2} + W^{-2} \right)^2
	\,,
\end{equation}
where we have used the abbreviations
\begin{align}
	\mathscr J_m
		&= \frac{J_m'(U)}{U J_m(U)}\,,
	&
	\mathscr K_m
		&= \frac{K_m'(W)}{W K_m(W)}\,,
\end{align}
cf.\ e.g.\ Ref.~\cite[Eq.~3.27 on p.~124]{Liu2005} and Ref.~\cite[Eq.~16.166 on p.~511]{Davis2009} (where $\epsilon_2/\epsilon_2$ should read $\epsilon_1/\epsilon_2$).
Because $\beta$ and $\omega$ enter this equation through $U$ and $W$ as arguments of Bessel functions, this is a transcendental equation which must be solved numerically. More precisely, by solving $\det \Pi_0 = 0$ for a prescribed vacuum wavelength $\lambda = 2 \pi/\omega$ and given waveguide parameters (i.e.\ core radius $\rho$ and refractive indices $n_1, n_2$), one can determine the effective refractive index $\bar n := \beta/\omega$ as a function of $\lambda$: $\bar n = \bar n(\lambda)$, which constitutes the waveguide dispersion relation. Subsequently, the coefficients $\bs \alpha$ can be determined from \eqref{eq:continuity alpha} up to an overall factor which describes the amplitude of the electromagnetic wave.

In general, there are multiple solutions to this equation for given waveguide parameters and given values of $m$ and $\omega$.
Here, we focus on \emph{single-mode} fibres (i.e.\ fibres with sufficiently thin cores, operated at suitable frequencies) where there is only one single solution for $m = +1$ and there are no solutions for higher values of $m$.

\paragraph{Central Mode}
For the central mode ($k = 0$), the continuity condition can be written as
\begin{equation}
	\Sigma_0^\pm \bs \sigma_{0,\pm}
	+ \Rho^\pm_0 \Xi \bs \alpha
	+ \half \Psi_0 \bs \alpha = 0\,.
\end{equation}
The first term describes the jumps arising from the homogeneous solution in \eqref{eq:Z perturbed parallel m±0}.
In the second term, which comes from the particular solution, the matrix $\Xi$ describes the discontinuity of the factor $\quarter (\zeta - 2 \rho^2 \beta^2)$ in \eqref{eq:Z perturbed parallel m±0}:
\begin{equation}
	\Xi
		= -\quarter \diag(2 \rho^2\beta^2 - U^2, 2 \rho^2\beta^2 + W^2,2 \rho^2\beta^2 - U^2, 2 \rho^2\beta^2 + W^2)\,.
\end{equation}
Finally, the last term contains all contributions from the $\chi$-terms in \eqref{eq:transverse components +}, \eqref{eq:transverse components -} and \eqref{eq:Z pl}:
\begin{equation}
	\label{eq:jmp Psi 0}
	\Psi_0	=
		\begin{Bmatrix}
			\half U_3^2/U^3	&	\half W_3^2/W^3		&	U_2^2/U^4	&	W_2^2/W^4	\\
			\beta	&	\omega	&	U		&	W
		\end{Bmatrix}_m\,,
\end{equation}
where
\begin{align}
	U_2^2
		&= + \rho^2(n_1^2 \omega^2 - 2\beta^2)\,,
	&
	U_3^2
		&= + \rho^2(n_1^2 \omega^2 - 3\beta^2)\,,
	\\
	W_2^2
		&= - \rho^2(n_2^2 \omega^2 - 2\beta^2)\,,
	&
	W_3^2
		&= - \rho^2(n_2^2 \omega^2 - 3\beta^2)\,.
\end{align}

At the considered level of accuracy, we may neglect terms of order $\Omega$ in $\Rho^\pm_0$.
This matrix contains the functions $\Gamma^\pm_\nu$ and $\Delta^\pm_\nu$ (via the matrix $\Lambda_\pm$). In \eqref{eq:integral Gamma def} and \eqref{eq:integral Delta def} we may thus replace $U_\pm$ and $W_\pm$ by $U$ and $W$ which yields the integrals
\begin{align}
	\Gamma_\nu(z)
		&= \frac{\pi}{2} \int_0^z J_\nu(U z')^2 z'\, \dd z'\,,
		\\
	\Delta_\nu(z)
		&= - \int_z^\infty K_\nu(W z')^2 z' \,\dd z'\,,
\end{align}
which also arise in similar calculations describing rotating waveguides. 
They evaluate to
\begin{align}
	\Gamma_\nu(z)
		&= \frac{\pi}{2} \frac{z^2}{2}\left(
			J_\nu(U z)^2 - J_{\nu - 1}(U z) J_{\nu + 1}(U z)
		\right)\,,
		\\
	\Delta_\nu(z)
		&= \frac{z^2}{2} \left(
			K_\nu(W z)^2 - K_{\nu - 1}(W z) K_{\nu + 1}(W z)
		\right)\,.
\end{align}
Expanding $\Sigma_{0,\pm} = \Pi_0 \pm \Omega \Sigma^\o1 + O(\Omega^2)$, we obtain the inverse
\begin{align}
	(\Sigma_0^\pm)^{-1}
		&= \pm \Omega^{-1} \Sigma' + O(\Omega^0)\,,
	\\
\shortintertext{where}
	\Sigma'
		&= \frac{1}{\trace(\Sigma^\o1 \adjugate \Pi_0)} \adjugate \Pi_0\,.
\end{align}
Here, $\adjugate \Pi_0$ denotes the adjugate of $\Pi_0$, which is the transpose of the cofactor matrix.
This leads to
\begin{equation}
	\bs\sigma_{0, \pm}
		= \mp \Omega^{-1} \Sigma'(
			\Rho_0 \Xi + \half \Psi
		) \bs \alpha
		+ O(\Omega) \,.
\end{equation}

Let us analyse this result in a bit more detail.
First, $\Sigma'$ can be re-expressed in a more transparent way by considering the denominator $\trace(\Sigma^\o1 \adjugate \Pi_0) = \p_{\omegaG} \det \Sigma_0^+$. Since $\Sigma_0^+$ is obtained from $\Pi_0$ by the substitution $\omega \to \omega + \omegaG$ and $\beta \to \beta + \t\k{_\pl} \omegaG$, one has
\begin{equation}
	\trace(\Sigma^\o1 \adjugate \Pi_0)
	= \p_\omega \det \Pi_0 + \t\k{_\pl} \p_\beta \det \Pi_0\,.
\end{equation}
The ratio of these partial derivatives has a physical interpretation: differentiating the defining equation for the dispersion relation $\det \Pi_0(\beta(\omega), \omega) = 0$ with respect to $\omega$ yields
\begin{equation}
	\beta'
		\equiv \frac{\p \beta}{\p \omega}
		= - \frac{\p_\omega \det \Pi_0}{\p_\beta \det \Pi_0}\,,
\end{equation}
so the ratio describes how the propagation constant $\beta$ changes with varying frequency $\omega$ (i.e.\ the reciprocal group velocity).
Equivalently, this can be expressed in terms of the effective refractive index $\bar n = \bar n(\omega)$ as
\begin{equation}
	\beta'
		= \bar n
		+ \omega \frac{\p \bar n}{\p \omega}\,.
\end{equation}
Hence, one obtains the alternative formula
\begin{equation}
	\Sigma'
		= - \frac{\p_\beta \det \Pi_0}{\beta' - \t\k{_\pl}} \adjugate \Pi_0\,.
\end{equation}
Next, using the definition of the adjugate and the fact that $\Pi_0$ is singular, we have $\Pi_0 \Sigma' = 0$, so $\operatorname{im} \Sigma' \subset \operatorname{ker} \Pi_0$. But since the kernel of $\Pi_0$ is one-dimensional and spanned by $\bs \alpha$, it is seen that $\bs \sigma_{0,\pm}$ is again proportional to $\bs \alpha$.
This leads to the general formula
\begin{equation}
	\label{eq:sigma phase}
	\bs \sigma_{0,\pm}
		= \mp \frac{1}{4 \Omega} \frac{c_1}{\beta' - \t\k{_\pl}} \bs \alpha\,,
\end{equation}
where $c_1$ is a coefficient depending on the waveguide parameters, which we compute numerically.
The overall factor $\mp 1/4$ was chosen in analogy to similar results in vacuum, where $c_1 = \beta' = 1$.
This equation essentially determines the phase induced by the gravitational wave, as is shown in detail in \Cref{s:phase shift}.
There, we also show that appropriate emission conditions ensure that no phase shift arises if the \GW\ is collinear with the symmetry axis of the waveguide.

\paragraph{Sidebands}

For the sidebands, we find the continuity conditions to be of the form
\begin{align}
	\Sigma^\pm_{+1} C_+ \bs \sigma_{+1,\pm}
		- \tfrac{i}{\sqrt 2} \Rho^\pm_{+1} C_{+1} \bs \alpha
		+ \half \Psi_{+1} \bs \alpha
		&= 0\,,
	\\
	\Sigma^\pm_{-1} C_- \bs \sigma_{-1,\pm}
		- \tfrac{i}{\sqrt 2} \Rho^\pm_{-1} C_{-1} \bs \alpha
		+ \half \Psi_{-1} \bs \alpha
		&= 0\,,
	\\
	\Sigma^\pm_{+2} (C_+)^2 \bs \sigma_{+2,\pm}
		+ \quarter \Rho^\pm_{+2} C_{+2} \bs \alpha
		+ \half \Psi_{+2} \bs \alpha
		&= 0\,,
	\\
	\Sigma^\pm_{-2} (C_-)^2 \bs \sigma_{-2,\pm}
		+ \quarter \Rho^\pm_{-2} C_{-2} \bs \alpha
		+ \half \Psi_{-2} \bs \alpha
		&= 0\,,
\end{align}
where the matrices $C_k$ account for the coefficients $\mp U$ and $-W$ in \eqref{eq:Bessel recursion}
\begin{align}
	C_{\pm 1}
		&= \diag\left( \mp U, - W, \mp U, - W, \right)\,,
		&
	C_{\pm 2}
		&= \diag\left( U^2, W^2, U^2, W^2 \right)\,,
\end{align}
and the $\Psi$-matrices are given by
\begin{align}
	\Psi_{\pm 1}
		&= \tfrac{1}{\sqrt 2}\left(
			2 \rho \beta\, \Psi_{\pm 1}^\o{a}
			+ \Psi_{\pm 1}^\o{b}
		\right)\,,
		\\
\shortintertext{where}
	\Psi_{\pm 1}^\o{a}
		&=
		\begin{vmatrix}
			\pm \frac{\rho \beta}{U^2} J_{m\pm1}'	&
			-   \frac{\rho \beta}{W^2} K_{m\pm1}'	&
			\pm i n_1 \frac{m\pm1}{U} \frac{\rho \omega}{U^2} J_{m\pm1}	&
			-   i n_2 \frac{m\pm1}{W} \frac{\rho \omega}{W^2} K_{m\pm1}
			\\
			0	&
			0	&
			0	&
			0	
		\end{vmatrix}\,,
		\\
	\Psi_{\pm 1}^\o{b}
		&=
		\begin{vmatrix}
			-   \frac{m\pm1}{U} J_{m\pm1}	&
			\mp \frac{m\pm1}{W} K_{m\pm1}	&
			0	&
			0	
			\\
			\pm i \frac{\rho \beta}{n_1^2 U} J_{m\pm1}	&
			+   i \frac{\rho \beta}{n_2^2 W} K_{m\pm1}	&
			+   \frac{\rho \omega}{n_1 U} J_{m\pm1}	&
			\pm \frac{\rho \omega}{n_2 W} K_{m\pm1}
		\end{vmatrix}\,,
	\\
\shortintertext{and}
	\Psi_{\pm 2}
		&= \frac{1}{2} \begin{vmatrix}
			\mp i \frac{\rho \beta}{U} J_{m \pm 1}	&
			- i \frac{\rho \beta}{W} K_{m \pm 1}		&
			0	&
			0	\\
			0	&
			0	&
			0	&
			0	
		\end{vmatrix}\,.
\end{align}
Here, the arguments of the Bessel functions have been suppressed for brevity: $J_\nu = J_\nu(U)$ and $K_\nu = K_\nu(W)$.

Since we are concerned with single-mode fibres, barring one exception which is discussed below, all coefficient matrices $\Sigma^\pm_k$ with $k \neq 0$ are invertible. This is because for vanishing $\Omega$ they reduce to the non-singular matrices $\Pi_k$ (with $k \neq 0$), and invertibility is maintained for sufficiently small $\Omega$ by continuity.
Thus, as a first approximation one can replace $\Sigma^\pm_k$ by $\Pi_k$ to obtain
\begin{align}
	\label{eq:sideband amplitudes 1}
	\text{for }k = \pm 1:
	\qquad
	\bs \sigma_{k,\pm}
		&= - C_k^{-1} \Pi_k^{-1}\left(
			\half \Psi_k - \frac{i}{\sqrt{2}} \rho \beta \Rho_k C_k 
		\right) \bs \alpha
		+ O(\Omega)\,,
	\\
	\label{eq:sideband amplitudes 2}
	\text{for }k = \pm 2:
	\qquad
	\bs \sigma_{k,\pm}
		&= - C_k^{-1} \Pi_k^{-1}\left(
			\half \Psi_k + \quarter \Rho_k C_k
		\right) \bs \alpha
		+ O(\Omega)\,,
\end{align}
which shows explicitly that the sideband amplitudes are of order $\varepsilon$ only, and thus much smaller than the corrections of the central mode of order $\varepsilon \omega/\omegaG$.

The exception to the above case is the following.
Since 
\begin{align}
	J_{-\nu}(z)
		&= (-1)^\nu J_{\nu}(z)\,,
	&
	K_{-\nu}(z)
		&= + K_{\nu}(z)\,,
\end{align}
see e.g.\ Ref.~\cite[eqns.~9.1.5 on p.~358 and 9.6.6 on p.~375]{Olver1964},
the dispersion relation \eqref{eq:dispersion det explicit} is invariant under $m \to -m$,
which implies that if $m + k = -m$, i.e.\ $k = - 2 m$, then $\Pi_k$ is singular again.
In our case, since $k$ ranges from $-2$ to $+2$, this arises only if $m = \pm 1$, so this effect \emph{does} occur in single-mode fibres.
Here, we restrict the discussion to the case $m = +1$, as the alternative $m = - 1$ is completely analogous.
For this special case $m = +1$ and $k = -2$, the matrices $\Sigma^\pm_{-2}$ are close to being singular:
\begin{equation}
	\Sigma^\pm_{-2} = \Pi_{-2} \pm \Omega\, \delta \Sigma_{-2} + O(\Omega^2)\,,
\end{equation}
which has the inverse
\begin{align}
	(\Sigma^\pm_{-2})^{-1}
		&= \pm \Omega^{-1} \Sigma'' + O(\Omega^0)\,,
		\\
\shortintertext{where}
	\Sigma''
		&= \frac{1}{\trace[(\delta \Sigma_{-2}) \adjugate(\Pi_{-2}) ]} \adjugate(\Pi_{-2})\,,
\end{align}
so that instead of \eqref{eq:sideband amplitudes 2} we obtain for $k = -2$
\begin{equation}
	C_{-2} \bs \sigma_{-2,\pm}
		= \mp \Omega^{-1} \Sigma'' \left(\half \Psi_{-2} + \quarter P_{-2} C_{-2}\right) \bs \alpha + O(\Omega^0)\,.
\end{equation}
Similar to $\Pi_{0}$ whose kernel is spanned by $\bs \alpha$, the kernel of $\Pi_{-2}$ is also one-dimensional and spanned by
\begin{equation}
	\bs \alpha^*
		= (+ \alpha_1, - \alpha_2, - \alpha_3, + \alpha_4)\,.
\end{equation}
Since the range of $\Sigma''$ is the kernel of $\Pi_{-2}$, the two vectors $C_{-2} \bs \sigma_{-2,\pm}$ are proportional to $\bs \alpha^*$, and we find, similar to the central mode
\begin{equation}
	\label{eq:sigma polarisation}
	C_{-2} \bs \sigma_{-2, \pm}
		= \pm a_{-2} \frac{\varDelta^2}{2 \Omega} \frac{c_2}{\beta' - \t\k{_\pl}} \bs \alpha^*\,,
\end{equation}
where $\beta'$ is the same constant as in the phase correction \eqref{eq:sigma phase}, and the coefficient $c_2$ is a new parameter, which we determine numerically.
Here, the relative index difference $\varDelta$ is defined as
\begin{equation}
	\label{eq:Delta def}
	\varDelta =
	1 - \frac{n_2}{n_1}\,,
\end{equation}
and we have factored out $\varDelta^2/2$ such that $c_2$ is typically of order unity (see the numerical examples below).

As shown explicitly in \Cref{s:polarisation}, this formula essentially describes the perturbation of the polarisation by the gravitational wave.

\section{Perturbation of the Phase}
\label{s:phase shift}

In this section, we implement boundary conditions which model a setup where monochromatic laser light of definite polarisation is injected into the fibre.
As we will see, contrary to the unperturbed case, this will require adding further terms to the field computed so far.
This leads to an expression for the phase correction which vanishes at the point where the light enters the fibre and grows with increasing distance. A similar construction for the eikonal in a non-dispersive regime can be found in Ref.~\cite[Sect.~2]{Sazhin1997}.

Neglecting boundary effects from the ends of the waveguide, the light sent into the waveguide (or coming out of it) is simply obtained by restricting the field to a definite value of $z$, where we consider $z = 0$ for simplicity.
Also, we restrict the discussion to the “large corrections”, i.e.\ to $k = 0$ and $k = - 2m$ where the perturbations were found to be of order $1/\Omega$. The remaining “small components” are expected to be experimentally irrelevant.

Recall that the coefficients of the central mode ($k = 0$) were given by
\begin{equation}
	\bs \sigma_{0, \pm} = \mp \frac{1}{4\Omega} \frac{c_1}{\beta' - \t\k{_\pl}} \bs \alpha\,.
\end{equation}
Since this is proportional to the unperturbed coefficients $\bs \alpha$, we may write the central mode ($k = 0$) of the \emph{overall} field in the form
\begin{align}
	\t Z{^{\o0}^a_0}
	+ \varepsilon \t Z{^{\o1}^a_0}
		&= \t{\mathfrak Z}{^a} \tilde \Phi\,,
	\\
\shortintertext{where}
	\t{\mathfrak Z}{^\pl}
		&= f_0(\alpha; r) \,,
	\\
	i \zeta \t{\mathfrak Z}{^\pm}
		&= \tfrac{1}{\sqrt 2} \rho (\beta \pm i j n \omega) f_\pm(\alpha; r) \,,
	\\
	\tilde \Phi
	&= 
	e^{i \psi^\o0} \left[
		1 - \frac{\varepsilon}{4 \Omega} \frac{a_0 c_1}{\beta' - \t\k{_\pl}}\left(
			e^{+i \kappa.x_\pl} - e^{-i \kappa.x_\pl}
		\right)
	\right]\,,
	\\
	\psi^\o0
		&= \omega t - \beta z + m \theta \,.
\end{align}
Evaluating the perturbation terms at $z = 0$, we find that this describes a situation where the incoming phase is modulated in a highly specific way by the gravitational wave. This is not the situation which we want to model so that correction terms are needed: instead of the phase modulation behaviour given by $\tilde \Phi$, we are concerned with the scenario where a polarised monochromatic plane wave (as emitted by a laser source) is injected into the fibre at $z = 0$.
Hence, we require the injected field to have a definite phase of $\psi = \omega t + m \theta$ (and definite polarisation, see the next section) at $z = 0$.

Such boundary conditions can be implemented by adding appropriate solutions of the unperturbed equations of slightly shifted frequencies.
Specifically, we add terms of the form
\begin{equation}
	\t*{\mathfrak z}{^a_\pm}
		= \t{\mathfrak Z}{^a} \exp(i(\omega_\pm t - (\beta \pm \omegaG \beta' )z + m \theta))\,,
\end{equation}
where the term $\beta' = \p \beta / \p \omega$ ensures that the unperturbed dispersion relation is satisfied.
(The shift in the frequency also causes a slight perturbation in the amplitude, but this is of order $\Omega$ and thus negligible here.)
The corrected overall field is thus
\begin{equation}
	\t Z{^a_0}
		= \t{\mathfrak Z}{^a} \tilde \Phi
		+ \frac{\varepsilon}{4 \Omega} \frac{a_0 c_1}{\beta' - \t\k{_\pl}} ( \t*{\mathfrak z}{^a_+} - \t*{\mathfrak z}{^a_-} )
		= \t{\mathfrak Z}{^a} \Phi
		\,,
\end{equation}
where
\begin{equation}
	\Phi
	=
	e^{i \psi^\o0} \left[
		1 - \frac{\varepsilon}{4 \Omega} \frac{a_0 c_1}{\beta' - \t\k{_\pl}}\left(
			e^{+i \omegaG t} (e^{-i \omegaG \t\k{_\pl} z} - e^{-i \omegaG \beta' z})
			-e^{-i \omegaG t} (e^{+i \omegaG \t\k{_\pl} z} - e^{+i \omegaG \beta' z})
		\right)
	\right]\,.
\end{equation}
Assuming the fibre to be much shorter than the \GW\ wavelength, i.e. $\omegaG z \ll 1$, one finally obtains
\begin{equation}
	\Phi
	\approx
	e^{i \psi^\o0}\left[
		1 -  \ihalf \varepsilon a_0 c_1 \cos(\omegaG t) \omega z
	\right]
	\approx
	e^{i \psi^\o0} \exp(
		- \ihalf \varepsilon a_0 c_1 \cos(\omegaG t) \omega z
	)\,,
\end{equation}
which can be written as a perturbation of the phase:
\begin{equation}
	\label{eq:result phase}
	\psi
		= \omega t - \beta z + m \theta
		- \half \varepsilon a_0 c_1 \omega z \cos(\omegaG t)\,.
\end{equation}
This shows that the phase perturbation grows linearly with the distance from the emission point and oscillates in time exactly as the gravitational wave does.

\section{Perturbation of the Polarisation}
\label{s:polarisation}

Having identified the perturbation of the central mode $k = 0$ as a correction to the phase,
we now show that the other “large term” with $m + k = - m$ corresponds to a perturbation of the polarisation.

To see this, we follow Ref.~\cite{Liu2005}, where it is shown that modes with real field patterns, i.e.\ an angular dependence of the form $\cos \theta$ or $\sin \theta$, correspond to linear polarisation. To describe such fields, one must take linear combinations of the solutions with $m = +1$ and $m = -1$, and we shall make the standard assumption of weak guidance $\varDelta \equiv (n_1 - n_2)/n_1 \ll 1$.
It should be noted that non-zero values of $\varDelta$ cause the modes not to be \emph{perfectly} linearly polarised, but these deviations are independent of $t$ and $z$ and can thus not be mistaken for gravitational wave signals.

Let us briefly review the construction in the unperturbed case. For brevity, we restrict the discussion to the field in the core, as the analysis of the cladding is completely analogous.
Assuming weak guidance, the longitudinal field components are
\begin{equation}
\begin{aligned}
	\text{for }m = +1:\qquad
	\t Z{^\pl} = (1 - i j) J_{+1}(U r/\rho) e^{i(\omega t - \beta z + \theta)}\,,\\
	\text{for }m = -1:\qquad
	\t Z{^\pl} = (1 + i j) J_{-1}(U r/\rho) e^{i(\omega t - \beta z - \theta)} \,,
\end{aligned}
\end{equation}
cf.\ Ref.~\cite[Section~3.2]{Liu2005}.
The overall normalisation is arbitrary and of no concern for the discussion here.
At the level of the amplitude vector $\bs \alpha$, this means
\begin{equation}
\begin{aligned}
	\text{for }m = +1:\qquad
	\alpha_1 + j \alpha_2
		&= 1 - i j\,,
	\\
	\text{for }m = -1:\qquad
	\alpha^*_1 + j \alpha^*_2
		&= 1 + i j\,.
\end{aligned}
\end{equation}
Plugging this into the expressions for the unperturbed field \eqref{eq:unperturbed pm}, setting $\beta = n \omega$ with $n \approx n_1$ and computing the Cartesian components via \eqref{eq:differential dx dy}, one obtains the so-called $\text{HE}$-modes
\begin{equation}
	\HE{1}{\pm1}:\qquad
	\begin{pmatrix}
		Z^x	&
		Z^y
	\end{pmatrix}
	= \mathscr A(r)
	\begin{pmatrix}
		\mp i - j	&
		1 \mp i j
	\end{pmatrix}
	\,,
\end{equation}
where $\mathscr A(r) = \frac{n \rho \omega}{U} J_0(U r /\rho)$. The second index distinguishes the cases $m = +1$ and $m = -1$.
The so-called LP-modes are then given by the linear combinations
\begin{equation}
\begin{aligned}
	\LP{0}{1}{x}
		&= \ihalf( \HE{1}{+1} - \HE{1}{-1} )\,,
		\\
	\LP{0}{1}{y}
		&= \half ( \HE{1}{+1} + \HE{1}{-1} )\,,
\end{aligned}
\end{equation}
whose $x$- and $y$-components are given by
\begin{equation}
	\label{eq:modes LP unperturbed}
\begin{aligned}
	\LP{0}{1}{x}:\qquad
	\begin{pmatrix}
		Z^x	&
		Z^y
	\end{pmatrix}
	= \mathscr A(r)
	\begin{pmatrix}
		+1	&
		+j
	\end{pmatrix}\,,
	\\
	\LP{0}{1}{y}:\qquad
	\begin{pmatrix}
		Z^x	&
		Z^y
	\end{pmatrix}
	= \mathscr A(r)
	\begin{pmatrix}
		-j	&
		+1
	\end{pmatrix}\,.
\end{aligned}
\end{equation}
These modes are thus linearly polarised in the $x$- and $y$-directions, respectively.

This construction can be carried over directly to the perturbed problem.
For this, we determine the coefficients of the “large sidebands” to be
\begin{align}
	\text{for }m = +1:\qquad
	C_{-2} \bs \sigma_{-2, \pm}
		&= \pm a_{-2} \frac{\varDelta^2}{2\Omega} \frac{c_2}{\beta' - \t\k{_\pl}} \bs \alpha^*\,,
	\\
	\text{for }m = -1:\qquad
	C_{+2} \bs \sigma_{+2, \pm}
		&= \pm a_{+2} \frac{\varDelta^2}{2\Omega} \frac{c_2}{\beta' - \t\k{_\pl}} \bs \alpha\,.
\end{align}
As for the central mode, we impose boundary conditions such that these sidebands do not contribute at $z = 0$.
We may thus repeat the argument from the previous section to find that these sidebands grow (for $\omegaG z \ll 1$) linearly with the distance $z$ and oscillate in time as $\cos(\omegaG t)$, which is described by the function
\begin{equation}
	\label{eq:polarisation function}
	\tilde f(t,z) := c_2 \varDelta^2 \cos(\omegaG t)\, \omega z\,.
\end{equation}
With this, one finds (up to the overall phase correction from before) that the perturbation of the longitudinal component of the field is obtained from the unperturbed expression by the substitution
\begin{align}
	(1- i j)J_{+1}e^{+i \theta}
		&\mapsto
		(1- i j)J_{+1}e^{+i \theta}	+ i \varepsilon a_{-2} \tilde f(t,z) (1+ i j)J_{+1}e^{-i \theta}
		\,,
		\\
	(1+ i j)J_{-1}e^{-i \theta}
		&\mapsto
		(1+ i j)J_{-1}e^{+i \theta}	+ i \varepsilon a_{+2} \tilde f(t,z) (1+ i j)J_{-1}e^{+i \theta}
		\,,
\end{align}
where the arguments of the Bessel functions have been suppressed for brevity: they are ${J_{\pm 1} = J_{\pm 1}(U r/\rho)}$.
Using \eqref{eq:A++} and \eqref{eq:A--}, the factor arising here evaluates to
\begin{equation}
	i a_{\mp 2} = \ihalf( \t A{_x_x} - \t A{_y_y}) \mp \t A{_x_y}\,.
\end{equation}
Defining the abbreviations
\begin{align}
	\xi
		&= \half( \t A{_x_x} - \t A{_y_y}) \tilde f(t, z)\,,
	&
	\eta
		&= \t A{_x_y} \tilde f(t,z)\,,
\end{align}
and forming the same linear combinations as in the unperturbed case, we find the \emph{perturbed} LP-modes
\begin{equation}
	\label{eq:modes LP perturbed}
\begin{aligned}
	\LPp{0}{1}{x}:\qquad
	\begin{pmatrix}
		Z^x	&
		Z^y
	\end{pmatrix}
	= \mathscr A(r)
	\begin{pmatrix}
		+(1 - i \varepsilon \xi ) + i j \varepsilon \eta		&
		+j (1 - i \varepsilon \xi ) - i \varepsilon \eta
	\end{pmatrix}\,,
	\\
	\LPp{0}{1}{y}:\qquad
	\begin{pmatrix}
		Z^x	&
		Z^y
	\end{pmatrix}
	= \mathscr A(r)
	\begin{pmatrix}
		-j (1 + i \varepsilon \xi) - i \varepsilon \eta	&
		+(1 + i \varepsilon \xi) - i j \varepsilon \eta
	\end{pmatrix}\,.
\end{aligned}
\end{equation}
Comparing with the unperturbed LP-modes \eqref{eq:modes LP unperturbed}, we find that the effect of the gravitational wave is the following transformation of the polarisation directions:
\begin{equation}
	\label{eq:result polarisation plane}
	\begin{pmatrix}
		\f_x\\
		\f_y
	\end{pmatrix}
	\mapsto
	\begin{pmatrix}
		1 - i \varepsilon \xi		&	- i \varepsilon \eta\\
		- i \varepsilon \eta		&	1 + i \varepsilon \xi
	\end{pmatrix}
	\begin{pmatrix}
		\f_x\\
		\f_y
	\end{pmatrix}\,.
\end{equation}
The modes of different polarisation $\f_x$ and $\f_y$ thus obtain slightly different phase shifts coming from $ 1 \mp i \varepsilon \xi \approx \exp(\mp i \varepsilon \xi )$, which arise whenever $\t A{_x_x} \neq \t A{_y_y}$. This can be interpreted as weak birefringence.

In contrast, the $\eta$-terms, which correspond to $\t A{_x_y} = \t A{_y_x}$, cannot be absorbed into the phase, as they describe weak deformations of linear polarisation into elliptic polarisation of alternating direction.

To obtain a definite interpretation in terms of polarisation states as measured by a \emph{physical} polarisation filter or polarising beam splitter, one would have to model such devices in the perturbed metric. Accordingly, one might think that the above stated transformation could be absorbed in a redefinition of the polarisation directions. However, such redefinitions of the basis \emph{cannot} alter the result significantly, as the perturbed basis is necessarily related to the unperturbed one by a \emph{real} transformation, which cannot absorb the \emph{complex} transformation coefficients found here. Hence, both the birefringence term $\xi$ and the deformation term $\eta$ are genuine physical effects and not merely coordinate artifacts.

For completeness, we also give the equivalent transformation in the complex basis $\mathfrak{f}_\pm = \tfrac{1}{\sqrt{2}}(\f_x \pm i \f_y)$, which describes circular polarisation:
\begin{equation}
	\label{eq:result polarisation circular}
	\begin{pmatrix}
		\mathfrak f_+\\
		\mathfrak f_-
	\end{pmatrix}
	\mapsto
	\begin{pmatrix}
		1	&	- \varepsilon(i \xi - \eta) \\
		-\varepsilon(i \xi + \eta) 	& 1
	\end{pmatrix}
	\begin{pmatrix}
		\mathfrak f_+\\
		\mathfrak f_-
	\end{pmatrix}\,.
\end{equation}
Also in this basis, the effect of the gravitational wave is to introduce a weak mixing of the two polarisation states which grows with the distance from $z = 0$ and oscillates in time.

\section{Numerical Examples and Angular Dependence}
\label{s:numerical examples}

Having found the first order effect of weak gravitational waves on the guided modes, let us give some numerical examples.
\begin{table}[ht]
	\centering
	\begin{tabular}{llll}
		\toprule
				 &	Quantity	&	Fibre 1		&	Fibre 2\\
		\midrule
		$n_1$
				& Core refractive index
				& $1.4712$
				& $1.4715$
				 \\
		$n_2$
				& Cladding refractive index
				& $1.4659$
				& $1.4648$
				 \\
		$\bar n$
				& Effective refractive index
				& $1.4682$
				& $1.4682$
				 \\
		$\rho$	
				& Core radius
				& $4.1\,\text{µm}$
				& $4.1\,\text{µm}$
				\\
		$\varDelta$
				& Relative Index Difference
				& $0.36\%$
				& $0.45\%$
				\\
		$\mathrm{NA}$
				& Numerical Aperture
				& $0.125$
				& $0.140$
				\\
		\bottomrule
	\end{tabular}
	\caption{Parameters of typical single-mode fibres at a vacuum wavelength $\lambda = 1500\,\mathrm{nm}$.}
	\label{tab:waveguide parameters}
\end{table}
\vspace{-0.5\baselineskip}
\Cref{tab:waveguide parameters} lists two sets of parameters for typical single-mode fibres (with~$m = \pm 1$) operated at a vacuum wavelength of $\lambda = 1500\,\mathrm{nm}$.
The first parameter set is the same as in Refs.~\cite{Beig_2018} and \cite{Mieling2020}, where the influence of Earth’s gravitational field and its rotation on waveguides was analysed.
Both fibres have the same core radius and effective refractive index $\bar n$, but their constituents have different refractive indices $n_1, n_2$. Hence, their values for the relative index difference $\varDelta = (n_1 - n_2)/n_1$ and the numerical aperture $\mathrm{NA} = \sqrt{n_1^2 - n_2^2}$ also differ.

\begin{table}[ht]
	\centering
	\begin{tabular}{llll}
		\toprule
				 &	Quantity	&	Fibre 1		&	Fibre 2\\
		\midrule
		$\beta'$
				& Inverse Group Velocity
				& $1.4716$
				& $1.4723$
				\\
		$c_1$
				& Phase Parameter
				& $1.4665$
				& $1.4662$
				 \\
		$c_2$
				& Polarisation Parameter
				& $0.3482$
				& $0.3219$
				 \\
		\bottomrule
	\end{tabular}
	\caption{Numerical results for the coefficients $c_1, c_2$ and $\beta'$.}
	\label{tab:numerical results}
\end{table}
\vspace{-0.5\baselineskip}
We have implemented our calculations in \emph{Wolfram Mathematica} and computed the quantities $c_1, c_2$ and $\beta'$ which arise in \eqref{eq:sigma phase} and \eqref{eq:sigma polarisation}. The results are given in \Cref{tab:numerical results}.
The inverse group velocity $\beta' = \p \beta/\p \omega$ is larger than the effective refractive index $\bar n$, which shows that the waveguide exhibits normal dispersion $\p \bar n / \p \omega > 0 \Leftrightarrow \p \bar n / \p \lambda < 0$.

The  “phase parameter” $c_1$ determines the perturbation of the phase via \eqref{eq:result phase}.
As can be seen from \Cref{tab:waveguide parameters}, this quantity is comparable to the effective refractive index $\bar n$ (or equally $n_1$ or $n_2$): the relative differences are $0.12\%$ and $0.14\%$, respectively, which is even smaller than the relative index difference $\varDelta$. For most applications, the approximation $c_1 \approx \bar n$ will thus be sufficient.

The “polarisation parameter” $c_2$ determines the perturbation of the polarisation according to \eqref{eq:result polarisation plane} and \eqref{eq:result polarisation circular} via the function $\tilde f$ defined in \eqref{eq:polarisation function}.
Here, we see the reason for factoring $\varDelta^2$ in \eqref{eq:sigma polarisation} and \eqref{eq:polarisation function}: for the two considered fibres, the values of $\varDelta^2$ differ by a factor of approximately $1.6$, but the values of $c_2$ are almost the same.

Together, the two parameters $c_1, c_2$ fully characterise the influence of the gravitational wave on the fibre modes derived here. To describe the angular dependence of these effects, we choose coordinates $x,y,z$ where $z$ is the distance along the symmetry axis of the waveguide and the wave-vector of the gravitational wave lies in the $xz$-plane.
The unit vector $\t\k{_\pl}$ in $\kappa = \omegaG(\dd t - \t\k{_i} \dd\t x{^i})$ and the \GW\ amplitude $\t A{_i_j}$ can then be parameterised by
\begin{align}
	(\t\k{_i})
		&= (\sin \vartheta, 0, \cos \vartheta)\,,
		\\
	(\t A{_{ij}})
		&= \begin{pmatrix}
			h_+ \cos^2 \vartheta
				& h_\times \cos \vartheta
				& -h_+ \cos \vartheta \sin \vartheta
			\\
			h_\times \cos \vartheta
				& - h_+
				& - h_\times \sin \vartheta
			\\
			- h_+ \cos \vartheta \sin \vartheta
				& - h_\times \sin \vartheta
				& h_+ \sin^2 \vartheta
		\end{pmatrix}
	\,,
\end{align}
where $h_+$ and $h_\times$ are real constants describing the two possible polarisation states of the gravitational wave.
Here, $\vartheta$ denotes the incidence angle of the gravitational wave, which is not to be confused with the angular coordinate of the cylindrical coordinate system $(r, \theta, z)$.
The terms relevant for our purposes are
\begin{align}
	\t A{_z_z}
		&= h_+ \sin^2 \vartheta\,,
	&
	\half(\t A{_x_x} - \t A{_y_y})
		&= h_+ \half(1 + \cos^2 \vartheta)\,,
	&
	\t A{_x_y}
		&= h_\times \cos \vartheta\,,
\end{align}
as they determine the phase shift, the birefringence effect, and the deformation of the electromagnetic polarisation, respectively. Their angular dependence is depicted in \Cref{fig:angular dependence 0}, where we plot only absolute values in order to obtain positive radii.

As expected, the phase shift vanishes whenever the gravitational wave propagates in the same or the opposite direction as the electromagnetic wave and is maximal when the two waves propagate in orthogonal directions.
The polarisation-dependent effects are maximal in the collinear case and are minimal for orthogonal propagation. Note that the deformation of linear polarisation to elliptic polarisation vanishes for orthogonal propagation, but the birefringence effect persists in this case. Moreover, the phase and birefringence terms have a definite sign (for fixed values of $\t h{_+}$ and $\t h{_\times}$) but the polarisation term $\t A{_x_y}$ can have arbitrary sign. For example, if $\t h{_\times}$ is positive, then the “right lobe” in \Cref{fig:angular dependence 0} corresponds to positive $\t A{_x_y}$, while the “left lobe” corresponds to negative values.

The order-of-magnitude of the overall effect caused by continuous \GWs\ emitted by a rotating neutron star can be estimated by
\begin{equation}
	\varepsilon \approx 10^{-24}
		\left(\frac{e}{10^{-6}}\right)
		\left(\frac{I}{10^{38}\,\mathrm{kg}\,\mathrm{m}^2}\right)
		\left(\frac{f}{\mathrm{kHz}}\right)^{2}
		\left(\frac{d}{\mathrm{kpc}}\right)^{-1}
\end{equation}
where $I$ is the moment of inertia about the rotation axis, $e$ the eccentricity parameter, ${f = \omegaG/2\pi}$ the emitted frequency, and $d$ is the distance to the source \cites{Jaranowski1998}[Sect.~4.2]{Maggiore2007}.
(As no continuous \GWs\ have been detected so far, no precise values for these quantities are known, but the normalisations indicate plausible values.)
Using $c_1 \approx \bar n \approx 1$, the amplitude of the phase perturbation is found to be approximately
\begin{equation*}
	\delta \psi \approx
		2 \times 10^{-6}\,{\text{µ}\mathrm{rad}} \times 
		\left(\frac{e}{10^{-6}}\right)
		\left(\frac{I}{10^{38}\,\mathrm{kg}\,\mathrm{m}^2}\right)
		\left(\frac{f}{\mathrm{kHz}}\right)^{2}
		\left(\frac{d}{\mathrm{kpc}}\right)^{-1}
		\left(\frac{\ell}{\mathrm{km}}\right)\,,
\end{equation*}
where  $\ell$ is the length of the optical fibre.
Relative to this phase perturbation, the amplitude perturbation effects are suppressed by an additional factor of $\Delta^2 \approx (0.5\%)^2 = 2.5 \times 10^{-5}$, so that we expect only the phase perturbation to be of experimental relevance. We leave an experimental feasibility assessment for future work, but to the best of our knowledge, this is beyond experimental reach in the foreseeable future.

\section{Comparison with Light Propagation in Vacuum}

Finally, we comment on the relation to standard results for light propagation in vacuum.
First, consider the phase \eqref{eq:summary phase} for $m = 0$. In vacuum, the dispersion relation reduces to $\beta = \omega$ and the phase parameter is $c_1 = 1$, so that the overall phase takes the form
\begin{equation}
	\label{eq:phase in vacuum}
	\psi_\text{vac} = \t k{_\mu} \t x{^\mu} - \half \varepsilon \t A{_z_z} \omega z \cos(\omegaG t)\,,
\end{equation}
where $k = \omega (\dd t - \dd z)$.
This is in agreement with the standard formula
\begin{equation}
	\psi^\o1
		= \frac{\t k{_\mu} \t k{_\nu}}{2 \omega^2} \int_{\omega z}^{0} \t h{^\mu^\nu}(\t x{^\rho} + s \t k{^\rho}/\omega^2)) \dd s\,,
\end{equation}
where $s \mapsto \t x{^\rho} + s \t k{^\rho}/\omega^2$ is the light ray emitted from the injection point ($z = 0$) at $s = - \omega z$ and reaching the observation point at $s = 0$, cf.\ Ref.~\cite[Eq.~(3.28a)]{Finn2009}.
For the particular case considered here, this integral evaluates to
\begin{equation}
	\psi^\o1
		= - \frac{\omega \t A{_z_z}}{\omegaG (1 - \t\k{_z})}[
			\sin(\kappa.x) - \sin(\kappa.x - \omegaG z (1 - \t\k{_z}))
		]\,,
\end{equation}
and expanding to leading order in $\omegaG z \ll 1$ reproduces \eqref{eq:phase in vacuum}.
(As already observed in Ref.~\cite[Sect.~5]{Cooperstock1993}, the second term is absent in Ref.~\cite[Eq.~(2.20)]{Lobo1992} because of different boundary conditions used.  We note that without this term the result is ill-behaved in the limit $\t\k{_z} \to 1$.)

Secondly, we note that $\Delta = (n_1 - n_2)/n_1$ in \eqref{eq:summary polarisation} vanishes in vacuum, so that no polarisation perturbations arise at the order considered. This is in agreement with the geometrical optics setting, where such effects arise only at next-to-leading order in $\omegaG/\omega$, which is negligible in the considered setup.

Thus, the formulae presented here indeed reproduce the standard results for vacuum in the limit $n_1 \to 1$ and $n_2 \to 1$.

\section{Conclusion}

We have solved perturbatively the Maxwell equations for a step-index waveguide in the presence of a weak gravitational wave (\GW) of low frequency.
The correction of the central mode was found to describe a perturbation of the phase as expressed by \eqref{eq:result phase}. 
The parameter $c_1$ arising here was found to be well approximated by the effective refractive index $\bar n$: the relative error is smaller than the relative index difference $\varDelta = (n_1 - n_2)/n_1$.

Additionally, the “large sideband” (where the azimuthal mode index $m = \pm 1$ is flipped) was identified as a perturbation of the electromagnetic wave polarisation. This perturbation for linearly polarised fields is given explicitly in \eqref{eq:result polarisation plane}, and for circular polarisation in \eqref{eq:result polarisation circular}.
Compared to the phase shift, this polarisation-dependent effect (describing birefringence and coupling of polarisation states) is suppressed by a factor of $\varDelta^2$.
However, contrary to the phase shift, this effect is also present when the gravitational wave propagates along the symmetry axis of the waveguide, as it is sensitive to the components $\half( \t A{_x_x} - \t A{_y_y})$ and $\t A{_x_y}$ of the gravitational wave amplitude.
Similar results for birefringence induced by gravitational waves were obtained for plane waves in infinitely extended media \cite{Iacopini1979,Pegoraro1980}, and also in vacuum for \emph{exact} plane gravitational wave backgrounds \cite{Bini2011}.

Further sidebands, where the azimuthal mode index is shifted by $\pm  1$ or $\pm 2$, were also computed (equations \eqref{eq:sideband amplitudes 1} and \eqref{eq:sideband amplitudes 2}) but found to be negligibly small compared to the two main effects just mentioned: they are suppressed by the frequency ratio $\Omega = \omegaG/\omega$.

Comparing with light propagation in vacuum, we find that the waveguide dispersion enters in intermediate steps of the calculation, e.g.\ in \eqref{eq:sigma phase} (which is structurally similar to equation (2.20) in Ref.~\cite{Lobo1992}) via the term $\beta' = \p \beta / \p \omega$. However, having implemented physically plausible boundary conditions modelling incoming radiation, we find that the dispersion term $\beta'$ cancelled in the final results for the phase correction \eqref{eq:result phase} and for the perturbation of the polarisation \eqref{eq:result polarisation plane}.

Considering the calculations presented here in the context of previous calculations which determined the influence of Earth’s gravitational field and its rotation on fibre-modes, we conclude that the phase shifts in all three problems are well described using semi-heuristic arguments based on the eikonal equation in a “fictitious optical metric” where the discontinuous function $n$ in \eqref{eq:optical metric def} is replaced by the constant value of the effective refractive index $\bar n$ (computed from the unperturbed problem). In all cases, the difference between such geometrical-optics estimates and the result obtained by solving the true field equations was essentially determined by the relative index difference $\varDelta \ll 1$.
In contrast, the perturbation of the polarisation found here goes beyond such simplistic models, and thus required a thorough analysis of Maxwell’s equations.

Apart from neglecting terms of order $\omegaG/\omega$ (low-frequency regime) and $\rho \omegaG$ (which is consistent since $\rho \omega \sim 1$), we have neglected the elastic response of the fibre to the gravitational wave, which influences light propagation therein due to the induced displacement of the core-cladding interface (where continuity conditions had to be imposed) and due to stress-induced changes of the optical properties of the materials.
Effects of the second kind were previously studied for transparent solid bars (elasto-optical antennas), see e.g.\ Refs.~\cite{Vinet1979, Boyer1980, Vinet1985, Cros1985}, but such calculations have not been carried over to optical fibres yet.
\section*{Acknowledgements}
I thank Piotr Chruściel, Christopher Hilweg and Stefan Palenta for helpful discussions and gratefully acknowledge funding via a fellowship of the Vienna Doctoral School in Physics~(VDSP), as well as support from the Austrian Science Fund (FWF) in the course of the project P34274.
\clearpage
\appendix
\section{Derivation of the Wave Equation}
\label{appendix:wave equation}

Here, we derive the wave equation for the electromagnetic field in the \GW\ metric considered above, neglecting terms of order $\varepsilon^2$.
Throughout this section, we raise and lower indices with the \emph{full} spatial metric $\t \gs{_i_j}$.
Equation~\eqref{eq:maxwell evolution covariant} can then be written as
\begin{equation}
	n \t {\dot Z}{^i}
		+ j \t \epsilon{^i^j^k} \t \nabla{_j} \t Z{_k}
		= 0\,,
\end{equation}
where $\nabla$ denotes the spatial covariant derivative.
Lowering the index, keeping in mind that the spatial metric is time dependent, one obtains the equivalent form
\begin{equation}
	n \t {\dot Z}{_i}
		- n \t {\dot \gs}{_i_j} \t Z{^k}
		+ j \t \epsilon{_i_j_k} \t \nabla{^j} \t Z{^k}
		= 0\,.
\end{equation}
Applying $n \p_0$ to the first equation and using this last equation, we find
\begin{equation}
	n^2 \t {\ddot Z}{^i}
		+ j \t \epsilon{^i^j^k} \t \nabla{_j} ( n \t {\dot \gs}{_k_l} \t Z{^l} - j \t \epsilon{_k_l_m} \t \nabla{^l} \t Z{^m})
		= 0\,.
\end{equation}
Here, no time derivatives of the volume form occur as the trace of the metric perturbation is time-independent, and no derivatives of the Christoffel symbols occur as the exterior derivative is independent of the connection.
Using $\t \nabla{_j} \t Z{^j} = 0$ and the Ricci identity, we have
\begin{equation}
	\t \epsilon{^i^j^k} \t \epsilon{_k_l_m} \t \nabla{_j} \t \nabla{^l} \t Z{^m}
		= \t \nabla{_j} \t \nabla{^i} \t Z{^j}
		- \t \nabla{_j} \t \nabla{^j} \t Z{^i}
		= \t R{^i_j} Z^j - \Delta \t Z{^i}\,,
\end{equation}
where $\t R{^i_j}$ is the \emph{spatial} Ricci tensor.
Inserting this into the previous equation then yields
\begin{equation}
	n^2 \t {\ddot Z}{^i}
		- \Delta \t Z{^i}
		+ \t R{^i_j} Z^j
		+ j \t \epsilon{^i^j^k} \t \nabla{_j} ( n \t {\dot \gs}{_k_l} \t Z{^l})
		= 0\,.
\end{equation}
\checkedtogether{24.8.}

\section{Continuity Conditions}
\label{appendix:continuity conditions}

The jumps of the various field components are linear combinations of the various parameters $\alpha$ and $\sigma$.
For example, if $w$ denotes any field component
\begin{equation}
	\jump{w}
		= a^1 \alpha_1 + a^2 \alpha_2 + s^1 \sigma_1 + s^2 \sigma_2\,,
\end{equation}
with some coefficients $a^1,a^2,s^1,s^2$, for which we shall simply write
\begin{equation}
	\jump{w}
		= (a^1, a^2)_{\alpha} + (s^1, s^2)_{\sigma}\,.
\end{equation}
To analyse the jumps of the real fields $E, D, B, H$, we separate the $j$-real and $j$-imaginary parts.
Accordingly, we write
\begin{equation}
	(a + j b, c + j d)_{\alpha}
		= [a + j b, c + j d, -b + j a, -d + j c]_{\alpha}
\end{equation}
to state explicitly which coefficients multiply the $j$-real parameters $\Re \alpha_1, \Re \alpha_2, \Im \alpha_1, \Im \alpha_2$.
The real and imaginary parts are then given by
\begin{align}
	\label{eq:jump re}
	\Re (a + j b, c + j d)_{\alpha}
		&= [a, c, -b, -d]_{\alpha}\,,
	\\
	\label{eq:jump im}
	\Im (a + j b, c + j d)_{\alpha}
		&= [b, d, a, c]_{\alpha}\,.
\end{align}
Collecting the above mentioned real parameters of $\alpha$ into one column vector
\begin{equation}
	  \alpha
		= \begin{pmatrix}
			\Re \alpha_1	&
			\Re \alpha_2	&
			\Im \alpha_1	&
			\Im \alpha_2
		\end{pmatrix}^{\textsf{T}}\,,
\end{equation}
and similarly for $\sigma$, we may express the dependence of $\jump{Z}$, as defined in \eqref{eq:jump vector}, in matrix form
\begin{equation}
	\jump{Z} = M \bs \alpha + \Sigma \bs \sigma\,,
\end{equation}
where $M$ and $\Sigma$ are $j$-real matrices of dimension $4 \times 4$. 
In the perturbed case, we may similarly collect the real parameters of the $\sigma$’s into additional column vectors and add their contributions with additional matrices.

These “continuity matrices” fully describe the jumps of the fields at the core-cladding interface and are hence useful to implement continuity conditions. We discuss their general form and compute the matrix for the unperturbed problem explicitly.

\subsection{General Structure of the Continuity Matrices}
\label{appendix:jmp symmetries}

Since the electric and magnetic fields are derived from the single complex field $Z$ defined in \eqref{eq:complex field def}, the continuity matrices have only eight independent components.
This can be seen from the equations \eqref{eq:jump re} and \eqref{eq:jump im}: if $w$ denotes any field component, then $\jump{\Re w}$ completely determines $\jump{\Im w}$.
More precisely, if the jumps of the real fields $\Re Z^r$ and $\Re Z_\pl$ are given
\begin{align}
	\jump{\Re Z^r}
		&= [a, b, c, d]_\alpha\,,
	&
	\jump{\Re Z_\pl}
		&= [e, f, g, h]_\alpha\,,
\end{align}
then the vector $\jump{Z}$, as defined in \eqref{eq:jump vector}, is fully determined:
\begin{equation}
	\begin{pmatrix}
		\jump{D^r}		\\
		\jump{E_\pl}	\\
		\jump{B^r}		\\
		\jump{H_\pl}	\\
	\end{pmatrix}
	= 
	\begin{pmatrix}
		a		&	b		&	c		&	d		\\
		e/n_1^2	&	f/n_2^2	&	g/n_1^2	&	h/n_2^2	\\
		-c/n_1	&	-d/n_2	&	a/n_1	&	b/n_2	\\
		-g/n_1	&	-h/n_2	&	e/n_1	&	f/n_2
	\end{pmatrix}
	\begin{pmatrix}
		\Re \alpha_1	\\
		\Re \alpha_2	\\
		\Im \alpha_1	\\
		\Im \alpha_2
	\end{pmatrix}\,.
\end{equation}
Thus, the first two rows (which describe $\jump{D^r} = \Re_j \jump{Z^r}$ and $\jump{E_\pl} = \Im_j \jump{n^{-2}Z_\pl}$) determine the entire matrix $M$ in $\jump{Z} = M \bs \alpha$. The form given here is in agreement with the abbreviating notation \eqref{eq:jmp shorthand} up to redefinition of the parameters.

\subsection{Continuity Matrix of the Unperturbed Problem}
\label{appendix:jmp unperturbed}
As an example, we calculate the continuity matrix for the unperturbed problem explicitly.
From \eqref{eq:unperturbed pm} we have
\begin{equation}
	i \zeta \t Z{^{\o0}^r}
		= \rho^2 \beta \p_r \t Z{^{\o0}^\pl} - i j \rho^2 \omega m n r^{-1} \t Z{^{\o0}^\pl}\,.
\end{equation}
Using \eqref{eq:unperturbed parallel}, the jumps of $\t Z{^{\o0}^\pl}$ and its radial derivative are readily found to be
\begin{align}
	\label{eq:jump Z0 unperturbed}
	\jump{ \t Z{^{\o0}^\pl} }
		&= ( J_m(U), - K_m(W) )\,,
	\\
	\jump{ \rho \p_r \t Z{^{\o0}^\pl} }
		&= ( U J_m'(U), - W K_m'(W) )\,.
\end{align}
Here, we omit the subscript $\alpha$, as the parameters $\sigma$ do not enter the unperturbed problem.
From this, we find
\begin{equation}
	\jump{i \zeta \t Z{^{\o0}^r}}
		= \rho \beta ( U J_m'(U), - W K_m'(W) )
		- i j \rho \omega m ( n_1 J_m(U), - n_2 K_m(W) )\,.
\end{equation}
Since $\zeta$ equals $+U^2$ in the core and $-W^2$ in the cladding, we deduce
\begin{equation}
	\jump{\t Z{^{\o0}^r}}
		= - i \rho \beta \left( \frac{1}{U} J_m'(U), \frac{1}{W} K_m'(W) \right)
		- j \rho \omega m \left( \frac{n_1}{U^2} J_m(U), \frac{n_2}{W^2} K_m(W) \right)\,,
\end{equation}
and taking the $j$-real part using \eqref{eq:jump re}, we obtain
\begin{equation}
	\jump{\t D{^{\o0}^r}}
		= \left[-i \frac{\rho \beta}{U} J_m'(U), -i \frac{\rho \beta}{W} K_m'(W),
			m n_1 \frac{\rho \omega}{U^2} J_m(U), m n_2 \frac{\rho \omega}{W^2} K_m(W) \right]\,.
\end{equation}
The jump of $E_\pl = n^{-2} \Re Z^\pl$ is obtained directly from \eqref{eq:jump Z0 unperturbed}:
\begin{equation}
	\jump{\t E{^{\o0}_\pl}}
		= [ n_1^{-2} J_m(U), n_2^{-2} K_m(W), 0, 0 ]\,.
\end{equation}
The last two equations determine the unperturbed discontinuity matrix and the result is in agreement with $\Pi_0$ given by equation~\eqref{eq:jmp Pi}.

\clearpage
\printbibliography[heading=bibintoc]

@InCollection{Olver1964,
  author    = {Olver, F. W. J.},
  booktitle = {{Handbook of Mathematical Functions with Formulas, Graphs, and Mathematical Tables}},
  title     = {{Bessel Functions of Integer Order}},
  year      = {1964},
  editor    = {Abramowitz, M. and Stegun, I. A.},
  pages     = {355--433},
  crossref  = {Abramowitz1964},
}

@Book{Abramowitz1964,
  author    = {Abramowitz, M. and Stegun, I. A.},
  publisher = {U.S. Department of Commerce},
  title     = {{Handbook of Mathematical Functions with Formulas, Graphs, and Mathematical Tables}},
  year      = {1964},
  address   = {Washington, DC},
  edition   = {Tenth Printing, December 1972, with corrections},
  number    = {55},
  series    = {National Bureau of Standards Applied Mathematics Series},
  issuance  = {monographic},
  lccn      = {64060036},
  source    = {DLC},
}

@Article{Beig_2018,
  author    = {Beig, R. and Chruściel, P. T. and Hilweg, C. and Kornreich, P. and Walther, P.},
  title     = {Weakly gravitating isotropic waveguides},
  journal   = {Classical and Quantum Gravity},
  year      = {2018},
  volume    = {35},
  number    = {24},
  pages     = {244001},
  month     = nov,
  doi       = {10.1088/1361-6382/aae873},
  publisher = {{IOP} Publishing},
}

@Article{Tang2000,
  author          = {Tang, M. and Li, F.},
  journal         = {Classical and Quantum Gravity},
  title           = {{Circular waveguide: a possible gravitational wave antenna}},
  year            = {2000},
  issn            = {0264-9381},
  month           = jun,
  number          = {12},
  pages           = {2449--2453},
  volume          = {17},
  doi             = {10.1088/0264-9381/17/12/316},
  url             = {https://iopscience.iop.org/article/10.1088/0264-9381/17/12/316},
}

@Article{Cruise2000,
  author          = {Cruise, A. M.},
  journal         = {Classical and Quantum Gravity},
  title           = {{An electromagnetic detector for very-high-frequency gravitational waves}},
  year            = {2000},
  issn            = {0264-9381},
  month           = jul,
  number          = {13},
  pages           = {2525--2530},
  volume          = {17},
  doi             = {10.1088/0264-9381/17/13/305},
  url             = {https://iopscience.iop.org/article/10.1088/0264-9381/17/13/305},
}

@Article{Cooperstock1968,
  author          = {Cooperstock, F. I.},
  journal         = {Annals of Physics},
  title           = {{The interaction between electromagnetic and gravitational waves}},
  year            = {1968},
  issn            = {00034916},
  month           = mar,
  number          = {1},
  pages           = {173--181},
  volume          = {47},
  doi             = {10.1016/0003-4916(68)90233-9},
  url             = {https://linkinghub.elsevier.com/retrieve/pii/0003491668902339},
}

@Article{Cooperstock1993,
  author          = {Cooperstock, F. I. and Faraoni, V.},
  journal         = {Classical and Quantum Gravity},
  title           = {{Laser-interferometric detectors of gravitational waves}},
  year            = {1993},
  issn            = {02649381},
  number          = {6},
  pages           = {1189--1199},
  volume          = {10},
  archiveprefix   = {arXiv},
  arxivid         = {astro-ph/9303018},
  doi             = {10.1088/0264-9381/10/6/016},
  eprint          = {9303018},
  primaryclass    = {astro-ph},
}

@Article{Mensky2009,
  author          = {Mensky, M. B. and Rudenko, V. N.},
  journal         = {Gravitation and Cosmology},
  title           = {{High-frequency gravitational wave detector with electromagnetic-gravitational resonance}},
  year            = {2009},
  issn            = {0202-2893},
  month           = apr,
  number          = {2},
  pages           = {167--170},
  volume          = {15},
  doi             = {10.1134/S0202289309020133},
  url             = {http://link.springer.com/10.1134/S0202289309020133},
}

@Article{Braginsky1972,
  author          = {Braginsky, V. B. and Mensky, M. B.},
  journal         = {General Relativity and Gravitation},
  title           = {{Gravitational-electromagnetic resonance}},
  year            = {1972},
  issn            = {0001-7701},
  month           = dec,
  number          = {4},
  pages           = {401--402},
  volume          = {3},
  doi             = {10.1007/BF00759177},
  url             = {http://link.springer.com/10.1007/BF00759177},
}

@Article{Lobo1992,
  author          = {Lobo, J. A.},
  journal         = {Classical and Quantum Gravity},
  title           = {{Effect of a weak plane GW on a light beam}},
  year            = {1992},
  issn            = {0264-9381},
  month           = may,
  number          = {5},
  pages           = {1385--1394},
  volume          = {9},
  doi             = {10.1088/0264-9381/9/5/019},
  url             = {https://iopscience.iop.org/article/10.1088/0264-9381/9/5/019},
}

@Article{Finn2009,
  author          = {Finn, L. S.},
  journal         = {Physical Review D},
  title           = {{Response of interferometric gravitational wave detectors}},
  year            = {2009},
  issn            = {1550-7998},
  month           = jan,
  number          = {2},
  pages           = {022002},
  volume          = {79},
  archiveprefix   = {arXiv},
  arxivid         = {0810.4529},
  doi             = {10.1103/PhysRevD.79.022002},
  eprint          = {0810.4529},
  url             = {https://link.aps.org/doi/10.1103/PhysRevD.79.022002},
}

@Article{Rakhmanov2009,
  author          = {Rakhmanov, M.},
  journal         = {Classical and Quantum Gravity},
  title           = {{On the round-trip time for a photon propagating in the field of a plane gravitational wave}},
  year            = {2009},
  issn            = {0264-9381},
  month           = aug,
  number          = {15},
  pages           = {155010},
  volume          = {26},
  doi             = {10.1088/0264-9381/26/15/155010},
  url             = {https://iopscience.iop.org/article/10.1088/0264-9381/26/15/155010},
}

@Book{Jackson1998,
  author    = {Jackson, J. D.},
  publisher = {John Wiley \& Sons, Inc.},
  title     = {{Classical Electrodynamics}},
  year      = {1998},
  edition   = {Third Edition},
  isbn      = {978-0-471-30932-1},
  pagetotal = {832},
}

@Article{Gordon1923,
  author    = {Gordon, W.},
  journal   = {Annalen der Physik},
  title     = {{Zur Lichtfortpflanzung nach der Relativitätstheorie}},
  year      = {1923},
  number    = {22},
  pages     = {421--456},
  volume    = {377},
  doi       = {10.1002/andp.19233772202},
  publisher = {Wiley},
}

@Book{Liu2005,
  author    = {Liu, J.-M.},
  publisher = {Cambridge University Press},
  title     = {{Photonic Devices}},
  year      = {2005},
  doi       = {10.1017/cbo9780511614255},
}

@Book{Davis2009,
  author    = {Davis, C. C.},
  publisher = {Cambridge University Press},
  title     = {{Lasers and Electro-optics}},
  year      = {2009},
  edition   = {2nd edition},
  doi       = {10.1017/cbo9781139016629},
}

@Article{Mieling2020,
  author    = {Mieling, T. B.},
  journal   = {Classical and Quantum Gravity},
  title     = {{On the influence of Earth's rotation on light propagation in waveguides}},
  year      = {2020},
  month     = oct,
  number    = {22},
  pages     = {225001},
  volume    = {37},
  doi       = {10.1088/1361-6382/ababb2},
  publisher = {{IOP} Publishing},
}

@Article{Iacopini1979,
  author    = {Iacopini, E. and Picasso, E. and Pegoraro, F. and Radicati, L. A.},
  journal   = {Physics Letters A},
  title     = {Birefringence induced by gravitational waves: A suggestion for a new detector},
  year      = {1979},
  month     = sep,
  number    = {2},
  pages     = {140--142},
  volume    = {73},
  doi       = {10.1016/0375-9601(79)90460-2},
  publisher = {Elsevier {BV}},
}

@Article{Pegoraro1980,
  author    = {Pegoraro, F. and Radicati, L. A.},
  journal   = {Journal of Physics A: Mathematical and General},
  title     = {Dielectric tensor and magnetic permeability in the weak field approximation of general relativity},
  year      = {1980},
  month     = jul,
  number    = {7},
  pages     = {2411--2421},
  volume    = {13},
  doi       = {10.1088/0305-4470/13/7/024},
  publisher = {{IOP} Publishing},
}

@Article{Bini2011,
  author    = {Bini, D. and Fortini, P. and Haney, M.and Ortolan, A.},
  journal   = {Classical and Quantum Gravity},
  title     = {Electromagnetic waves in gravitational wave spacetimes},
  year      = {2011},
  month     = nov,
  number    = {23},
  pages     = {235007},
  volume    = {28},
  doi       = {10.1088/0264-9381/28/23/235007},
  publisher = {{IOP} Publishing},
}

@Book{Born2019,
  author    = {Born, M. and Wolf, E.},
  publisher = {Cambridge University Press},
  title     = {{Principles of Optics}},
  year      = {2019},
  edition   = {60th Anniversary Edition},
  doi       = {10.1017/9781108769914},
}

@Article{Vinet1985,
  author    = {Vinet, J.-Y.},
  journal   = {Annales de Physique},
  title     = {Elasto optical antennas},
  year      = {1985},
  number    = {3},
  pages     = {253--261},
  volume    = {10},
  doi       = {10.1051/anphys:01985001003025300},
  publisher = {{EDP} Sciences},
}

@Article{Cros1985,
  author    = {du Cros, F. T.},
  journal   = {Annales de Physique},
  title     = {Electromagnetic and elasto-optical systems for the reception or generation of gravitational radiation},
  year      = {1985},
  number    = {3},
  pages     = {263--286},
  volume    = {10},
  doi       = {10.1051/anphys:01985001003026300},
  publisher = {{EDP} Sciences},
}

@Article{Vinet1979,
  author    = {Vinet, J.-Y.},
  journal   = {Annales de l'I.H.P. Physique th\'eorique},
  title     = {Elasto-optical detection of gravitational waves},
  year      = {1979},
  number    = {3},
  pages     = {251--262},
  volume    = {30},
  publisher = {Gauthier-Villars},
  url       = {http://www.numdam.org/item/AIHPA_1979__30_3_251_0},
}

@Article{Boyer1980,
  author    = {Boyer, G. R. and Lamouroux, B. F. and Prade, B. S. and Vinet, J.-Y.},
  journal   = {Applied Optics},
  title     = {Elastooptical antenna for detection of gravitational radiation},
  year      = {1980},
  month     = feb,
  number    = {3},
  pages     = {382},
  volume    = {19},
  doi       = {10.1364/ao.19.000382},
  publisher = {The Optical Society},
}

@Article{Rudenko1980,
  author    = {Rudenko, V. N. and Sazhin, M. V.},
  journal   = {Soviet Journal of Quantum Electronics},
  title     = {Laser interferometer as a gravitational wave detector},
  year      = {1980},
  month     = nov,
  number    = {11},
  pages     = {1366--1373},
  volume    = {10},
  doi       = {10.1070/qe1980v010n11abeh010312},
  publisher = {{IOP} Publishing},
}

@Book{Hodgson2005,
  author    = {Hodgson, N. and Weber, H.},
  publisher = {Springer, New York},
  title     = {{Laser Resonators and Beam Propagation}},
  year      = {2005},
  edition   = {Second Edition},
  isbn      = {978-0-387-40078-5},
  number    = {108},
  series    = {Springer Series in Optical Sciences},
  doi       = {10.1007/b106789},
}

@Book{Damask2004,
  author    = {Damask, J. N.},
  publisher = {Springer, New York},
  title     = {{Polarization Optics in Telecommunications}},
  year      = {2004},
  isbn      = {978-0-387-22493-0},
  number    = {101},
  series    = {Springer Series in Optical Sciences},
  doi       = {10.1007/b137386},
}

@TechReport{Linsay1983,
  author      = {Linsay, P. and Saulson, P. and Weiss, R. and Whitcomb, S.},
  institution = {Prepared for the National Science Foundation},
  title       = {{A Study of a Long Baseline Gravitational Wave Antenna System}},
  year        = {1983},
  type        = {techreport},
  url         = {https://dcc.ligo.org/LIGO-T830001/public},
}

@PhdThesis{Adhikari2004,
  author = {Adhikari, R.},
  school = {Massachusetts Institute of Technology},
  title  = {Sensitivity and noise analysis of 4 km laser interferometric gravitational wave antennae},
  year   = {2004},
  month  = jul,
  type   = {phdthesis},
  url    = {https://dspace.mit.edu/handle/1721.1/28646},
}

@Article{Cornish2009,
  author    = {Neil J. Cornish},
  journal   = {Physical Review D},
  title     = {Alternative derivation of the response of interferometric gravitational wave detectors},
  year      = {2009},
  month     = oct,
  number    = {8},
  volume    = {80},
  doi       = {10.1103/physrevd.80.087101},
  publisher = {American Physical Society ({APS})},
}

@Article{Sazhin1997,
  author    = {Sazhin, M. V. and Markova, S. N.},
  journal   = {Physics Letters A},
  title     = {Optical resonator in the field of gravitational waves},
  year      = {1997},
  month     = aug,
  number    = {1-2},
  pages     = {43--48},
  volume    = {233},
  doi       = {10.1016/s0375-9601(97)00436-2},
  publisher = {Elsevier {BV}},
}

@Book{Saulson1994,
  author    = {Saulson, P. R.},
  publisher = {World Scientific},
  title     = {{Fundamentals of Interferometric Gravitational Wave Detectors}},
  year      = {1994},
  isbn      = {981-02-1820-6},
  doi       = {10.1142/2410},
}

@Article{Koop2014,
  author    = {Koop, M. J. and Finn, L. S.},
  journal   = {Physical Review D},
  title     = {Physical response of light-time gravitational wave detectors},
  year      = {2014},
  month     = sep,
  number    = {6},
  volume    = {90},
  doi       = {10.1103/physrevd.90.062002},
  publisher = {American Physical Society ({APS})},
}

@Article{Forward1978,
  author    = {Forward, R. L.},
  journal   = {Physical Review D},
  title     = {Wideband laser-interferometer graviational-radiation experiment},
  year      = {1978},
  month     = jan,
  number    = {2},
  pages     = {379--390},
  volume    = {17},
  doi       = {10.1103/physrevd.17.379},
  publisher = {American Physical Society ({APS})},
}

@Article{Schutz1987,
  author    = {Schutz, B. F. and Tinto, M.},
  journal   = {Monthly Notices of the Royal Astronomical Society},
  title     = {{Antenna patterns of interferometric detectors of gravitational waves - I. Linearly polarized waves}},
  year      = {1987},
  month     = jan,
  number    = {1},
  pages     = {131--154},
  volume    = {224},
  doi       = {10.1093/mnras/224.1.131},
  publisher = {Oxford University Press ({OUP})},
}

@Article{Faraoni2007,
  author    = {Faraoni, V.},
  journal   = {General Relativity and Gravitation},
  title     = {A common misconception about {LIGO} detectors of gravitational waves},
  year      = {2007},
  month     = mar,
  number    = {5},
  pages     = {677--684},
  volume    = {39},
  doi       = {10.1007/s10714-007-0415-5},
  publisher = {Springer Science and Business Media {LLC}},
}

@Article{Abbott2016,
  author    = {Abbott, B. P. and Abbott, R. and Abbott, T. D. and Abernathy, M. R. and Acernese, F. and Ackley, K. and Adams, C. and Adams, T. and Addesso, P. and Adhikari, R. X. and Adya, V. B. and Affeldt, C. and Agathos, M. and Agatsuma, K. and Aggarwal, N. and Aguiar, O. D. and Aiello, L. and Ain, A. and Ajith, P. and Allen, B. and Allocca, A. and Altin, P. A. and Anderson, S. B. and Anderson, W. G. and Arai, K. and Arain, M. A. and Araya, M. C. and Arceneaux, C. C. and Areeda, J. S. and Arnaud, N. and Arun, K. G. and Ascenzi, S. and Ashton, G. and Ast, M. and Aston, S. M. and Astone, P. and Aufmuth, P. and Aulbert, C. and Babak, S. and Bacon, P. and Bader, M. K. M. and Baker, P. T. and Baldaccini, F. and Ballardin, G. and Ballmer, S. W. and Barayoga, J. C. and Barclay, S. E. and Barish, B. C. and Barker, D. and Barone, F. and Barr, B. and Barsotti, L. and Barsuglia, M. and Barta, D. and Bartlett, J. and Barton, M. A. and Bartos, I. and Bassiri, R. and Basti, A. and Batch, J. C. and Baune, C. and Bavigadda, V. and Bazzan, M. and Behnke, B. and Bejger, M. and Belczynski, C. and Bell, A. S. and Bell, C. J. and Berger, B. K. and Bergman, J. and Bergmann, G. and Berry, C. P. L. and Bersanetti, D. and Bertolini, A. and Betzwieser, J. and Bhagwat, S. and Bhandare, R. and Bilenko, I. A. and Billingsley, G. and Birch, J. and Birney, R. and Birnholtz, O. and Biscans, S. and Bisht, A. and Bitossi, M. and Biwer, C. and Bizouard, M. A. and Blackburn, J. K. and Blair, C. D. and Blair, D. G. and Blair, R. M. and Bloemen, S. and Bock, O. and Bodiya, T. P. and Boer, M. and Bogaert, G. and Bogan, C. and Bohe, A. and Bojtos, P. and Bond, C. and Bondu, F. and Bonnand, R. and Boom, B. A. and Bork, R. and Boschi, V. and Bose, S. and Bouffanais, Y. and Bozzi, A. and Bradaschia, C. and Brady, P. R. and Braginsky, V. B. and Branchesi, M. and Brau, J. E. and Briant, T. and Brillet, A. and Brinkmann, M. and Brisson, V. and Brockill, P. and Brooks, A. F. and Brown, D. A. and Brown, D. D. and Brown, N. M. and Buchanan, C. C. and Buikema, A. and Bulik, T. and Bulten, H. J. and Buonanno, A. and Buskulic, D. and Buy, C. and Byer, R. L. and Cabero, M. and Cadonati, L. and Cagnoli, G. and Cahillane, C. and Bustillo, J. Calder{\'{o}}n and Callister, T. and Calloni, E. and Camp, J. B. and Cannon, K. C. and Cao, J. and Capano, C. D. and Capocasa, E. and Carbognani, F. and Caride, S. and Diaz, J. Casanueva and Casentini, C. and Caudill, S. and Cavagli{\`{a}}, M. and Cavalier, F. and Cavalieri, R. and Cella, G. and Cepeda, C. B. and Baiardi, L. Cerboni and Cerretani, G. and Cesarini, E. and Chakraborty, R. and Chalermsongsak, T. and Chamberlin, S. J. and Chan, M. and Chao, S. and Charlton, P. and Chassande-Mottin, E. and Chen, H. Y. and Chen, Y. and Cheng, C. and Chincarini, A. and Chiummo, A. and Cho, H. S. and Cho, M. and Chow, J. H. and Christensen, N. and Chu, Q. and Chua, S. and Chung, S. and Ciani, G. and Clara, F. and Clark, J. A. and Cleva, F. and Coccia, E. and Cohadon, P.-F. and Colla, A. and Collette, C. G. and Cominsky, L. and Constancio, M. and Conte, A. and Conti, L. and Cook, D. and Corbitt, T. R. and Cornish, N. and Corsi, A. and Cortese, S. and Costa, C. A. and Coughlin, M. W. and Coughlin, S. B. and Coulon, J.-P. and Countryman, S. T. and Couvares, P. and Cowan, E. E. and Coward, D. M. and Cowart, M. J. and Coyne, D. C. and Coyne, R. and Craig, K. and Creighton, J. D. E. and Creighton, T. D. and Cripe, J. and Crowder, S. G. and Cruise, A. M. and Cumming, A. and Cunningham, L. and Cuoco, E. and Canton, T. Dal and Danilishin, S. L. and D'Antonio, S. and Danzmann, K. and Darman, N. S. and Costa, C. F. Da Silva and Dattilo, V. and Dave, I. and Daveloza, H. P. and Davier, M. and Davies, G. S. and Daw, E. J. and Day, R. and De, S. and DeBra, D. and Debreczeni, G. and Degallaix, J. and Laurentis, M. De and Del{\'{e}}glise, S. and Pozzo, W. Del and Denker, T. and Dent, T. and Dereli, H. and Dergachev, V. and DeRosa, R. T. and Rosa, R. De and DeSalvo, R. and Dhurandhar, S. and D{\'{\i}}az, M. C. and Fiore, L. Di and Giovanni, M. Di and Lieto, A. Di and Pace, S. Di and Palma, I. Di and Virgilio, A. Di and Dojcinoski, G. and Dolique, V. and Donovan, F. and Dooley, K. L. and Doravari, S. and Douglas, R. and Downes, T. P. and Drago, M. and Drever, R. W. P. and Driggers, J. C. and Du, Z. and Ducrot, M. and Dwyer, S. E. and Edo, T. B. and Edwards, M. C. and Effler, A. and Eggenstein, H.-B. and Ehrens, P. and Eichholz, J. and Eikenberry, S. S. and Engels, W. and Essick, R. C. and Etzel, T. and Evans, M. and Evans, T. M. and Everett, R. and Factourovich, M. and Fafone, V. and Fair, H. and Fairhurst, S. and Fan, X. and Fang, Q. and Farinon, S. and Farr, B. and Farr, W. M. and Favata, M. and Fays, M. and Fehrmann, H. and Fejer, M. M. and Feldbaum, D. and Ferrante, I. and Ferreira, E. C. and Ferrini, F. and Fidecaro, F. and Finn, L. S. and Fiori, I. and Fiorucci, D. and Fisher, R. P. and Flaminio, R. and Fletcher, M. and Fong, H. and Fournier, J.-D. and Franco, S. and Frasca, S. and Frasconi, F. and Frede, M. and Frei, Z. and Freise, A. and Frey, R. and Frey, V. and Fricke, T. T. and Fritschel, P. and Frolov, V. V. and Fulda, P. and Fyffe, M. and Gabbard, H. A. G. and Gair, J. R. and Gammaitoni, L. and Gaonkar, S. G. and Garufi, F. and Gatto, A. and Gaur, G. and Gehrels, N. and Gemme, G. and Gendre, B. and Genin, E. and Gennai, A. and George, J. and Gergely, L. and Germain, V. and Ghosh, Abhirup and Ghosh, Archisman and Ghosh, S. and Giaime, J. A. and Giardina, K. D. and Giazotto, A. and Gill, K. and Glaefke, A. and Gleason, J. R. and Goetz, E. and Goetz, R. and Gondan, L. and Gonz{\'{a}}lez, G. and Castro, J. M. Gonzalez and Gopakumar, A. and Gordon, N. A. and Gorodetsky, M. L. and Gossan, S. E. and Gosselin, M. and Gouaty, R. and Graef, C. and Graff, P. B. and Granata, M. and Grant, A. and Gras, S. and Gray, C. and Greco, G. and Green, A. C. and Greenhalgh, R. J. S. and Groot, P. and Grote, H. and Grunewald, S. and Guidi, G. M. and Guo, X. and Gupta, A. and Gupta, M. K. and Gushwa, K. E. and Gustafson, E. K. and Gustafson, R. and Hacker, J. J. and Hall, B. R. and Hall, E. D. and Hammond, G. and Haney, M. and Hanke, M. M. and Hanks, J. and Hanna, C. and Hannam, M. D. and Hanson, J. and Hardwick, T. and Harms, J. and Harry, G. M. and Harry, I. W. and Hart, M. J. and Hartman, M. T. and Haster, C.-J. and Haughian, K. and Healy, J. and Heefner, J. and Heidmann, A. and Heintze, M. C. and Heinzel, G. and Heitmann, H. and Hello, P. and Hemming, G. and Hendry, M. and Heng, I. S. and Hennig, J. and Heptonstall, A. W. and Heurs, M. and Hild, S. and Hoak, D. and Hodge, K. A. and Hofman, D. and Hollitt, S. E. and Holt, K. and Holz, D. E. and Hopkins, P. and Hosken, D. J. and Hough, J. and Houston, E. A. and Howell, E. J. and Hu, Y. M. and Huang, S. and Huerta, E. A. and Huet, D. and Hughey, B. and Husa, S. and Huttner, S. H. and Huynh-Dinh, T. and Idrisy, A. and Indik, N. and Ingram, D. R. and Inta, R. and Isa, H. N. and Isac, J.-M. and Isi, M. and Islas, G. and Isogai, T. and Iyer, B. R. and Izumi, K. and Jacobson, M. B. and Jacqmin, T. and Jang, H. and Jani, K. and Jaranowski, P. and Jawahar, S. and Jim{\'{e}}nez-Forteza, F. and Johnson, W. W. and Johnson-McDaniel, N. K. and Jones, D. I. and Jones, R. and Jonker, R. J. G. and Ju, L. and Haris, K. and Kalaghatgi, C. V. and Kalogera, V. and Kandhasamy, S. and Kang, G. and Kanner, J. B. and Karki, S. and Kasprzack, M. and Katsavounidis, E. and Katzman, W. and Kaufer, S. and Kaur, T. and Kawabe, K. and Kawazoe, F. and K{\'{e}}f{\'{e}}lian, F. and Kehl, M. S. and Keitel, D. and Kelley, D. B. and Kells, W. and Kennedy, R. and Keppel, D. G. and Key, J. S. and Khalaidovski, A. and Khalili, F. Y. and Khan, I. and Khan, S. and Khan, Z. and Khazanov, E. A. and Kijbunchoo, N. and Kim, C. and Kim, J. and Kim, K. and Kim, Nam-Gyu and Kim, Namjun and Kim, Y.-M. and King, E. J. and King, P. J. and Kinzel, D. L. and Kissel, J. S. and Kleybolte, L. and Klimenko, S. and Koehlenbeck, S. M. and Kokeyama, K. and Koley, S. and Kondrashov, V. and Kontos, A. and Koranda, S. and Korobko, M. and Korth, W. Z. and Kowalska, I. and Kozak, D. B. and Kringel, V. and Krishnan, B. and Kr{\'{o}}lak, A. and Krueger, C. and Kuehn, G. and Kumar, P. and Kumar, R. and Kuo, L. and Kutynia, A. and Kwee, P. and Lackey, B. D. and Landry, M. and Lange, J. and Lantz, B. and Lasky, P. D. and Lazzarini, A. and Lazzaro, C. and Leaci, P. and Leavey, S. and Lebigot, E. O. and Lee, C. H. and Lee, H. K. and Lee, H. M. and Lee, K. and Lenon, A. and Leonardi, M. and Leong, J. R. and Leroy, N. and Letendre, N. and Levin, Y. and Levine, B. M. and Li, T. G. F. and Libson, A. and Littenberg, T. B. and Lockerbie, N. A. and Logue, J. and Lombardi, A. L. and London, L. T. and Lord, J. E. and Lorenzini, M. and Loriette, V. and Lormand, M. and Losurdo, G. and Lough, J. D. and Lousto, C. O. and Lovelace, G. and Lück, H. and Lundgren, A. P. and Luo, J. and Lynch, R. and Ma, Y. and MacDonald, T. and Machenschalk, B. and MacInnis, M. and Macleod, D. M. and Maga{\~{n}}a-Sandoval, F. and Magee, R. M. and Mageswaran, M. and Majorana, E. and Maksimovic, I. and Malvezzi, V. and Man, N. and Mandel, I. and Mandic, V. and Mangano, V. and Mansell, G. L. and Manske, M. and Mantovani, M. and Marchesoni, F. and Marion, F. and M{\'{a}}rka, S. and M{\'{a}}rka, Z. and Markosyan, A. S. and Maros, E. and Martelli, F. and Martellini, L. and Martin, I. W. and Martin, R. M. and Martynov, D. V. and Marx, J. N. and Mason, K. and Masserot, A. and Massinger, T. J. and Masso-Reid, M. and Matichard, F. and Matone, L. and Mavalvala, N. and Mazumder, N. and Mazzolo, G. and McCarthy, R. and McClelland, D. E. and McCormick, S. and McGuire, S. C. and McIntyre, G. and McIver, J. and McManus, D. J. and McWilliams, S. T. and Meacher, D. and Meadors, G. D. and Meidam, J. and Melatos, A. and Mendell, G. and Mendoza-Gandara, D. and Mercer, R. A. and Merilh, E. and Merzougui, M. and Meshkov, S. and Messenger, C. and Messick, C. and Meyers, P. M. and Mezzani, F. and Miao, H. and Michel, C. and Middleton, H. and Mikhailov, E. E. and Milano, L. and Miller, J. and Millhouse, M. and Minenkov, Y. and Ming, J. and Mirshekari, S. and Mishra, C. and Mitra, S. and Mitrofanov, V. P. and Mitselmakher, G. and Mittleman, R. and Moggi, A. and Mohan, M. and Mohapatra, S. R. P. and Montani, M. and Moore, B. C. and Moore, C. J. and Moraru, D. and Moreno, G. and Morriss, S. R. and Mossavi, K. and Mours, B. and Mow-Lowry, C. M. and Mueller, C. L. and Mueller, G. and Muir, A. W. and Mukherjee, Arunava and Mukherjee, D. and Mukherjee, S. and Mukund, N. and Mullavey, A. and Munch, J. and Murphy, D. J. and Murray, P. G. and Mytidis, A. and Nardecchia, I. and Naticchioni, L. and Nayak, R. K. and Necula, V. and Nedkova, K. and Nelemans, G. and Neri, M. and Neunzert, A. and Newton, G. and Nguyen, T. T. and Nielsen, A. B. and Nissanke, S. and Nitz, A. and Nocera, F. and Nolting, D. and Normandin, M. E. N. and Nuttall, L. K. and Oberling, J. and Ochsner, E. and O'Dell, J. and Oelker, E. and Ogin, G. H. and Oh, J. J. and Oh, S. H. and Ohme, F. and Oliver, M. and Oppermann, P. and Oram, Richard J. and O'Reilly, B. and O'Shaughnessy, R. and Ott, C. D. and Ottaway, D. J. and Ottens, R. S. and Overmier, H. and Owen, B. J. and Pai, A. and Pai, S. A. and Palamos, J. R. and Palashov, O. and Palomba, C. and Pal-Singh, A. and Pan, H. and Pan, Y. and Pankow, C. and Pannarale, F. and Pant, B. C. and Paoletti, F. and Paoli, A. and Papa, M. A. and Paris, H. R. and Parker, W. and Pascucci, D. and Pasqualetti, A. and Passaquieti, R. and Passuello, D. and Patricelli, B. and Patrick, Z. and Pearlstone, B. L. and Pedraza, M. and Pedurand, R. and Pekowsky, L. and Pele, A. and Penn, S. and Perreca, A. and Pfeiffer, H. P. and Phelps, M. and Piccinni, O. and Pichot, M. and Pickenpack, M. and Piergiovanni, F. and Pierro, V. and Pillant, G. and Pinard, L. and Pinto, I. M. and Pitkin, M. and Poeld, J. H. and Poggiani, R. and Popolizio, P. and Post, A. and Powell, J. and Prasad, J. and Predoi, V. and Premachandra, S. S. and Prestegard, T. and Price, L. R. and Prijatelj, M. and Principe, M. and Privitera, S. and Prix, R. and Prodi, G. A. and Prokhorov, L. and Puncken, O. and Punturo, M. and Puppo, P. and Pürrer, M. and Qi, H. and Qin, J. and Quetschke, V. and Quintero, E. A. and Quitzow-James, R. and Raab, F. J. and Rabeling, D. S. and Radkins, H. and Raffai, P. and Raja, S. and Rakhmanov, M. and Ramet, C. R. and Rapagnani, P. and Raymond, V. and Razzano, M. and Re, V. and Read, J. and Reed, C. M. and Regimbau, T. and Rei, L. and Reid, S. and Reitze, D. H. and Rew, H. and Reyes, S. D. and Ricci, F. and Riles, K. and Robertson, N. A. and Robie, R. and Robinet, F. and Rocchi, A. and Rolland, L. and Rollins, J. G. and Roma, V. J. and Romano, J. D. and Romano, R. and Romanov, G. and Romie, J. H. and Rosi{\'{n}}ska, D. and Rowan, S. and Rüdiger, A. and Ruggi, P. and Ryan, K. and Sachdev, S. and Sadecki, T. and Sadeghian, L. and Salconi, L. and Saleem, M. and Salemi, F. and Samajdar, A. and Sammut, L. and Sampson, L. M. and Sanchez, E. J. and Sandberg, V. and Sandeen, B. and Sanders, G. H. and Sanders, J. R. and Sassolas, B. and Sathyaprakash, B. S. and Saulson, P. R. and Sauter, O. and Savage, R. L. and Sawadsky, A. and Schale, P. and Schilling, R. and Schmidt, J. and Schmidt, P. and Schnabel, R. and Schofield, R. M. S. and Schönbeck, A. and Schreiber, E. and Schuette, D. and Schutz, B. F. and Scott, J. and Scott, S. M. and Sellers, D. and Sengupta, A. S. and Sentenac, D. and Sequino, V. and Sergeev, A. and Serna, G. and Setyawati, Y. and Sevigny, A. and Shaddock, D. A. and Shaffer, T. and Shah, S. and Shahriar, M. S. and Shaltev, M. and Shao, Z. and Shapiro, B. and Shawhan, P. and Sheperd, A. and Shoemaker, D. H. and Shoemaker, D. M. and Siellez, K. and Siemens, X. and Sigg, D. and Silva, A. D. and Simakov, D. and Singer, A. and Singer, L. P. and Singh, A. and Singh, R. and Singhal, A. and Sintes, A. M. and Slagmolen, B. J. J. and Smith, J. R. and Smith, M. R. and Smith, N. D. and Smith, R. J. E. and Son, E. J. and Sorazu, B. and Sorrentino, F. and Souradeep, T. and Srivastava, A. K. and Staley, A. and Steinke, M. and Steinlechner, J. and Steinlechner, S. and Steinmeyer, D. and Stephens, B. C. and Stevenson, S. P. and Stone, R. and Strain, K. A. and Straniero, N. and Stratta, G. and Strauss, N. A. and Strigin, S. and Sturani, R. and Stuver, A. L. and Summerscales, T. Z. and Sun, L. and Sutton, P. J. and Swinkels, B. L. and Szczepa{\'{n}}czyk, M. J. and Tacca, M. and Talukder, D. and Tanner, D. B. and T{\'{a}}pai, M. and Tarabrin, S. P. and Taracchini, A. and Taylor, R. and Theeg, T. and Thirugnanasambandam, M. P. and Thomas, E. G. and Thomas, M. and Thomas, P. and Thorne, K. A. and Thorne, K. S. and Thrane, E. and Tiwari, S. and Tiwari, V. and Tokmakov, K. V. and Tomlinson, C. and Tonelli, M. and Torres, C. V. and Torrie, C. I. and Töyrä, D. and Travasso, F. and Traylor, G. and Trifir{\`{o}}, D. and Tringali, M. C. and Trozzo, L. and Tse, M. and Turconi, M. and Tuyenbayev, D. and Ugolini, D. and Unnikrishnan, C. S. and Urban, A. L. and Usman, S. A. and Vahlbruch, H. and Vajente, G. and Valdes, G. and Vallisneri, M. and van Bakel, N. and van Beuzekom, M. and van den Brand, J. F. J. and Broeck, C. Van Den and Vander-Hyde, D. C. and van der Schaaf, L. and van Heijningen, J. V. and van Veggel, A. A. and Vardaro, M. and Vass, S. and Vas{\'{u}}th, M. and Vaulin, R. and Vecchio, A. and Vedovato, G. and Veitch, J. and Veitch, P. J. and Venkateswara, K. and Verkindt, D. and Vetrano, F. and Vicer{\'{e}}, A. and Vinciguerra, S. and Vine, D. J. and Vinet, J.-Y. and Vitale, S. and Vo, T. and Vocca, H. and Vorvick, C. and Voss, D. and Vousden, W. D. and Vyatchanin, S. P. and Wade, A. R. and Wade, L. E. and Wade, M. and Waldman, S. J. and Walker, M. and Wallace, L. and Walsh, S. and Wang, G. and Wang, H. and Wang, M. and Wang, X. and Wang, Y. and Ward, H. and Ward, R. L. and Warner, J. and Was, M. and Weaver, B. and Wei, L.-W. and Weinert, M. and Weinstein, A. J. and Weiss, R. and Welborn, T. and Wen, L. and We{\ss}els, P. and Westphal, T. and Wette, K. and Whelan, J. T. and Whitcomb, S. E. and White, D. J. and Whiting, B. F. and Wiesner, K. and Wilkinson, C. and Willems, P. A. and Williams, L. and Williams, R. D. and Williamson, A. R. and Willis, J. L. and Willke, B. and Wimmer, M. H. and Winkelmann, L. and Winkler, W. and Wipf, C. C. and Wiseman, A. G. and Wittel, H. and Woan, G. and Worden, J. and Wright, J. L. and Wu, G. and Yablon, J. and Yakushin, I. and Yam, W. and Yamamoto, H. and Yancey, C. C. and Yap, M. J. and Yu, H. and Yvert, M. and Zadro{\.{z}}ny, A. and Zangrando, L. and Zanolin, M. and Zendri, J.-P. and Zevin, M. and Zhang, F. and Zhang, L. and Zhang, M. and Zhang, Y. and Zhao, C. and Zhou, M. and Zhou, Z. and Zhu, X. J. and Zucker, M. E. and Zuraw, S. E. and Zweizig, J.},
  journal   = {Physical Review Letters},
  title     = {{Observation of Gravitational Waves from a Binary Black Hole Merger}},
  year      = {2016},
  month     = feb,
  number    = {6},
  volume    = {116},
  doi       = {10.1103/physrevlett.116.061102},
  publisher = {American Physical Society ({APS})},
}

@Article{Abbott2017,
  author    = {Abbott, B. P. and Abbott, R. and Abbott, T. D. and Acernese, F. and Ackley, K. and Adams, C. and Adams, T. and Addesso, P. and Adhikari, R. X. and Adya, V. B. and Affeldt, C. and Afrough, M. and Agarwal, B. and Agathos, M. and Agatsuma, K. and Aggarwal, N. and Aguiar, O. D. and Aiello, L. and Ain, A. and Ajith, P. and Allen, B. and Allen, G. and Allocca, A. and Altin, P. A. and Amato, A. and Ananyeva, A. and Anderson, S. B. and Anderson, W. G. and Angelova, S. V. and Antier, S. and Appert, S. and Arai, K. and Araya, M. C. and Areeda, J. S. and Arnaud, N. and Arun, K. G. and Ascenzi, S. and Ashton, G. and Ast, M. and Aston, S. M. and Astone, P. and Atallah, D. V. and Aufmuth, P. and Aulbert, C. and AultONeal, K. and Austin, C. and Avila-Alvarez, A. and Babak, S. and Bacon, P. and Bader, M. K. M. and Bae, S. and Baker, P. T. and Baldaccini, F. and Ballardin, G. and Ballmer, S. W. and Banagiri, S. and Barayoga, J. C. and Barclay, S. E. and Barish, B. C. and Barker, D. and Barkett, K. and Barone, F. and Barr, B. and Barsotti, L. and Barsuglia, M. and Barta, D. and Barthelmy, S. D. and Bartlett, J. and Bartos, I. and Bassiri, R. and Basti, A. and Batch, J. C. and Bawaj, M. and Bayley, J. C. and Bazzan, M. and B{\'{e}}csy, B. and Beer, C. and Bejger, M. and Belahcene, I. and Bell, A. S. and Berger, B. K. and Bergmann, G. and Bero, J. J. and Berry, C. P. L. and Bersanetti, D. and Bertolini, A. and Betzwieser, J. and Bhagwat, S. and Bhandare, R. and Bilenko, I. A. and Billingsley, G. and Billman, C. R. and Birch, J. and Birney, R. and Birnholtz, O. and Biscans, S. and Biscoveanu, S. and Bisht, A. and Bitossi, M. and Biwer, C. and Bizouard, M. A. and Blackburn, J. K. and Blackman, J. and Blair, C. D. and Blair, D. G. and Blair, R. M. and Bloemen, S. and Bock, O. and Bode, N. and Boer, M. and Bogaert, G. and Bohe, A. and Bondu, F. and Bonilla, E. and Bonnand, R. and Boom, B. A. and Bork, R. and Boschi, V. and Bose, S. and Bossie, K. and Bouffanais, Y. and Bozzi, A. and Bradaschia, C. and Brady, P. R. and Branchesi, M. and Brau, J. E. and Briant, T. and Brillet, A. and Brinkmann, M. and Brisson, V. and Brockill, P. and Broida, J. E. and Brooks, A. F. and Brown, D. A. and Brown, D. D. and Brunett, S. and Buchanan, C. C. and Buikema, A. and Bulik, T. and Bulten, H. J. and Buonanno, A. and Buskulic, D. and Buy, C. and Byer, R. L. and Cabero, M. and Cadonati, L. and Cagnoli, G. and Cahillane, C. and Bustillo, J. Calder{\'{o}}n and Callister, T. A. and Calloni, E. and Camp, J. B. and Canepa, M. and Canizares, P. and Cannon, K. C. and Cao, H. and Cao, J. and Capano, C. D. and Capocasa, E. and Carbognani, F. and Caride, S. and Carney, M. F. and Diaz, J. Casanueva and Casentini, C. and Caudill, S. and Cavagli{\`{a}}, M. and Cavalier, F. and Cavalieri, R. and Cella, G. and Cepeda, C. B. and Cerd{\'{a}}-Dur{\'{a}}n, P. and Cerretani, G. and Cesarini, E. and Chamberlin, S. J. and Chan, M. and Chao, S. and Charlton, P. and Chase, E. and Chassande-Mottin, E. and Chatterjee, D. and Chatziioannou, K. and Cheeseboro, B. D. and Chen, H. Y. and Chen, X. and Chen, Y. and Cheng, H.-P. and Chia, H. and Chincarini, A. and Chiummo, A. and Chmiel, T. and Cho, H. S. and Cho, M. and Chow, J. H. and Christensen, N. and Chu, Q. and Chua, A. J. K. and Chua, S. and Chung, A. K. W. and Chung, S. and Ciani, G. and Ciolfi, R. and Cirelli, C. E. and Cirone, A. and Clara, F. and Clark, J. A. and Clearwater, P. and Cleva, F. and Cocchieri, C. and Coccia, E. and Cohadon, P.-F. and Cohen, D. and Colla, A. and Collette, C. G. and Cominsky, L. R. and Constancio, M. and Conti, L. and Cooper, S. J. and Corban, P. and Corbitt, T. R. and Cordero-Carri{\'{o}}n, I. and Corley, K. R. and Cornish, N. and Corsi, A. and Cortese, S. and Costa, C. A. and Coughlin, M. W. and Coughlin, S. B. and Coulon, J.-P. and Countryman, S. T. and Couvares, P. and Covas, P. B. and Cowan, E. E. and Coward, D. M. and Cowart, M. J. and Coyne, D. C. and Coyne, R. and Creighton, J. D. E. and Creighton, T. D. and Cripe, J. and Crowder, S. G. and Cullen, T. J. and Cumming, A. and Cunningham, L. and Cuoco, E. and Canton, T. Dal and D{\'{a}}lya, G. and Danilishin, S. L. and D'Antonio, S. and Danzmann, K. and Dasgupta, A. and Costa, C. F. Da Silva and Dattilo, V. and Dave, I. and Davier, M. and Davis, D. and Daw, E. J. and Day, B. and De, S. and DeBra, D. and Degallaix, J. and Laurentis, M. De and Del{\'{e}}glise, S. and Pozzo, W. Del and Demos, N. and Denker, T. and Dent, T. and Pietri, R. De and Dergachev, V. and Rosa, R. De and DeRosa, R. T. and Rossi, C. De and DeSalvo, R. and de Varona, O. and Devenson, J. and Dhurandhar, S. and D{\'{\i}}az, M. C. and Fiore, L. Di and Giovanni, M. Di and Girolamo, T. Di and Lieto, A. Di and Pace, S. Di and Palma, I. Di and Renzo, F. Di and Doctor, Z. and Dolique, V. and Donovan, F. and Dooley, K. L. and Doravari, S. and Dorrington, I. and Douglas, R. and {\'{A}}lvarez, M. Dovale and Downes, T. P. and Drago, M. and Dreissigacker, C. and Driggers, J. C. and Du, Z. and Ducrot, M. and Dupej, P. and Dwyer, S. E. and Edo, T. B. and Edwards, M. C. and Effler, A. and Eggenstein, H.-B. and Ehrens, P. and Eichholz, J. and Eikenberry, S. S. and Eisenstein, R. A. and Essick, R. C. and Estevez, D. and Etienne, Z. B. and Etzel, T. and Evans, M. and Evans, T. M. and Factourovich, M. and Fafone, V. and Fair, H. and Fairhurst, S. and Fan, X. and Farinon, S. and Farr, B. and Farr, W. M. and Fauchon-Jones, E. J. and Favata, M. and Fays, M. and Fee, C. and Fehrmann, H. and Feicht, J. and Fejer, M. M. and Fernandez-Galiana, A. and Ferrante, I. and Ferreira, E. C. and Ferrini, F. and Fidecaro, F. and Finstad, D. and Fiori, I. and Fiorucci, D. and Fishbach, M. and Fisher, R. P. and Fitz-Axen, M. and Flaminio, R. and Fletcher, M. and Fong, H. and Font, J. A. and Forsyth, P. W. F. and Forsyth, S. S. and Fournier, J.-D. and Frasca, S. and Frasconi, F. and Frei, Z. and Freise, A. and Frey, R. and Frey, V. and Fries, E. M. and Fritschel, P. and Frolov, V. V. and Fulda, P. and Fyffe, M. and Gabbard, H. and Gadre, B. U. and Gaebel, S. M. and Gair, J. R. and Gammaitoni, L. and Ganija, M. R. and Gaonkar, S. G. and Garcia-Quiros, C. and Garufi, F. and Gateley, B. and Gaudio, S. and Gaur, G. and Gayathri, V. and Gehrels, N. and Gemme, G. and Genin, E. and Gennai, A. and George, D. and George, J. and Gergely, L. and Germain, V. and Ghonge, S. and Ghosh, Abhirup and Ghosh, Archisman and Ghosh, S. and Giaime, J. A. and Giardina, K. D. and Giazotto, A. and Gill, K. and Glover, L. and Goetz, E. and Goetz, R. and Gomes, S. and Goncharov, B. and Gonz{\'{a}}lez, G. and Castro, J. M. Gonzalez and Gopakumar, A. and Gorodetsky, M. L. and Gossan, S. E. and Gosselin, M. and Gouaty, R. and Grado, A. and Graef, C. and Granata, M. and Grant, A. and Gras, S. and Gray, C. and Greco, G. and Green, A. C. and Gretarsson, E. M. and Groot, P. and Grote, H. and Grunewald, S. and Gruning, P. and Guidi, G. M. and Guo, X. and Gupta, A. and Gupta, M. K. and Gushwa, K. E. and Gustafson, E. K. and Gustafson, R. and Halim, O. and Hall, B. R. and Hall, E. D. and Hamilton, E. Z. and Hammond, G. and Haney, M. and Hanke, M. M. and Hanks, J. and Hanna, C. and Hannam, M. D. and Hannuksela, O. A. and Hanson, J. and Hardwick, T. and Harms, J. and Harry, G. M. and Harry, I. W. and Hart, M. J. and Haster, C.-J. and Haughian, K. and Healy, J. and Heidmann, A. and Heintze, M. C. and Heitmann, H. and Hello, P. and Hemming, G. and Hendry, M. and Heng, I. S. and Hennig, J. and Heptonstall, A. W. and Heurs, M. and Hild, S. and Hinderer, T. and Hoak, D. and Hofman, D. and Holt, K. and Holz, D. E. and Hopkins, P. and Horst, C. and Hough, J. and Houston, E. A. and Howell, E. J. and Hu, Y. M. and Huerta, E. A. and Huet, D. and Hughey, B. and Husa, S. and Huttner, S. H. and Huynh-Dinh, T. and Indik, N. and Inta, R. and Intini, G. and Isa, H. N. and Isac, J.-M. and Isi, M. and Iyer, B. R. and Izumi, K. and Jacqmin, T. and Jani, K. and Jaranowski, P. and Jawahar, S. and Jim{\'{e}}nez-Forteza, F. and Johnson, W. W. and Johnson-McDaniel, N. K. and Jones, D. I. and Jones, R. and Jonker, R. J. G. and Ju, L. and Junker, J. and Kalaghatgi, C. V. and Kalogera, V. and Kamai, B. and Kandhasamy, S. and Kang, G. and Kanner, J. B. and Kapadia, S. J. and Karki, S. and Karvinen, K. S. and Kasprzack, M. and Katolik, M. and Katsavounidis, E. and Katzman, W. and Kaufer, S. and Kawabe, K. and K{\'{e}}f{\'{e}}lian, F. and Keitel, D. and Kemball, A. J. and Kennedy, R. and Kent, C. and Key, J. S. and Khalili, F. Y. and Khan, I. and Khan, S. and Khan, Z. and Khazanov, E. A. and Kijbunchoo, N. and Kim, Chunglee and Kim, J. C. and Kim, K. and Kim, W. and Kim, W. S. and Kim, Y.-M. and Kimbrell, S. J. and King, E. J. and King, P. J. and Kinley-Hanlon, M. and Kirchhoff, R. and Kissel, J. S. and Kleybolte, L. and Klimenko, S. and Knowles, T. D. and Koch, P. and Koehlenbeck, S. M. and Koley, S. and Kondrashov, V. and Kontos, A. and Korobko, M. and Korth, W. Z. and Kowalska, I. and Kozak, D. B. and Krämer, C. and Kringel, V. and Krishnan, B. and Kr{\'{o}}lak, A. and Kuehn, G. and Kumar, P. and Kumar, R. and Kumar, S. and Kuo, L. and Kutynia, A. and Kwang, S. and Lackey, B. D. and Lai, K. H. and Landry, M. and Lang, R. N. and Lange, J. and Lantz, B. and Lanza, R. K. and Lartaux-Vollard, A. and Lasky, P. D. and Laxen, M. and Lazzarini, A. and Lazzaro, C. and Leaci, P. and Leavey, S. and Lee, C. H. and Lee, H. K. and Lee, H. M. and Lee, H. W. and Lee, K. and Lehmann, J. and Lenon, A. and Leonardi, M. and Leroy, N. and Letendre, N. and Levin, Y. and Li, T. G. F. and Linker, S. D. and Littenberg, T. B. and Liu, J. and Lo, R. K. L. and Lockerbie, N. A. and London, L. T. and Lord, J. E. and Lorenzini, M. and Loriette, V. and Lormand, M. and Losurdo, G. and Lough, J. D. and Lousto, C. O. and Lovelace, G. and Lück, H. and Lumaca, D. and Lundgren, A. P. and Lynch, R. and Ma, Y. and Macas, R. and Macfoy, S. and Machenschalk, B. and MacInnis, M. and Macleod, D. M. and Hernandez, I. Maga{\~{n}}a and Maga{\~{n}}a-Sandoval, F. and Zertuche, L. Maga{\~{n}}a and Magee, R. M. and Majorana, E. and Maksimovic, I. and Man, N. and Mandic, V. and Mangano, V. and Mansell, G. L. and Manske, M. and Mantovani, M. and Marchesoni, F. and Marion, F. and M{\'{a}}rka, S. and M{\'{a}}rka, Z. and Markakis, C. and Markosyan, A. S. and Markowitz, A. and Maros, E. and Marquina, A. and Marsh, P. and Martelli, F. and Martellini, L. and Martin, I. W. and Martin, R. M. and Martynov, D. V. and Mason, K. and Massera, E. and Masserot, A. and Massinger, T. J. and Masso-Reid, M. and Mastrogiovanni, S. and Matas, A. and Matichard, F. and Matone, L. and Mavalvala, N. and Mazumder, N. and McCarthy, R. and McClelland, D. E. and McCormick, S. and McCuller, L. and McGuire, S. C. and McIntyre, G. and McIver, J. and McManus, D. J. and McNeill, L. and McRae, T. and McWilliams, S. T. and Meacher, D. and Meadors, G. D. and Mehmet, M. and Meidam, J. and Mejuto-Villa, E. and Melatos, A. and Mendell, G. and Mercer, R. A. and Merilh, E. L. and Merzougui, M. and Meshkov, S. and Messenger, C. and Messick, C. and Metzdorff, R. and Meyers, P. M. and Miao, H. and Michel, C. and Middleton, H. and Mikhailov, E. E. and Milano, L. and Miller, A. L. and Miller, B. B. and Miller, J. and Millhouse, M. and Milovich-Goff, M. C. and Minazzoli, O. and Minenkov, Y. and Ming, J. and Mishra, C. and Mitra, S. and Mitrofanov, V. P. and Mitselmakher, G. and Mittleman, R. and Moffa, D. and Moggi, A. and Mogushi, K. and Mohan, M. and Mohapatra, S. R. P. and Montani, M. and Moore, C. J. and Moraru, D. and Moreno, G. and Morisaki, S. and Morriss, S. R. and Mours, B. and Mow-Lowry, C. M. and Mueller, G. and Muir, A. W. and Mukherjee, A. and Mukherjee, D. and Mukherjee, S. and Mukund, N. and Mullavey, A. and Munch, J. and Mu{\~{n}}iz, E. A. and Muratore, M. and Murray, P. G. and Napier, K. and Nardecchia, I. and Naticchioni, L. and Nayak, R. K. and Neilson, J. and Nelemans, G. and Nelson, T. J. N. and Nery, M. and Neunzert, A. and Nevin, L. and Newport, J. M. and Newton, G. and Ng, K. K. Y. and Nguyen, T. T. and Nichols, D. and Nielsen, A. B. and Nissanke, S. and Nitz, A. and Noack, A. and Nocera, F. and Nolting, D. and North, C. and Nuttall, L. K. and Oberling, J. and O'Dea, G. D. and Ogin, G. H. and Oh, J. J. and Oh, S. H. and Ohme, F. and Okada, M. A. and Oliver, M. and Oppermann, P. and Oram, R. J. and O'Reilly, B. and Ormiston, R. and Ortega, L. F. and O'Shaughnessy, R. and Ossokine, S. and Ottaway, D. J. and Overmier, H. and Owen, B. J. and Pace, A. E. and Page, J. and Page, M. A. and Pai, A. and Pai, S. A. and Palamos, J. R. and Palashov, O. and Palomba, C. and Pal-Singh, A. and Pan, Howard and Pan, Huang-Wei and Pang, B. and Pang, P. T. H. and Pankow, C. and Pannarale, F. and Pant, B. C. and Paoletti, F. and Paoli, A. and Papa, M. A. and Parida, A. and Parker, W. and Pascucci, D. and Pasqualetti, A. and Passaquieti, R. and Passuello, D. and Patil, M. and Patricelli, B. and Pearlstone, B. L. and Pedraza, M. and Pedurand, R. and Pekowsky, L. and Pele, A. and Penn, S. and Perez, C. J. and Perreca, A. and Perri, L. M. and Pfeiffer, H. P. and Phelps, M. and Piccinni, O. J. and Pichot, M. and Piergiovanni, F. and Pierro, V. and Pillant, G. and Pinard, L. and Pinto, I. M. and Pirello, M. and Pitkin, M. and Poe, M. and Poggiani, R. and Popolizio, P. and Porter, E. K. and Post, A. and Powell, J. and Prasad, J. and Pratt, J. W. W. and Pratten, G. and Predoi, V. and Prestegard, T. and Prijatelj, M. and Principe, M. and Privitera, S. and Prix, R. and Prodi, G. A. and Prokhorov, L. G. and Puncken, O. and Punturo, M. and Puppo, P. and Pürrer, M. and Qi, H. and Quetschke, V. and Quintero, E. A. and Quitzow-James, R. and Raab, F. J. and Rabeling, D. S. and Radkins, H. and Raffai, P. and Raja, S. and Rajan, C. and Rajbhandari, B. and Rakhmanov, M. and Ramirez, K. E. and Ramos-Buades, A. and Rapagnani, P. and Raymond, V. and Razzano, M. and Read, J. and Regimbau, T. and Rei, L. and Reid, S. and Reitze, D. H. and Ren, W. and Reyes, S. D. and Ricci, F. and Ricker, P. M. and Rieger, S. and Riles, K. and Rizzo, M. and Robertson, N. A. and Robie, R. and Robinet, F. and Rocchi, A. and Rolland, L. and Rollins, J. G. and Roma, V. J. and Romano, J. D. and Romano, R. and Romel, C. L. and Romie, J. H. and Rosi{\'{n}}ska, D. and Ross, M. P. and Rowan, S. and Rüdiger, A. and Ruggi, P. and Rutins, G. and Ryan, K. and Sachdev, S. and Sadecki, T. and Sadeghian, L. and Sakellariadou, M. and Salconi, L. and Saleem, M. and Salemi, F. and Samajdar, A. and Sammut, L. and Sampson, L. M. and Sanchez, E. J. and Sanchez, L. E. and Sanchis-Gual, N. and Sandberg, V. and Sanders, J. R. and Sassolas, B. and Sathyaprakash, B. S. and Saulson, P. R. and Sauter, O. and Savage, R. L. and Sawadsky, A. and Schale, P. and Scheel, M. and Scheuer, J. and Schmidt, J. and Schmidt, P. and Schnabel, R. and Schofield, R. M. S. and Schönbeck, A. and Schreiber, E. and Schuette, D. and Schulte, B. W. and Schutz, B. F. and Schwalbe, S. G. and Scott, J. and Scott, S. M. and Seidel, E. and Sellers, D. and Sengupta, A. S. and Sentenac, D. and Sequino, V. and Sergeev, A. and Shaddock, D. A. and Shaffer, T. J. and Shah, A. A. and Shahriar, M. S. and Shaner, M. B. and Shao, L. and Shapiro, B. and Shawhan, P. and Sheperd, A. and Shoemaker, D. H. and Shoemaker, D. M. and Siellez, K. and Siemens, X. and Sieniawska, M. and Sigg, D. and Silva, A. D. and Singer, L. P. and Singh, A. and Singhal, A. and Sintes, A. M. and Slagmolen, B. J. J. and Smith, B. and Smith, J. R. and Smith, R. J. E. and Somala, S. and Son, E. J. and Sonnenberg, J. A. and Sorazu, B. and Sorrentino, F. and Souradeep, T. and Spencer, A. P. and Srivastava, A. K. and Staats, K. and Staley, A. and Steinke, M. and Steinlechner, J. and Steinlechner, S. and Steinmeyer, D. and Stevenson, S. P. and Stone, R. and Stops, D. J. and Strain, K. A. and Stratta, G. and Strigin, S. E. and Strunk, A. and Sturani, R. and Stuver, A. L. and Summerscales, T. Z. and Sun, L. and Sunil, S. and Suresh, J. and Sutton, P. J. and Swinkels, B. L. and Szczepa{\'{n}}czyk, M. J. and Tacca, M. and Tait, S. C. and Talbot, C. and Talukder, D. and Tanner, D. B. and T{\'{a}}pai, M. and Taracchini, A. and Tasson, J. D. and Taylor, J. A. and Taylor, R. and Tewari, S. V. and Theeg, T. and Thies, F. and Thomas, E. G. and Thomas, M. and Thomas, P. and Thorne, K. A. and Thrane, E. and Tiwari, S. and Tiwari, V. and Tokmakov, K. V. and Toland, K. and Tonelli, M. and Tornasi, Z. and Torres-Forn{\'{e}}, A. and Torrie, C. I. and Töyrä, D. and Travasso, F. and Traylor, G. and Trinastic, J. and Tringali, M. C. and Trozzo, L. and Tsang, K. W. and Tse, M. and Tso, R. and Tsukada, L. and Tsuna, D. and Tuyenbayev, D. and Ueno, K. and Ugolini, D. and Unnikrishnan, C. S. and Urban, A. L. and Usman, S. A. and Vahlbruch, H. and Vajente, G. and Valdes, G. and Vallisneri, M. and van Bakel, N. and van Beuzekom, M. and van den Brand, J. F. J. and Broeck, C. Van Den and Vander-Hyde, D. C. and van der Schaaf, L. and van Heijningen, J. V. and van Veggel, A. A. and Vardaro, M. and Varma, V. and Vass, S. and Vas{\'{u}}th, M. and Vecchio, A. and Vedovato, G. and Veitch, J. and Veitch, P. J. and Venkateswara, K. and Venugopalan, G. and Verkindt, D. and Vetrano, F. and Vicer{\'{e}}, A. and Viets, A. D. and Vinciguerra, S. and Vine, D. J. and Vinet, J.-Y. and Vitale, S. and Vo, T. and Vocca, H. and Vorvick, C. and Vyatchanin, S. P. and Wade, A. R. and Wade, L. E. and Wade, M. and Walet, R. and Walker, M. and Wallace, L. and Walsh, S. and Wang, G. and Wang, H. and Wang, J. Z. and Wang, W. H. and Wang, Y. F. and Ward, R. L. and Warner, J. and Was, M. and Watchi, J. and Weaver, B. and Wei, L.-W. and Weinert, M. and Weinstein, A. J. and Weiss, R. and Wen, L. and Wessel, E. K. and We{\ss}els, P. and Westerweck, J. and Westphal, T. and Wette, K. and Whelan, J. T. and Whitcomb, S. E. and Whiting, B. F. and Whittle, C. and Wilken, D. and Williams, D. and Williams, R. D. and Williamson, A. R. and Willis, J. L. and Willke, B. and Wimmer, M. H. and Winkler, W. and Wipf, C. C. and Wittel, H. and Woan, G. and Woehler, J. and Wofford, J. and Wong, K. W. K. and Worden, J. and Wright, J. L. and Wu, D. S. and Wysocki, D. M. and Xiao, S. and Yamamoto, H. and Yancey, C. C. and Yang, L. and Yap, M. J. and Yazback, M. and Yu, Hang and Yu, Haocun and Yvert, M. and Zadro{\.{z}}ny, A. and Zanolin, M. and Zelenova, T. and Zendri, J.-P. and Zevin, M. and Zhang, L. and Zhang, M. and Zhang, T. and Zhang, Y.-H. and Zhao, C. and Zhou, M. and Zhou, Z. and Zhu, S. J. and Zhu, X. J. and Zimmerman, A. B. and Zucker, M. E. and Zweizig, J.},
  journal   = {Physical Review Letters},
  title     = {{GW170814: A Three-Detector Observation of Gravitational Waves from a Binary Black Hole Coalescence}},
  year      = {2017},
  month     = oct,
  number    = {14},
  volume    = {119},
  doi       = {10.1103/physrevlett.119.141101},
  publisher = {American Physical Society ({APS})},
}

@Article{Abbott2017a,
  author    = {Abbott, B. P. and Abbott, R. and Abbott, T. D. and Acernese, F. and Ackley, K. and Adams, C. and Adams, T. and Addesso, P. and Adhikari, R. X. and Adya, V. B. and Affeldt, C. and Afrough, M. and Agarwal, B. and Agathos, M. and Agatsuma, K. and Aggarwal, N. and Aguiar, O. D. and Aiello, L. and Ain, A. and Ajith, P. and Allen, B. and Allen, G. and Allocca, A. and Altin, P. A. and Amato, A. and Ananyeva, A. and Anderson, S. B. and Anderson, W. G. and Angelova, S. V. and Antier, S. and Appert, S. and Arai, K. and Araya, M. C. and Areeda, J. S. and Arnaud, N. and Arun, K. G. and Ascenzi, S. and Ashton, G. and Ast, M. and Aston, S. M. and Astone, P. and Atallah, D. V. and Aufmuth, P. and Aulbert, C. and AultONeal, K. and Austin, C. and Avila-Alvarez, A. and Babak, S. and Bacon, P. and Bader, M. K. M. and Bae, S. and Bailes, M. and Baker, P. T. and Baldaccini, F. and Ballardin, G. and Ballmer, S. W. and Banagiri, S. and Barayoga, J. C. and Barclay, S. E. and Barish, B. C. and Barker, D. and Barkett, K. and Barone, F. and Barr, B. and Barsotti, L. and Barsuglia, M. and Barta, D. and Barthelmy, S. D. and Bartlett, J. and Bartos, I. and Bassiri, R. and Basti, A. and Batch, J. C. and Bawaj, M. and Bayley, J. C. and Bazzan, M. and B{\'{e}}csy, B. and Beer, C. and Bejger, M. and Belahcene, I. and Bell, A. S. and Berger, B. K. and Bergmann, G. and Bernuzzi, S. and Bero, J. J. and Berry, C. P. L. and Bersanetti, D. and Bertolini, A. and Betzwieser, J. and Bhagwat, S. and Bhandare, R. and Bilenko, I. A. and Billingsley, G. and Billman, C. R. and Birch, J. and Birney, R. and Birnholtz, O. and Biscans, S. and Biscoveanu, S. and Bisht, A. and Bitossi, M. and Biwer, C. and Bizouard, M. A. and Blackburn, J. K. and Blackman, J. and Blair, C. D. and Blair, D. G. and Blair, R. M. and Bloemen, S. and Bock, O. and Bode, N. and Boer, M. and Bogaert, G. and Bohe, A. and Bondu, F. and Bonilla, E. and Bonnand, R. and Boom, B. A. and Bork, R. and Boschi, V. and Bose, S. and Bossie, K. and Bouffanais, Y. and Bozzi, A. and Bradaschia, C. and Brady, P. R. and Branchesi, M. and Brau, J. E. and Briant, T. and Brillet, A. and Brinkmann, M. and Brisson, V. and Brockill, P. and Broida, J. E. and Brooks, A. F. and Brown, D. A. and Brown, D. D. and Brunett, S. and Buchanan, C. C. and Buikema, A. and Bulik, T. and Bulten, H. J. and Buonanno, A. and Buskulic, D. and Buy, C. and Byer, R. L. and Cabero, M. and Cadonati, L. and Cagnoli, G. and Cahillane, C. and Bustillo, J. Calder{\'{o}}n and Callister, T. A. and Calloni, E. and Camp, J. B. and Canepa, M. and Canizares, P. and Cannon, K. C. and Cao, H. and Cao, J. and Capano, C. D. and Capocasa, E. and Carbognani, F. and Caride, S. and Carney, M. F. and Carullo, G. and Diaz, J. Casanueva and Casentini, C. and Caudill, S. and Cavagli{\`{a}}, M. and Cavalier, F. and Cavalieri, R. and Cella, G. and Cepeda, C. B. and Cerd{\'{a}}-Dur{\'{a}}n, P. and Cerretani, G. and Cesarini, E. and Chamberlin, S. J. and Chan, M. and Chao, S. and Charlton, P. and Chase, E. and Chassande-Mottin, E. and Chatterjee, D. and Chatziioannou, K. and Cheeseboro, B. D. and Chen, H. Y. and Chen, X. and Chen, Y. and Cheng, H.-P. and Chia, H. and Chincarini, A. and Chiummo, A. and Chmiel, T. and Cho, H. S. and Cho, M. and Chow, J. H. and Christensen, N. and Chu, Q. and Chua, A. J. K. and Chua, S. and Chung, A. K. W. and Chung, S. and Ciani, G. and Ciolfi, R. and Cirelli, C. E. and Cirone, A. and Clara, F. and Clark, J. A. and Clearwater, P. and Cleva, F. and Cocchieri, C. and Coccia, E. and Cohadon, P.-F. and Cohen, D. and Colla, A. and Collette, C. G. and Cominsky, L. R. and Constancio, M. and Conti, L. and Cooper, S. J. and Corban, P. and Corbitt, T. R. and Cordero-Carri{\'{o}}n, I. and Corley, K. R. and Cornish, N. and Corsi, A. and Cortese, S. and Costa, C. A. and Coughlin, M. W. and Coughlin, S. B. and Coulon, J.-P. and Countryman, S. T. and Couvares, P. and Covas, P. B. and Cowan, E. E. and Coward, D. M. and Cowart, M. J. and Coyne, D. C. and Coyne, R. and Creighton, J. D. E. and Creighton, T. D. and Cripe, J. and Crowder, S. G. and Cullen, T. J. and Cumming, A. and Cunningham, L. and Cuoco, E. and Canton, T. Dal and D{\'{a}}lya, G. and Danilishin, S. L. and D'Antonio, S. and Danzmann, K. and Dasgupta, A. and Costa, C. F. Da Silva and Dattilo, V. and Dave, I. and Davier, M. and Davis, D. and Daw, E. J. and Day, B. and De, S. and DeBra, D. and Degallaix, J. and Laurentis, M. De and Del{\'{e}}glise, S. and Pozzo, W. Del and Demos, N. and Denker, T. and Dent, T. and Pietri, R. De and Dergachev, V. and Rosa, R. De and DeRosa, R. T. and Rossi, C. De and DeSalvo, R. and de Varona, O. and Devenson, J. and Dhurandhar, S. and D{\'{\i}}az, M. C. and Dietrich, T. and Fiore, L. Di and Giovanni, M. Di and Girolamo, T. Di and Lieto, A. Di and Pace, S. Di and Palma, I. Di and Renzo, F. Di and Doctor, Z. and Dolique, V. and Donovan, F. and Dooley, K. L. and Doravari, S. and Dorrington, I. and Douglas, R. and {\'{A}}lvarez, M. Dovale and Downes, T. P. and Drago, M. and Dreissigacker, C. and Driggers, J. C. and Du, Z. and Ducrot, M. and Dudi, R. and Dupej, P. and Dwyer, S. E. and Edo, T. B. and Edwards, M. C. and Effler, A. and Eggenstein, H.-B. and Ehrens, P. and Eichholz, J. and Eikenberry, S. S. and Eisenstein, R. A. and Essick, R. C. and Estevez, D. and Etienne, Z. B. and Etzel, T. and Evans, M. and Evans, T. M. and Factourovich, M. and Fafone, V. and Fair, H. and Fairhurst, S. and Fan, X. and Farinon, S. and Farr, B. and Farr, W. M. and Fauchon-Jones, E. J. and Favata, M. and Fays, M. and Fee, C. and Fehrmann, H. and Feicht, J. and Fejer, M. M. and Fernandez-Galiana, A. and Ferrante, I. and Ferreira, E. C. and Ferrini, F. and Fidecaro, F. and Finstad, D. and Fiori, I. and Fiorucci, D. and Fishbach, M. and Fisher, R. P. and Fitz-Axen, M. and Flaminio, R. and Fletcher, M. and Fong, H. and Font, J. A. and Forsyth, P. W. F. and Forsyth, S. S. and Fournier, J.-D. and Frasca, S. and Frasconi, F. and Frei, Z. and Freise, A. and Frey, R. and Frey, V. and Fries, E. M. and Fritschel, P. and Frolov, V. V. and Fulda, P. and Fyffe, M. and Gabbard, H. and Gadre, B. U. and Gaebel, S. M. and Gair, J. R. and Gammaitoni, L. and Ganija, M. R. and Gaonkar, S. G. and Garcia-Quiros, C. and Garufi, F. and Gateley, B. and Gaudio, S. and Gaur, G. and Gayathri, V. and Gehrels, N. and Gemme, G. and Genin, E. and Gennai, A. and George, D. and George, J. and Gergely, L. and Germain, V. and Ghonge, S. and Ghosh, Abhirup and Ghosh, Archisman and Ghosh, S. and Giaime, J. A. and Giardina, K. D. and Giazotto, A. and Gill, K. and Glover, L. and Goetz, E. and Goetz, R. and Gomes, S. and Goncharov, B. and Gonz{\'{a}}lez, G. and Castro, J. M. Gonzalez and Gopakumar, A. and Gorodetsky, M. L. and Gossan, S. E. and Gosselin, M. and Gouaty, R. and Grado, A. and Graef, C. and Granata, M. and Grant, A. and Gras, S. and Gray, C. and Greco, G. and Green, A. C. and Gretarsson, E. M. and Groot, P. and Grote, H. and Grunewald, S. and Gruning, P. and Guidi, G. M. and Guo, X. and Gupta, A. and Gupta, M. K. and Gushwa, K. E. and Gustafson, E. K. and Gustafson, R. and Halim, O. and Hall, B. R. and Hall, E. D. and Hamilton, E. Z. and Hammond, G. and Haney, M. and Hanke, M. M. and Hanks, J. and Hanna, C. and Hannam, M. D. and Hannuksela, O. A. and Hanson, J. and Hardwick, T. and Harms, J. and Harry, G. M. and Harry, I. W. and Hart, M. J. and Haster, C.-J. and Haughian, K. and Healy, J. and Heidmann, A. and Heintze, M. C. and Heitmann, H. and Hello, P. and Hemming, G. and Hendry, M. and Heng, I. S. and Hennig, J. and Heptonstall, A. W. and Heurs, M. and Hild, S. and Hinderer, T. and Ho, W. C. G. and Hoak, D. and Hofman, D. and Holt, K. and Holz, D. E. and Hopkins, P. and Horst, C. and Hough, J. and Houston, E. A. and Howell, E. J. and Hreibi, A. and Hu, Y. M. and Huerta, E. A. and Huet, D. and Hughey, B. and Husa, S. and Huttner, S. H. and Huynh-Dinh, T. and Indik, N. and Inta, R. and Intini, G. and Isa, H. N. and Isac, J.-M. and Isi, M. and Iyer, B. R. and Izumi, K. and Jacqmin, T. and Jani, K. and Jaranowski, P. and Jawahar, S. and Jim{\'{e}}nez-Forteza, F. and Johnson, W. W. and Johnson-McDaniel, N. K. and Jones, D. I. and Jones, R. and Jonker, R. J. G. and Ju, L. and Junker, J. and Kalaghatgi, C. V. and Kalogera, V. and Kamai, B. and Kandhasamy, S. and Kang, G. and Kanner, J. B. and Kapadia, S. J. and Karki, S. and Karvinen, K. S. and Kasprzack, M. and Kastaun, W. and Katolik, M. and Katsavounidis, E. and Katzman, W. and Kaufer, S. and Kawabe, K. and K{\'{e}}f{\'{e}}lian, F. and Keitel, D. and Kemball, A. J. and Kennedy, R. and Kent, C. and Key, J. S. and Khalili, F. Y. and Khan, I. and Khan, S. and Khan, Z. and Khazanov, E. A. and Kijbunchoo, N. and Kim, Chunglee and Kim, J. C. and Kim, K. and Kim, W. and Kim, W. S. and Kim, Y.-M. and Kimbrell, S. J. and King, E. J. and King, P. J. and Kinley-Hanlon, M. and Kirchhoff, R. and Kissel, J. S. and Kleybolte, L. and Klimenko, S. and Knowles, T. D. and Koch, P. and Koehlenbeck, S. M. and Koley, S. and Kondrashov, V. and Kontos, A. and Korobko, M. and Korth, W. Z. and Kowalska, I. and Kozak, D. B. and Krämer, C. and Kringel, V. and Krishnan, B. and Kr{\'{o}}lak, A. and Kuehn, G. and Kumar, P. and Kumar, R. and Kumar, S. and Kuo, L. and Kutynia, A. and Kwang, S. and Lackey, B. D. and Lai, K. H. and Landry, M. and Lang, R. N. and Lange, J. and Lantz, B. and Lanza, R. K. and Larson, S. L. and Lartaux-Vollard, A. and Lasky, P. D. and Laxen, M. and Lazzarini, A. and Lazzaro, C. and Leaci, P. and Leavey, S. and Lee, C. H. and Lee, H. K. and Lee, H. M. and Lee, H. W. and Lee, K. and Lehmann, J. and Lenon, A. and Leon, E. and Leonardi, M. and Leroy, N. and Letendre, N. and Levin, Y. and Li, T. G. F. and Linker, S. D. and Littenberg, T. B. and Liu, J. and Liu, X. and Lo, R. K. L. and Lockerbie, N. A. and London, L. T. and Lord, J. E. and Lorenzini, M. and Loriette, V. and Lormand, M. and Losurdo, G. and Lough, J. D. and Lousto, C. O. and Lovelace, G. and Lück, H. and Lumaca, D. and Lundgren, A. P. and Lynch, R. and Ma, Y. and Macas, R. and Macfoy, S. and Machenschalk, B. and MacInnis, M. and Macleod, D. M. and Hernandez, I. Maga{\~{n}}a and Maga{\~{n}}a-Sandoval, F. and Zertuche, L. Maga{\~{n}}a and Magee, R. M. and Majorana, E. and Maksimovic, I. and Man, N. and Mandic, V. and Mangano, V. and Mansell, G. L. and Manske, M. and Mantovani, M. and Marchesoni, F. and Marion, F. and M{\'{a}}rka, S. and M{\'{a}}rka, Z. and Markakis, C. and Markosyan, A. S. and Markowitz, A. and Maros, E. and Marquina, A. and Marsh, P. and Martelli, F. and Martellini, L. and Martin, I. W. and Martin, R. M. and Martynov, D. V. and Marx, J. N. and Mason, K. and Massera, E. and Masserot, A. and Massinger, T. J. and Masso-Reid, M. and Mastrogiovanni, S. and Matas, A. and Matichard, F. and Matone, L. and Mavalvala, N. and Mazumder, N. and McCarthy, R. and McClelland, D. E. and McCormick, S. and McCuller, L. and McGuire, S. C. and McIntyre, G. and McIver, J. and McManus, D. J. and McNeill, L. and McRae, T. and McWilliams, S. T. and Meacher, D. and Meadors, G. D. and Mehmet, M. and Meidam, J. and Mejuto-Villa, E. and Melatos, A. and Mendell, G. and Mercer, R. A. and Merilh, E. L. and Merzougui, M. and Meshkov, S. and Messenger, C. and Messick, C. and Metzdorff, R. and Meyers, P. M. and Miao, H. and Michel, C. and Middleton, H. and Mikhailov, E. E. and Milano, L. and Miller, A. L. and Miller, B. B. and Miller, J. and Millhouse, M. and Milovich-Goff, M. C. and Minazzoli, O. and Minenkov, Y. and Ming, J. and Mishra, C. and Mitra, S. and Mitrofanov, V. P. and Mitselmakher, G. and Mittleman, R. and Moffa, D. and Moggi, A. and Mogushi, K. and Mohan, M. and Mohapatra, S. R. P. and Molina, I. and Montani, M. and Moore, C. J. and Moraru, D. and Moreno, G. and Morisaki, S. and Morriss, S. R. and Mours, B. and Mow-Lowry, C. M. and Mueller, G. and Muir, A. W. and Mukherjee, Arunava and Mukherjee, D. and Mukherjee, S. and Mukund, N. and Mullavey, A. and Munch, J. and Mu{\~{n}}iz, E. A. and Muratore, M. and Murray, P. G. and Nagar, A. and Napier, K. and Nardecchia, I. and Naticchioni, L. and Nayak, R. K. and Neilson, J. and Nelemans, G. and Nelson, T. J. N. and Nery, M. and Neunzert, A. and Nevin, L. and Newport, J. M. and Newton, G. and Ng, K. K. Y. and Nguyen, P. and Nguyen, T. T. and Nichols, D. and Nielsen, A. B. and Nissanke, S. and Nitz, A. and Noack, A. and Nocera, F. and Nolting, D. and North, C. and Nuttall, L. K. and Oberling, J. and O'Dea, G. D. and Ogin, G. H. and Oh, J. J. and Oh, S. H. and Ohme, F. and Okada, M. A. and Oliver, M. and Oppermann, P. and Oram, Richard J. and O'Reilly, B. and Ormiston, R. and Ortega, L. F. and O'Shaughnessy, R. and Ossokine, S. and Ottaway, D. J. and Overmier, H. and Owen, B. J. and Pace, A. E. and Page, J. and Page, M. A. and Pai, A. and Pai, S. A. and Palamos, J. R. and Palashov, O. and Palomba, C. and Pal-Singh, A. and Pan, Howard and Pan, Huang-Wei and Pang, B. and Pang, P. T. H. and Pankow, C. and Pannarale, F. and Pant, B. C. and Paoletti, F. and Paoli, A. and Papa, M. A. and Parida, A. and Parker, W. and Pascucci, D. and Pasqualetti, A. and Passaquieti, R. and Passuello, D. and Patil, M. and Patricelli, B. and Pearlstone, B. L. and Pedraza, M. and Pedurand, R. and Pekowsky, L. and Pele, A. and Penn, S. and Perez, C. J. and Perreca, A. and Perri, L. M. and Pfeiffer, H. P. and Phelps, M. and Piccinni, O. J. and Pichot, M. and Piergiovanni, F. and Pierro, V. and Pillant, G. and Pinard, L. and Pinto, I. M. and Pirello, M. and Pitkin, M. and Poe, M. and Poggiani, R. and Popolizio, P. and Porter, E. K. and Post, A. and Powell, J. and Prasad, J. and Pratt, J. W. W. and Pratten, G. and Predoi, V. and Prestegard, T. and Prijatelj, M. and Principe, M. and Privitera, S. and Prix, R. and Prodi, G. A. and Prokhorov, L. G. and Puncken, O. and Punturo, M. and Puppo, P. and Pürrer, M. and Qi, H. and Quetschke, V. and Quintero, E. A. and Quitzow-James, R. and Raab, F. J. and Rabeling, D. S. and Radkins, H. and Raffai, P. and Raja, S. and Rajan, C. and Rajbhandari, B. and Rakhmanov, M. and Ramirez, K. E. and Ramos-Buades, A. and Rapagnani, P. and Raymond, V. and Razzano, M. and Read, J. and Regimbau, T. and Rei, L. and Reid, S. and Reitze, D. H. and Ren, W. and Reyes, S. D. and Ricci, F. and Ricker, P. M. and Rieger, S. and Riles, K. and Rizzo, M. and Robertson, N. A. and Robie, R. and Robinet, F. and Rocchi, A. and Rolland, L. and Rollins, J. G. and Roma, V. J. and Romano, J. D. and Romano, R. and Romel, C. L. and Romie, J. H. and Rosi{\'{n}}ska, D. and Ross, M. P. and Rowan, S. and Rüdiger, A. and Ruggi, P. and Rutins, G. and Ryan, K. and Sachdev, S. and Sadecki, T. and Sadeghian, L. and Sakellariadou, M. and Salconi, L. and Saleem, M. and Salemi, F. and Samajdar, A. and Sammut, L. and Sampson, L. M. and Sanchez, E. J. and Sanchez, L. E. and Sanchis-Gual, N. and Sandberg, V. and Sanders, J. R. and Sassolas, B. and Sathyaprakash, B. S. and Saulson, P. R. and Sauter, O. and Savage, R. L. and Sawadsky, A. and Schale, P. and Scheel, M. and Scheuer, J. and Schmidt, J. and Schmidt, P. and Schnabel, R. and Schofield, R. M. S. and Schönbeck, A. and Schreiber, E. and Schuette, D. and Schulte, B. W. and Schutz, B. F. and Schwalbe, S. G. and Scott, J. and Scott, S. M. and Seidel, E. and Sellers, D. and Sengupta, A. S. and Sentenac, D. and Sequino, V. and Sergeev, A. and Shaddock, D. A. and Shaffer, T. J. and Shah, A. A. and Shahriar, M. S. and Shaner, M. B. and Shao, L. and Shapiro, B. and Shawhan, P. and Sheperd, A. and Shoemaker, D. H. and Shoemaker, D. M. and Siellez, K. and Siemens, X. and Sieniawska, M. and Sigg, D. and Silva, A. D. and Singer, L. P. and Singh, A. and Singhal, A. and Sintes, A. M. and Slagmolen, B. J. J. and Smith, B. and Smith, J. R. and Smith, R. J. E. and Somala, S. and Son, E. J. and Sonnenberg, J. A. and Sorazu, B. and Sorrentino, F. and Souradeep, T. and Spencer, A. P. and Srivastava, A. K. and Staats, K. and Staley, A. and Steinke, M. and Steinlechner, J. and Steinlechner, S. and Steinmeyer, D. and Stevenson, S. P. and Stone, R. and Stops, D. J. and Strain, K. A. and Stratta, G. and Strigin, S. E. and Strunk, A. and Sturani, R. and Stuver, A. L. and Summerscales, T. Z. and Sun, L. and Sunil, S. and Suresh, J. and Sutton, P. J. and Swinkels, B. L. and Szczepa{\'{n}}czyk, M. J. and Tacca, M. and Tait, S. C. and Talbot, C. and Talukder, D. and Tanner, D. B. and T{\'{a}}pai, M. and Taracchini, A. and Tasson, J. D. and Taylor, J. A. and Taylor, R. and Tewari, S. V. and Theeg, T. and Thies, F. and Thomas, E. G. and Thomas, M. and Thomas, P. and Thorne, K. A. and Thorne, K. S. and Thrane, E. and Tiwari, S. and Tiwari, V. and Tokmakov, K. V. and Toland, K. and Tonelli, M. and Tornasi, Z. and Torres-Forn{\'{e}}, A. and Torrie, C. I. and Töyrä, D. and Travasso, F. and Traylor, G. and Trinastic, J. and Tringali, M. C. and Trozzo, L. and Tsang, K. W. and Tse, M. and Tso, R. and Tsukada, L. and Tsuna, D. and Tuyenbayev, D. and Ueno, K. and Ugolini, D. and Unnikrishnan, C. S. and Urban, A. L. and Usman, S. A. and Vahlbruch, H. and Vajente, G. and Valdes, G. and Vallisneri, M. and van Bakel, N. and van Beuzekom, M. and van den Brand, J. F. J. and Broeck, C. Van Den and Vander-Hyde, D. C. and van der Schaaf, L. and van Heijningen, J. V. and van Veggel, A. A. and Vardaro, M. and Varma, V. and Vass, S. and Vas{\'{u}}th, M. and Vecchio, A. and Vedovato, G. and Veitch, J. and Veitch, P. J. and Venkateswara, K. and Venugopalan, G. and Verkindt, D. and Vetrano, F. and Vicer{\'{e}}, A. and Viets, A. D. and Vinciguerra, S. and Vine, D. J. and Vinet, J.-Y. and Vitale, S. and Vo, T. and Vocca, H. and Vorvick, C. and Vyatchanin, S. P. and Wade, A. R. and Wade, L. E. and Wade, M. and Walet, R. and Walker, M. and Wallace, L. and Walsh, S. and Wang, G. and Wang, H. and Wang, J. Z. and Wang, W. H. and Wang, Y. F. and Ward, R. L. and Warner, J. and Was, M. and Watchi, J. and Weaver, B. and Wei, L.-W. and Weinert, M. and Weinstein, A. J. and Weiss, R. and Wen, L. and Wessel, E. K. and We{\ss}els, P. and Westerweck, J. and Westphal, T. and Wette, K. and Whelan, J. T. and Whitcomb, S. E. and Whiting, B. F. and Whittle, C. and Wilken, D. and Williams, D. and Williams, R. D. and Williamson, A. R. and Willis, J. L. and Willke, B. and Wimmer, M. H. and Winkler, W. and Wipf, C. C. and Wittel, H. and Woan, G. and Woehler, J. and Wofford, J. and Wong, K. W. K. and Worden, J. and Wright, J. L. and Wu, D. S. and Wysocki, D. M. and Xiao, S. and Yamamoto, H. and Yancey, C. C. and Yang, L. and Yap, M. J. and Yazback, M. and Yu, Hang and Yu, Haocun and Yvert, M. and Zadro{\.{z}}ny, A. and Zanolin, M. and Zelenova, T. and Zendri, J.-P. and Zevin, M. and Zhang, L. and Zhang, M. and Zhang, T. and Zhang, Y.-H. and Zhao, C. and Zhou, M. and Zhou, Z. and Zhu, S. J. and Zhu, X. J. and Zimmerman, A. B. and Zucker, M. E. and Zweizig, J.},
  journal   = {Physical Review Letters},
  title     = {{GW170817: Observation of Gravitational Waves from a Binary Neutron Star Inspiral}},
  year      = {2017},
  month     = oct,
  number    = {16},
  volume    = {119},
  doi       = {10.1103/physrevlett.119.161101},
  publisher = {American Physical Society ({APS})},
}

@Article{Abbott2020,
  author    = {Abbott, R. and Abbott, T. D. and Abraham, S. and Acernese, F. and Ackley, K. and Adams, C. and Adhikari, R. X. and Adya, V. B. and Affeldt, C. and Agathos, M. and Agatsuma, K. and Aggarwal, N. and Aguiar, O. D. and Aich, A. and Aiello, L. and Ain, A. and Ajith, P. and Akcay, S. and Allen, G. and Allocca, A. and Altin, P. A. and Amato, A. and Anand, S. and Ananyeva, A. and Anderson, S. B. and Anderson, W. G. and Angelova, S. V. and Ansoldi, S. and Antier, S. and Appert, S. and Arai, K. and Araya, M. C. and Areeda, J. S. and Ar{\`{e}}ne, M. and Arnaud, N. and Aronson, S. M. and Arun, K. G. and Asali, Y. and Ascenzi, S. and Ashton, G. and Aston, S. M. and Astone, P. and Aubin, F. and Aufmuth, P. and AultONeal, K. and Austin, C. and Avendano, V. and Babak, S. and Bacon, P. and Badaracco, F. and Bader, M. K. M. and Bae, S. and Baer, A. M. and Baird, J. and Baldaccini, F. and Ballardin, G. and Ballmer, S. W. and Bals, A. and Balsamo, A. and Baltus, G. and Banagiri, S. and Bankar, D. and Bankar, R. S. and Barayoga, J. C. and Barbieri, C. and Barish, B. C. and Barker, D. and Barkett, K. and Barneo, P. and Barone, F. and Barr, B. and Barsotti, L. and Barsuglia, M. and Barta, D. and Bartlett, J. and Bartos, I. and Bassiri, R. and Basti, A. and Bawaj, M. and Bayley, J. C. and Bazzan, M. and B{\'{e}}csy, B. and Bejger, M. and Belahcene, I. and Bell, A. S. and Beniwal, D. and Benjamin, M. G. and Benkel, R. and Bentley, J. D. and Bergamin, F. and Berger, B. K. and Bergmann, G. and Bernuzzi, S. and Berry, C. P. L. and Bersanetti, D. and Bertolini, A. and Betzwieser, J. and Bhandare, R. and Bhandari, A. V. and Bidler, J. and Biggs, E. and Bilenko, I. A. and Billingsley, G. and Birney, R. and Birnholtz, O. and Biscans, S. and Bischi, M. and Biscoveanu, S. and Bisht, A. and Bissenbayeva, G. and Bitossi, M. and Bizouard, M. A. and Blackburn, J. K. and Blackman, J. and Blair, C. D. and Blair, D. G. and Blair, R. M. and Bobba, F. and Bode, N. and Boer, M. and Boetzel, Y. and Bogaert, G. and Bondu, F. and Bonilla, E. and Bonnand, R. and Booker, P. and Boom, B. A. and Bork, R. and Boschi, V. and Bose, S. and Bossilkov, V. and Bosveld, J. and Bouffanais, Y. and Bozzi, A. and Bradaschia, C. and Brady, P. R. and Bramley, A. and Branchesi, M. and Brau, J. E. and Breschi, M. and Briant, T. and Briggs, J. H. and Brighenti, F. and Brillet, A. and Brinkmann, M. and Brito, R. and Brockill, P. and Brooks, A. F. and Brooks, J. and Brown, D. D. and Brunett, S. and Bruno, G. and Bruntz, R. and Buikema, A. and Bulik, T. and Bulten, H. J. and Buonanno, A. and Buskulic, D. and Byer, R. L. and Cabero, M. and Cadonati, L. and Cagnoli, G. and Cahillane, C. and Bustillo, J. Calder{\'{o}}n and Callaghan, J. D. and Callister, T. A. and Calloni, E. and Camp, J. B. and Canepa, M. and Cannon, K. C. and Cao, H. and Cao, J. and Carapella, G. and Carbognani, F. and Caride, S. and Carney, M. F. and Carullo, G. and Diaz, J. Casanueva and Casentini, C. and Casta{\~{n}}eda, J. and Caudill, S. and Cavagli{\`{a}}, M. and Cavalier, F. and Cavalieri, R. and Cella, G. and Cerd{\'{a}}-Dur{\'{a}}n, P. and Cesarini, E. and Chaibi, O. and Chakravarti, K. and Chan, C. and Chan, M. and Chao, S. and Charlton, P. and Chase, E. A. and Chassande-Mottin, E. and Chatterjee, D. and Chaturvedi, M. and Chatziioannou, K. and Chen, H. Y. and Chen, X. and Chen, Y. and Cheng, H.-P. and Cheong, C. K. and Chia, H. Y. and Chiadini, F. and Chierici, R. and Chincarini, A. and Chiummo, A. and Cho, G. and Cho, H. S. and Cho, M. and Christensen, N. and Chu, Q. and Chua, S. and Chung, K. W. and Chung, S. and Ciani, G. and Ciecielag, P. and Cie{\'{s}}lar, M. and Ciobanu, A. A. and Ciolfi, R. and Cipriano, F. and Cirone, A. and Clara, F. and Clark, J. A. and Clearwater, P. and Clesse, S. and Cleva, F. and Coccia, E. and Cohadon, P.-F. and Cohen, D. and Colleoni, M. and Collette, C. G. and Collins, C. and Colpi, M. and Constancio, M. and Conti, L. and Cooper, S. J. and Corban, P. and Corbitt, T. R. and Cordero-Carri{\'{o}}n, I. and Corezzi, S. and Corley, K. R. and Cornish, N. and Corre, D. and Corsi, A. and Cortese, S. and Costa, C. A. and Cotesta, R. and Coughlin, M. W. and Coughlin, S. B. and Coulon, J.-P. and Countryman, S. T. and Couvares, P. and Covas, P. B. and Coward, D. M. and Cowart, M. J. and Coyne, D. C. and Coyne, R. and Creighton, J. D. E. and Creighton, T. D. and Cripe, J. and Croquette, M. and Crowder, S. G. and Cudell, J.-R. and Cullen, T. J. and Cumming, A. and Cummings, R. and Cunningham, L. and Cuoco, E. and Curylo, M. and Canton, T. Dal and D{\'{a}}lya, G. and Dana, A. and Daneshgaran-Bajastani, L. M. and D'Angelo, B. and Danilishin, S. L. and D'Antonio, S. and Danzmann, K. and Darsow-Fromm, C. and Dasgupta, A. and Datrier, L. E. H. and Dattilo, V. and Dave, I. and Davier, M. and Davies, G. S. and Davis, D. and Daw, E. J. and DeBra, D. and Deenadayalan, M. and Degallaix, J. and Laurentis, M. De and Del{\'{e}}glise, S. and Delfavero, M. and Lillo, N. De and Pozzo, W. Del and DeMarchi, L. M. and D'Emilio, V. and Demos, N. and Dent, T. and Pietri, R. De and Rosa, R. De and Rossi, C. De and DeSalvo, R. and de Varona, O. and Dhurandhar, S. and D{\'{\i}}az, M. C. and Diaz-Ortiz, M. and Dietrich, T. and Fiore, L. Di and Fronzo, C. Di and Giorgio, C. Di and Giovanni, F. Di and Giovanni, M. Di and Girolamo, T. Di and Lieto, A. Di and Ding, B. and Pace, S. Di and Palma, I. Di and Renzo, F. Di and Divakarla, A. K. and Dmitriev, A. and Doctor, Z. and Donovan, F. and Dooley, K. L. and Doravari, S. and Dorrington, I. and Downes, T. P. and Drago, M. and Driggers, J. C. and Du, Z. and Ducoin, J.-G. and Dupej, P. and Durante, O. and D'Urso, D. and Dwyer, S. E. and Easter, P. J. and Eddolls, G. and Edelman, B. and Edo, T. B. and Edy, O. and Effler, A. and Ehrens, P. and Eichholz, J. and Eikenberry, S. S. and Eisenmann, M. and Eisenstein, R. A. and Ejlli, A. and Errico, L. and Essick, R. C. and Estelles, H. and Estevez, D. and Etienne, Z. B. and Etzel, T. and Evans, M. and Evans, T. M. and Ewing, B. E. and Fafone, V. and Fairhurst, S. and Fan, X. and Farinon, S. and Farr, B. and Farr, W. M. and Fauchon-Jones, E. J. and Favata, M. and Fays, M. and Fazio, M. and Feicht, J. and Fejer, M. M. and Feng, F. and Fenyvesi, E. and Ferguson, D. L. and Fernandez-Galiana, A. and Ferrante, I. and Ferreira, E. C. and Ferreira, T. A. and Fidecaro, F. and Fiori, I. and Fiorucci, D. and Fishbach, M. and Fisher, R. P. and Fittipaldi, R. and Fitz-Axen, M. and Fiumara, V. and Flaminio, R. and Floden, E. and Flynn, E. and Fong, H. and Font, J. A. and Forsyth, P. W. F. and Fournier, J.-D. and Frasca, S. and Frasconi, F. and Frei, Z. and Freise, A. and Frey, R. and Frey, V. and Fritschel, P. and Frolov, V. V. and Fronz{\`{e}}, G. and Fulda, P. and Fyffe, M. and Gabbard, H. A. and Gadre, B. U. and Gaebel, S. M. and Gair, J. R. and Galaudage, S. and Ganapathy, D. and Ganguly, A. and Gaonkar, S. G. and Garc{\'{\i}}a-Quir{\'{o}}s, C. and Garufi, F. and Gateley, B. and Gaudio, S. and Gayathri, V. and Gemme, G. and Genin, E. and Gennai, A. and George, D. and George, J. and Gergely, L. and Ghonge, S. and Ghosh, Abhirup and Ghosh, Archisman and Ghosh, S. and Giacomazzo, B. and Giaime, J. A. and Giardina, K. D. and Gibson, D. R. and Gier, C. and Gill, K. and Glanzer, J. and Gniesmer, J. and Godwin, P. and Goetz, E. and Goetz, R. and Gohlke, N. and Goncharov, B. and Gonz{\'{a}}lez, G. and Gopakumar, A. and Gossan, S. E. and Gosselin, M. and Gouaty, R. and Grace, B. and Grado, A. and Granata, M. and Grant, A. and Gras, S. and Grassia, P. and Gray, C. and Gray, R. and Greco, G. and Green, A. C. and Green, R. and Gretarsson, E. M. and Griggs, H. L. and Grignani, G. and Grimaldi, A. and Grimm, S. J. and Grote, H. and Grunewald, S. and Gruning, P. and Guidi, G. M. and Guimaraes, A. R. and Guix{\'{e}}, G. and Gulati, H. K. and Guo, Y. and Gupta, A. and Gupta, Anchal and Gupta, P. and Gustafson, E. K. and Gustafson, R. and Haegel, L. and Halim, O. and Hall, E. D. and Hamilton, E. Z. and Hammond, G. and Haney, M. and Hanke, M. M. and Hanks, J. and Hanna, C. and Hannam, M. D. and Hannuksela, O. A. and Hansen, T. J. and Hanson, J. and Harder, T. and Hardwick, T. and Haris, K. and Harms, J. and Harry, G. M. and Harry, I. W. and Hasskew, R. K. and Haster, C.-J. and Haughian, K. and Hayes, F. J. and Healy, J. and Heidmann, A. and Heintze, M. C. and Heinze, J. and Heitmann, H. and Hellman, F. and Hello, P. and Hemming, G. and Hendry, M. and Heng, I. S. and Hennes, E. and Hennig, J. and Heurs, M. and Hild, S. and Hinderer, T. and Hoback, S. Y. and Hochheim, S. and Hofgard, E. and Hofman, D. and Holgado, A. M. and Holland, N. A. and Holt, K. and Holz, D. E. and Hopkins, P. and Horst, C. and Hough, J. and Howell, E. J. and Hoy, C. G. and Huang, Y. and Hübner, M. T. and Huerta, E. A. and Huet, D. and Hughey, B. and Hui, V. and Husa, S. and Huttner, S. H. and Huxford, R. and Huynh-Dinh, T. and Idzkowski, B. and Iess, A. and Inchauspe, H. and Ingram, C. and Intini, G. and Isac, J.-M. and Isi, M. and Iyer, B. R. and Jacqmin, T. and Jadhav, S. J. and Jadhav, S. P. and James, A. L. and Jani, K. and Janthalur, N. N. and Jaranowski, P. and Jariwala, D. and Jaume, R. and Jenkins, A. C. and Jiang, J. and Johns, G. R. and Johnson-McDaniel, N. K. and Jones, A. W. and Jones, D. I. and Jones, J. D. and Jones, P. and Jones, R. and Jonker, R. J. G. and Ju, L. and Junker, J. and Kalaghatgi, C. V. and Kalogera, V. and Kamai, B. and Kandhasamy, S. and Kang, G. and Kanner, J. B. and Kapadia, S. J. and Karki, S. and Kashyap, R. and Kasprzack, M. and Kastaun, W. and Katsanevas, S. and Katsavounidis, E. and Katzman, W. and Kaufer, S. and Kawabe, K. and K{\'{e}}f{\'{e}}lian, F. and Keitel, D. and Keivani, A. and Kennedy, R. and Key, J. S. and Khadka, S. and Khalili, F. Y. and Khan, I. and Khan, S. and Khan, Z. A. and Khazanov, E. A. and Khetan, N. and Khursheed, M. and Kijbunchoo, N. and Kim, Chunglee and Kim, G. J. and Kim, J. C. and Kim, K. and Kim, W. and Kim, W. S. and Kim, Y.-M. and Kimball, C. and King, P. J. and Kinley-Hanlon, M. and Kirchhoff, R. and Kissel, J. S. and Kleybolte, L. and Klimenko, S. and Knowles, T. D. and Knyazev, E. and Koch, P. and Koehlenbeck, S. M. and Koekoek, G. and Koley, S. and Kondrashov, V. and Kontos, A. and Koper, N. and Korobko, M. and Korth, W. Z. and Kovalam, M. and Kozak, D. B. and Kringel, V. and Krishnendu, N. V. and Kr{\'{o}}lak, A. and Krupinski, N. and Kuehn, G. and Kumar, A. and Kumar, P. and Kumar, Rahul and Kumar, Rakesh and Kumar, S. and Kuo, L. and Kutynia, A. and Lackey, B. D. and Laghi, D. and Lalande, E. and Lam, T. L. and Lamberts, A. and Landry, M. and Lane, B. B. and Lang, R. N. and Lange, J. and Lantz, B. and Lanza, R. K. and Rosa, I. La and Lartaux-Vollard, A. and Lasky, P. D. and Laxen, M. and Lazzarini, A. and Lazzaro, C. and Leaci, P. and Leavey, S. and Lecoeuche, Y. K. and Lee, C. H. and Lee, H. M. and Lee, H. W. and Lee, J. and Lee, K. and Lehmann, J. and Leroy, N. and Letendre, N. and Levin, Y. and Li, A. K. Y. and Li, J. and Li, K. and Li, T. G. F. and Li, X. and Linde, F. and Linker, S. D. and Linley, J. N. and Littenberg, T. B. and Liu, J. and Liu, X. and Llorens-Monteagudo, M. and Lo, R. K. L. and Lockwood, A. and London, L. T. and Longo, A. and Lorenzini, M. and Loriette, V. and Lormand, M. and Losurdo, G. and Lough, J. D. and Lousto, C. O. and Lovelace, G. and Lück, H. and Lumaca, D. and Lundgren, A. P. and Ma, Y. and Macas, R. and Macfoy, S. and MacInnis, M. and Macleod, D. M. and MacMillan, I. A. O. and Macquet, A. and Hernandez, I. Maga{\~{n}}a and Maga{\~{n}}a-Sandoval, F. and Magee, R. M. and Majorana, E. and Maksimovic, I. and Malik, A. and Man, N. and Mandic, V. and Mangano, V. and Mansell, G. L. and Manske, M. and Mantovani, M. and Mapelli, M. and Marchesoni, F. and Marion, F. and M{\'{a}}rka, S. and M{\'{a}}rka, Z. and Markakis, C. and Markosyan, A. S. and Markowitz, A. and Maros, E. and Marquina, A. and Marsat, S. and Martelli, F. and Martin, I. W. and Martin, R. M. and Martinez, V. and Martynov, D. V. and Masalehdan, H. and Mason, K. and Massera, E. and Masserot, A. and Massinger, T. J. and Masso-Reid, M. and Mastrogiovanni, S. and Matas, A. and Matichard, F. and Mavalvala, N. and Maynard, E. and McCann, J. J. and McCarthy, R. and McClelland, D. E. and McCormick, S. and McCuller, L. and McGuire, S. C. and McIsaac, C. and McIver, J. and McManus, D. J. and McRae, T. and McWilliams, S. T. and Meacher, D. and Meadors, G. D. and Mehmet, M. and Mehta, A. K. and Villa, E. Mejuto and Melatos, A. and Mendell, G. and Mercer, R. A. and Mereni, L. and Merfeld, K. and Merilh, E. L. and Merritt, J. D. and Merzougui, M. and Meshkov, S. and Messenger, C. and Messick, C. and Metzdorff, R. and Meyers, P. M. and Meylahn, F. and Mhaske, A. and Miani, A. and Miao, H. and Michaloliakos, I. and Michel, C. and Middleton, H. and Milano, L. and Miller, A. L. and Miller, S. and Millhouse, M. and Mills, J. C. and Milotti, E. and Milovich-Goff, M. C. and Minazzoli, O. and Minenkov, Y. and Mishkin, A. and Mishra, C. and Mistry, T. and Mitra, S. and Mitrofanov, V. P. and Mitselmakher, G. and Mittleman, R. and Mo, G. and Mogushi, K. and Mohapatra, S. R. P. and Mohite, S. R. and Molina-Ruiz, M. and Mondin, M. and Montani, M. and Moore, C. J. and Moraru, D. and Morawski, F. and Moreno, G. and Morisaki, S. and Mours, B. and Mow-Lowry, C. M. and Mozzon, S. and Muciaccia, F. and Mukherjee, Arunava and Mukherjee, D. and Mukherjee, S. and Mukherjee, Subroto and Mukund, N. and Mullavey, A. and Munch, J. and Mu{\~{n}}iz, E. A. and Murray, P. G. and Nagar, A. and Nardecchia, I. and Naticchioni, L. and Nayak, R. K. and Neil, B. F. and Neilson, J. and Nelemans, G. and Nelson, T. J. N. and Nery, M. and Neunzert, A. and Ng, K. Y. and Ng, S. and Nguyen, C. and Nguyen, P. and Nichols, D. and Nichols, S. A. and Nissanke, S. and Nocera, F. and Noh, M. and North, C. and Nothard, D. and Nuttall, L. K. and Oberling, J. and O'Brien, B. D. and Oganesyan, G. and Ogin, G. H. and Oh, J. J. and Oh, S. H. and Ohme, F. and Ohta, H. and Okada, M. A. and Oliver, M. and Olivetto, C. and Oppermann, P. and Oram, Richard J. and O'Reilly, B. and Ormiston, R. G. and Ortega, L. F. and O'Shaughnessy, R. and Ossokine, S. and Osthelder, C. and Ottaway, D. J. and Overmier, H. and Owen, B. J. and Pace, A. E. and Pagano, G. and Page, M. A. and Pagliaroli, G. and Pai, A. and Pai, S. A. and Palamos, J. R. and Palashov, O. and Palomba, C. and Pan, H. and Panda, P. K. and Pang, P. T. H. and Pankow, C. and Pannarale, F. and Pant, B. C. and Paoletti, F. and Paoli, A. and Parida, A. and Parker, W. and Pascucci, D. and Pasqualetti, A. and Passaquieti, R. and Passuello, D. and Patricelli, B. and Payne, E. and Pearlstone, B. L. and Pechsiri, T. C. and Pedersen, A. J. and Pedraza, M. and Pele, A. and Penn, S. and Perego, A. and Perez, C. J. and P{\'{e}}rigois, C. and Perreca, A. and Perri{\`{e}}s, S. and Petermann, J. and Pfeiffer, H. P. and Phelps, M. and Phukon, K. S. and Piccinni, O. J. and Pichot, M. and Piendibene, M. and Piergiovanni, F. and Pierro, V. and Pillant, G. and Pinard, L. and Pinto, I. M. and Piotrzkowski, K. and Pirello, M. and Pitkin, M. and Plastino, W. and Poggiani, R. and Pong, D. Y. T. and Ponrathnam, S. and Popolizio, P. and Porter, E. K. and Powell, J. and Prajapati, A. K. and Prasai, K. and Prasanna, R. and Pratten, G. and Prestegard, T. and Principe, M. and Prodi, G. A. and Prokhorov, L. and Punturo, M. and Puppo, P. and Pürrer, M. and Qi, H. and Quetschke, V. and Quinonez, P. J. and Raab, F. J. and Raaijmakers, G. and Radkins, H. and Radulesco, N. and Raffai, P. and Rafferty, H. and Raja, S. and Rajan, C. and Rajbhandari, B. and Rakhmanov, M. and Ramirez, K. E. and Ramos-Buades, A. and Rana, Javed and Rao, K. and Rapagnani, P. and Raymond, V. and Razzano, M. and Read, J. and Regimbau, T. and Rei, L. and Reid, S. and Reitze, D. H. and Rettegno, P. and Ricci, F. and Richardson, C. J. and Richardson, J. W. and Ricker, P. M. and Riemenschneider, G. and Riles, K. and Rizzo, M. and Robertson, N. A. and Robinet, F. and Rocchi, A. and Rodriguez-Soto, R. D. and Rolland, L. and Rollins, J. G. and Roma, V. J. and Romanelli, M. and Romano, R. and Romel, C. L. and Romero-Shaw, I. M. and Romie, J. H. and Rose, C. A. and Rose, D. and Rose, K. and Rosi{\'{n}}ska, D. and Rosofsky, S. G. and Ross, M. P. and Rowan, S. and Rowlinson, S. J. and Roy, P. K. and Roy, Santosh and Roy, Soumen and Ruggi, P. and Rutins, G. and Ryan, K. and Sachdev, S. and Sadecki, T. and Sakellariadou, M. and Salafia, O. S. and Salconi, L. and Saleem, M. and Samajdar, A. and Sanchez, E. J. and Sanchez, L. E. and Sanchis-Gual, N. and Sanders, J. R. and Santiago, K. A. and Santos, E. and Sarin, N. and Sassolas, B. and Sathyaprakash, B. S. and Sauter, O. and Savage, R. L. and Savant, V. and Sawant, D. and Sayah, S. and Schaetzl, D. and Schale, P. and Scheel, M. and Scheuer, J. and Schmidt, P. and Schnabel, R. and Schofield, R. M. S. and Schönbeck, A. and Schreiber, E. and Schulte, B. W. and Schutz, B. F. and Schwarm, O. and Schwartz, E. and Scott, J. and Scott, S. M. and Seidel, E. and Sellers, D. and Sengupta, A. S. and Sennett, N. and Sentenac, D. and Sequino, V. and Sergeev, A. and Setyawati, Y. and Shaddock, D. A. and Shaffer, T. and Shahriar, M. S. and Sharifi, S. and Sharma, A. and Sharma, P. and Shawhan, P. and Shen, H. and Shikauchi, M. and Shink, R. and Shoemaker, D. H. and Shoemaker, D. M. and Shukla, K. and ShyamSundar, S. and Siellez, K. and Sieniawska, M. and Sigg, D. and Singer, L. P. and Singh, D. and Singh, N. and Singha, A. and Singhal, A. and Sintes, A. M. and Sipala, V. and Skliris, V. and Slagmolen, B. J. J. and Slaven-Blair, T. J. and Smetana, J. and Smith, J. R. and Smith, R. J. E. and Somala, S. and Son, E. J. and Soni, S. and Sorazu, B. and Sordini, V. and Sorrentino, F. and Souradeep, T. and Sowell, E. and Spencer, A. P. and Spera, M. and Srivastava, A. K. and Srivastava, V. and Staats, K. and Stachie, C. and Standke, M. and Steer, D. A. and Steinke, M. and Steinlechner, J. and Steinlechner, S. and Steinmeyer, D. and Stevenson, S. and Stocks, D. and Stops, D. J. and Stover, M. and Strain, K. A. and Stratta, G. and Strunk, A. and Sturani, R. and Stuver, A. L. and Sudhagar, S. and Sudhir, V. and Summerscales, T. Z. and Sun, L. and Sunil, S. and Sur, A. and Suresh, J. and Sutton, P. J. and Swinkels, B. L. and Szczepa{\'{n}}czyk, M. J. and Tacca, M. and Tait, S. C. and Talbot, C. and Tanasijczuk, A. J. and Tanner, D. B. and Tao, D. and T{\'{a}}pai, M. and Tapia, A. and Martin, E. N. Tapia San and Tasson, J. D. and Taylor, R. and Tenorio, R. and Terkowski, L. and Thirugnanasambandam, M. P. and Thomas, M. and Thomas, P. and Thompson, J. E. and Thondapu, S. R. and Thorne, K. A. and Thrane, E. and Tinsman, C. L. and Saravanan, T. R. and Tiwari, Shubhanshu and Tiwari, S. and Tiwari, V. and Toland, K. and Tonelli, M. and Tornasi, Z. and Torres-Forn{\'{e}}, A. and Torrie, C. I. and e Melo, I. Tosta and Töyrä, D. and Trail, E. A. and Travasso, F. and Traylor, G. and Tringali, M. C. and Tripathee, A. and Trovato, A. and Trudeau, R. J. and Tsang, K. W. and Tse, M. and Tso, R. and Tsukada, L. and Tsuna, D. and Tsutsui, T. and Turconi, M. and Ubhi, A. S. and Udall, R. and Ueno, K. and Ugolini, D. and Unnikrishnan, C. S. and Urban, A. L. and Usman, S. A. and Utina, A. C. and Vahlbruch, H. and Vajente, G. and Valdes, G. and Valentini, M. and van Bakel, N. and van Beuzekom, M. and van den Brand, J. F. J. and Broeck, C. Van Den and Vander-Hyde, D. C. and van der Schaaf, L. and Heijningen, J. V. Van and van Veggel, A. A. and Vardaro, M. and Varma, V. and Vass, S. and Vas{\'{u}}th, M. and Vecchio, A. and Vedovato, G. and Veitch, J. and Veitch, P. J. and Venkateswara, K. and Venugopalan, G. and Verkindt, D. and Veske, D. and Vetrano, F. and Vicer{\'{e}}, A. and Viets, A. D. and Vinciguerra, S. and Vine, D. J. and Vinet, J.-Y. and Vitale, S. and Vivanco, Francisco Hernandez and Vo, T. and Vocca, H. and Vorvick, C. and Vyatchanin, S. P. and Wade, A. R. and Wade, L. E. and Wade, M. and Walet, R. and Walker, M. and Wallace, G. S. and Wallace, L. and Walsh, S. and Wang, J. Z. and Wang, S. and Wang, W. H. and Ward, R. L. and Warden, Z. A. and Warner, J. and Was, M. and Watchi, J. and Weaver, B. and Wei, L.-W. and Weinert, M. and Weinstein, A. J. and Weiss, R. and Wellmann, F. and Wen, L. and We{\ss}els, P. and Westhouse, J. W. and Wette, K. and Whelan, J. T. and Whiting, B. F. and Whittle, C. and Wilken, D. M. and Williams, D. and Willis, J. L. and Willke, B. and Winkler, W. and Wipf, C. C. and Wittel, H. and Woan, G. and Woehler, J. and Wofford, J. K. and Wong, C. and Wright, J. L. and Wu, D. S. and Wysocki, D. M. and Xiao, L. and Yamamoto, H. and Yang, L. and Yang, Y. and Yang, Z. and Yap, M. J. and Yazback, M. and Yeeles, D. W. and Yu, Hang and Yu, Haocun and Yuen, S. H. R. and Zadro{\.{z}}ny, A. K. and Zadro{\.{z}}ny, A. and Zanolin, M. and Zelenova, T. and Zendri, J.-P. and Zevin, M. and Zhang, J. and Zhang, L. and Zhang, T. and Zhao, C. and Zhao, G. and Zhou, M. and Zhou, Z. and Zhu, X. J. and Zimmerman, A. B. and Zucker, M. E. and Zweizig, J.},
  journal   = {Physical Review D},
  title     = {{GW}190412: Observation of a binary-black-hole coalescence with asymmetric masses},
  year      = {2020},
  month     = aug,
  number    = {4},
  volume    = {102},
  doi       = {10.1103/physrevd.102.043015},
  publisher = {American Physical Society ({APS})},
}

@Article{Dergachev2020b,
  author    = {Dergachev, V. and Papa, M. A.},
  journal   = {Physical Review Letters},
  title     = {{Results from the First All-Sky Search for Continuous Gravitational Waves from Small-Ellipticity Sources}},
  year      = {2020},
  month     = oct,
  number    = {17},
  pages     = {171101},
  volume    = {125},
  doi       = {10.1103/physrevlett.125.171101},
  publisher = {American Physical Society ({APS})},
}

@Article{Dergachev2020a,
  author    = {Dergachev, V. and Papa, M. A.},
  journal   = {Physical Review D},
  title     = {{Results from an extended Falcon all-sky survey for continuous gravitational waves}},
  year      = {2020},
  month     = jan,
  number    = {2},
  pages     = {022001},
  volume    = {101},
  doi       = {10.1103/physrevd.101.022001},
  publisher = {American Physical Society ({APS})},
}

@Book{Maggiore2007,
  author    = {Maggiore, M.},
  publisher = {Oxford University Press},
  title     = {{Gravitational Waves. Volume 1}},
  year      = {2007},
  isbn      = {0198570740},
  month     = oct,
  ean       = {9780198570745},
}

@Article{Jaranowski1998,
  author    = {Jaranowski, P. and Królak, A. and Schutz, B. F.},
  journal   = {Physical Review D},
  title     = {Data analysis of gravitational-wave signals from spinning neutron stars: The signal and its detection},
  year      = {1998},
  month     = aug,
  number    = {6},
  pages     = {063001},
  volume    = {58},
  doi       = {10.1103/physrevd.58.063001},
  publisher = {American Physical Society ({APS})},
}
\end{document}